\documentclass[12pt,a4paper]{article}
\usepackage{jheplikemod,tocloft,amsmath,textcomp,mathrsfs,mathtools,graphicx,bbm,epigraph,tikz}
\usepackage[bb=libus]{mathalpha}
\usepackage[T1]{fontenc}
\usepackage{lmodern}
\hypersetup{pdftitle={Maximal Weight Masters Two Loops},pdfcreator={},linkcolor=[rgb]{0.15,0.35,0.75},colorlinks=true,citecolor=[rgb]{0.675,0,0.2},urlcolor=[rgb]{0.15,0.35,0.65}}
\graphicspath{{figures/}}
\let\olditemize\itemize\renewcommand{\itemize}{\vspace{-9pt}\olditemize\setlength{\itemsep}{0.5pt}\setlength{\parskip}{0pt}\setlength{\parsep}{-2pt}}
\let\oldenumerate\enumerate\renewcommand{\enumerate}{\vspace{-4pt}\oldenumerate\setlength{\itemsep}{1pt}\setlength{\parskip}{0pt}\setlength{\parsep}{0pt}}
\makeatletter\renewcommand\section{\addtocontents{toc}{\protect\addvspace{-2.25\p@}}\@startsection {section}{1}{\z@}{-0.0ex \@plus .2ex \@minus 0.2ex}{1ex \@plus.1ex\@minus .5ex}{\normalfont\large\bfseries}}
\renewcommand\subsection{\addtocontents{toc}{\protect\addvspace{-2.5\p@}}\@startsection {subsection}{1}{\z@}{0.5ex \@plus .2ex \@minus 0.2ex}{0.75ex \@plus.1ex\@minus .5ex}{\normalfont\bfseries}}
\renewcommand\subsubsection{\addtocontents{toc}{\protect\addvspace{-2.5\p@}}\@startsection {subsubsection}{1}{\z@}{0.5ex \@plus .2ex \@minus 0.2ex}{0.75ex \@plus.1ex\@minus .5ex}{\normalfont\bfseries}}
\definecolor{rindou1}{rgb}{0.4431,0.2862,0.7960}
\definecolor{rindou2}{rgb}{0.0078,0.1215,0.4392}
\definecolor{lapis}{rgb}{0.0.0470,0.2941,0.5568}
\definecolor{emerald}{rgb}{0.31, 0.78, 0.47}
\definecolor{pinegreen}{rgb}{0.0, 0.47, 0.44}
\definecolor{jade}{rgb}{0.0, 0.66, 0.42}
\definecolor{teal}{rgb}{0.0, 0.5, 0.5}
\definecolor{hblue}{rgb}{0,0,0.575}
\definecolor{hred}{rgb}{0.575,0.0,0.225}
\definecolor{hgreen}{rgb}{0.0,0.4,0.2}
\definecolor{hteal}{rgb}{0.0,0.445,0.6451}
\definecolor{totalCount}{rgb}{0,0,0.575}
\definecolor{topCount}{rgb}{0.575,0.0,0.225}
\definecolor{contactCount}{rgb}{0.3,0.4,0.575}
\definecolor{topCountOdd}{rgb}{0.775,0.4,0.425}
\definecolor{dim}{rgb}{0.55,0.55,0.55}
\definecolor{deemph}{rgb}{0.25,0.25,0.25}
\colorlet{contactCount}{hteal}
\colorlet{topCountOdd}{topCount!45!white}
\colorlet{contactCountOdd}{contactCount!45!white}
\colorlet{oddCount}{black!45!white}
\colorlet{totalCountOdd}{totalCount!45!white}
\colorlet{contactCountEven}{contactCount!75!black}
\colorlet{totalCountEven}{totalCount!75!black}
\colorlet{topCountEven}{topCount!75!black}
\definecolor{varcolor}{rgb}{0.1,0.55,0.25}
\definecolor{functioncolor}{rgb}{0.1,0.35,0.75}

\newcounter{topologyIndex}
\newcounter{topology}
\newcommand{\addintegers}[2]{\inteval{#1 + #2}}
\newcommand{\eq}[1]{\vspace{-0.5pt}\begin{equation}#1\vspace{-0.5pt}\end{equation}}

\newcommand{\fwbox}[2]{\text{\makebox[#1][c]{$\hspace{-150pt}\displaystyle#2\hspace{-150pt}$}}}
\newcommand{\fwboxL}[2]{\text{\makebox[#1][l]{$#2$}}}
\newcommand{\fwboxR}[2]{\text{\makebox[#1][r]{$#2$}}}
\newcommand{\equivR}{\fwbox{14.5pt}{\hspace{-0pt}\fwboxR{0pt}{\raisebox{0.47pt}{\hspace{1.25pt}:\hspace{-4pt}}}=\fwboxL{0pt}{}}}
\newcommand{\equivL}{\fwbox{14.5pt}{\fwboxR{0pt}{}=\fwboxL{0pt}{\raisebox{0.47pt}{\hspace{-4pt}:\hspace{1.25pt}}}}}
\newcommand{\fig}[3]{\raisebox{#1}{$\begin{array}{@{}c@{}}\includegraphics[scale=#2]{#3}\end{array}$}}

\newcommand{\bigger}[1]{\raisebox{-0.95pt}{\scalebox{1.25}{$#1$}}}

\renewcommand{\phi}{\varphi}
\renewcommand{\bar}{\overline}
\renewcommand{\hat}{\widehat}

\renewcommand{\tilde}{\widetilde}
\newcommand{\ab}[1]{\langle #1\rangle}

\newcommand{\newcap}{\mathrm{\raisebox{0.75pt}{{$\,\bigcap\,$}}}}
\newcommand{\tcap}{\scalebox{0.9}{$\!\newcap\!$}}

\newcommand{\x}[2]{{\color{black}(}\hspace{-0.85pt}{\color{black}#1}\hspace{-0.25pt}{\color{black}|}\hspace{-0.25pt}{\color{black}#2}\hspace{-0.85pt}{\color{black})}}
\newcommand{\dbar}{\fwboxL{7.2pt}{\raisebox{4.5pt}{\fwboxL{0pt}{\scalebox{1.5}[0.75]{\hspace{1.25pt}\text{-}}}}d}}
\renewcommand{\r}[1]{{\color{hred}#1}}
\renewcommand{\b}[1]{{\color{hblue}#1}}
\newcommand{\g}[1]{{\color{hgreen}#1}}
\renewcommand{\t}[1]{{\color{hteal}#1}}

\renewcommand{\brace}[1]{\hspace{-2pt}{\texttt{[}}#1{\texttt{]}}\hspace{-1pt}}
\newcommand{\pattern}{{\color{varcolor}\rule[-1.05pt]{7.5pt}{.65pt}}}
\newcommand{\patternTwo}{{\color{varcolor}\rule[-1.05pt]{12pt}{.65pt}}}

\renewcommand{\inf}{{X}}
\newcommand{\indices}[2]{{\hspace{-0.0pt}}^{\smash{#1}}_{\phantom{\smash{#1}}\smash{#2}}}
\newcommand{\fun}[1]{\text{\texttt{\textbf{\texttt{\color{functioncolor}#1}}}}}
\newcommand{\uscore}{\rule[-1.05pt]{7.5pt}{.75pt}}
\newcommand{\var}[1]{{{\color{varcolor}{\sl#1}}}}

\newcommand{\cfig}[1]{\begin{array}{@{}c@{}}\includegraphics[scale=1]{#1}\end{array}}

\newcommand{\rankBreakdownTable}[5]{\begin{array}{@{}c@{}c@{$\;$}c@{}c@{}c@{}c@{}c@{}c@{}}
\fwbox{52pt}{\begin{array}{@{}c@{}}\text{{\footnotesize master}}\\[-9pt]\text{{\footnotesize numerators}}\\[-9pt]~\\[-5pt]\end{array}}&\fwboxR{24pt}{\begin{array}{@{}c@{}}~\\[-9pt]\text{{\small {\color{totalCount}rank}}}\\[-9pt]~\\[-5pt]\end{array}}
&\fwbox{16pt}{\begin{array}{@{}c@{}}
\text{{\tiny\phantom{-}{\color{topCount}top-}}}\\[-11pt]\text{{\tiny {\color{topCount}level}}}\end{array}}&\fwbox{5pt}{\text{{\tiny{+}}}}&\fwbox{22pt}{\begin{array}{@{}c@{}}\text{{\tiny\phantom{-}{\color{contactCount}contact}}}\\[-11pt]\text{{\tiny {\color{contactCount}terms}}}\end{array}}\\[-3pt]\cline{3-5}\\[-18pt]
\fwbox{52pt}{#1}&\fwbox{24pt}{{\color{totalCount}\addintegers{#2}{\addintegers{#3}{\addintegers{#4}{#5}}}}}&\fwboxL{0pt}{\rule[-2pt]{0.5pt}{11pt}\hspace{-1pt}}\fwbox{16pt}{{\color{topCount}\addintegers{#2}{#4}}}&\fwbox{5pt}{+}&\fwbox{22pt}{{\color{contactCount}\addintegers{#3}{#5}}}\fwboxR{0pt}{\hspace{-1pt}\rule[-2pt]{0.45pt}{11pt}}&&\multicolumn{1}{c}{}
\\[-3pt]
\cline{3-7}\\[-18pt]
&&\fwboxL{0pt}{\rule[-2pt]{0.5pt}{11pt}\hspace{-1pt}}\fwbox{16pt}{{\color{topCountEven}#2}}&\fwbox{5pt}{+}&\fwbox{22pt}{{\color{contactCountEven}#3}}\fwboxR{0pt}{\hspace{-1pt}\rule[-2pt]{0.45pt}{11pt}}&\fwbox{16pt}{{\color{totalCountEven}\addintegers{#2}{#3}}}&\fwboxL{26pt}{\begin{array}{@{}c@{}}\\[-21pt]\text{{\tiny\phantom{-}{\color{totalCountEven}parity-}}}\\[-12pt]\text{{\tiny {\color{totalCountEven}even}}}\end{array}\raisebox{0.4pt}{\hspace{-3pt}\text{{\tiny\color{totalCountEven}\phantom{}}}}\hspace{4.5pt}{\color{black}\hspace{-0.0pt}\rule[-2.185pt]{0.45pt}{11.305pt}}}\\[-6pt]
&&\fwboxL{0pt}{\rule[-2pt]{0.5pt}{11pt}\hspace{-1pt}}\fwbox{16pt}{{\color{topCountOdd}#4}}&\fwbox{5pt}{{\color{oddCount}+}}&\fwbox{22pt}{{\color{contactCountOdd}#5}}\fwboxR{0pt}{\hspace{-1pt}\rule[-2pt]{0.5pt}{11pt}}&\fwbox{16pt}{{\color{totalCountOdd}\addintegers{#4}{#5}}}&\fwboxL{26pt}{\begin{array}{@{}c@{}}\\[-21pt]\text{{\tiny\phantom{-}{\color{totalCountOdd}parity-}}}\\[-12pt]\text{{\tiny {\color{totalCountOdd}odd}}}\end{array}\raisebox{0.4pt}{\hspace{-3pt}\text{{\tiny\color{totalCountOdd}\phantom{}}}}\hspace{4.5pt}{\color{black}\rule[-2.185pt]{0.45pt}{11.305pt}}}\\[-3pt]\cline{3-7}
\end{array}}

\newcommand{\dataTableTri}[1]{\stepcounter{topologyIndex}\setcounter{topology}{#1}
\renewcommand{\theequation}{$\hspace{-0pt}I^{\fwbox{15pt}{\![\r{3}]}}_{\fwbox{15pt}{\rule{0pt}{6.5pt}\r{\arabic{topology}}}}\hspace{-2pt}$}
\vspace{-10pt}\eq{\hspace{-80pt}\hspace{-36pt}\raisebox{-2pt}{$\cfig{3_#1}$}\hspace{5pt}\raisebox{-3.pt}{$\cfig{3_#1_table}$}\hspace{20pt}\hspace{-80pt}\vspace{-28pt}\label{int3|#1}}
\renewcommand{\theequation}{$\Omega^{\fwbox{15pt}{\![\r{3}]}}_{\fwbox{15pt}{\rule{0pt}{6.5pt}\r{\arabic{topology}}}}\hspace{-2pt}$}
~\\[10pt]\noindent{\textbf{\b{Defining Contours}:}}\vspace{-20pt}\eq{\hspace{5pt}\begin{array}{@{}l@{}}\begin{array}{@{}c@{}c@{}c@{}c@{}c@{}c}~\\[-15pt]
\fwboxR{0pt}{\raisebox{2pt}{$\left\{\rule{0pt}{22pt}\right.$}\hspace{-8.5pt}}\hspace{5pt}\cfig{3_#1_1}\hspace{2pt}\fwboxL{0pt}{\hspace{-4.5pt}\raisebox{2pt}{$\left.\rule{0pt}{22pt}\right\}$}}\\[-10pt]
\phantom{{}_{1}}\hspace{2pt}\,\Omega_1\\[-7pt]
\end{array}
\end{array}\vspace{-0pt}\label{omega3|#1}}}

\newcommand{\topLabel}{}
\newcommand{\topLabelM}{}
\newcommand{\topLabelE}{}

\newcommand{\dataTableQAB}[4]{
\renewcommand{\theequation}{$\hspace{-0pt}I^{[\r{2}\g{1}\b{2}]}_{\fwbox{20pt}{\rule{0pt}{6.5pt}\r{#1}\g{#2}\b{#3}\g{#4}}}\hspace{-2pt}$}
\vspace{-10pt}\eq{\hspace{-80pt}\hspace{-36pt}\raisebox{-2pt}{$\cfig{212_#1#2#3#4}$}\hspace{5pt}\raisebox{-3.pt}{$\cfig{212_#1#2#3#4_table}$}\hspace{20pt}\hspace{-80pt}\vspace{-28pt}\label{int212|#1#2#3#4}}
\renewcommand{\theequation}{$\Omega^{[\r{2}\g{1}\b{2}]}_{\fwbox{20pt}{\rule{0pt}{6.5pt}\r{#1}\g{#2}\b{#3}\g{#4
}}}\hspace{-2pt}$}
~\\[6pt]\noindent{\textbf{\b{Defining Contours}:}}\vspace{-20pt}\eq{\hspace{5pt}\begin{array}{@{}l@{}}\begin{array}{@{}c@{}c@{}c@{}c@{}c@{}c}~\\[-15pt]
\fwboxR{0pt}{\raisebox{2pt}{$\left\{\rule{0pt}{14pt}\right.$}\hspace{-8.5pt}}\hspace{2pt}\cfig{212_#1#2#3#4_1}\hspace{-2pt}\fwboxL{0pt}{\hspace{-4.5pt}\raisebox{2pt}{$\left.\rule{0pt}{14pt}\right\}$}}\\[-10pt]
\phantom{{}_{1}}\hspace{2pt}\,\Omega_1\\[-7pt]
\end{array}
\end{array}\vspace{-0pt}\label{omega212|#1#2#3#4}}}

\newcommand{\dataTableQA}[4]{\ifthenelse{\equal{#2}{1}}{\renewcommand{\topLabelM}{\g{\smash{\hspace{0.4pt}\bar{\hspace{-0.4pt}1\hspace{-0.4pt}}\hspace{0.4pt}}}}}{\renewcommand{\topLabelM}{\g{#2}}}
\ifthenelse{\equal{#4}{1}}{\renewcommand{\topLabelE}{\g{\smash{\hspace{0.4pt}\bar{\hspace{-0.4pt}1\hspace{-0.4pt}}\hspace{0.4pt}}}}}{\renewcommand{\topLabelE}{\g{#4}}}\renewcommand{\topLabel}{\fwbox{20pt}{\rule{0pt}{6.5pt}\r{#1}\topLabelM\b{#3}\topLabelE}}
\renewcommand{\theequation}{$\hspace{-0pt}I^{\fwbox{20pt}{\!\![\r{2}\g{1}\b{2}]}}_{\topLabel}\hspace{-1pt}$}
\vspace{-12pt}\eq{\hspace{-80pt}\hspace{-36pt}\raisebox{-2pt}{$\cfig{212_#1#2#3#4}$}\hspace{5pt}\raisebox{-3.pt}{$\cfig{212_#1#2#3#4_table}$}\hspace{20pt}\hspace{-80pt}\vspace{-28pt}\label{int212|#1#2#3#4}}
\renewcommand{\theequation}{$\hspace{-0pt}\Omega^{\fwbox{20pt}{\!\![\r{2}\g{1}\b{2}]}}_{\topLabel}\hspace{-1pt}$}
~\\[6pt]\noindent{\textbf{\b{Defining Contours}:}}\vspace{-24pt}\eq{\hspace{5pt}\begin{array}{@{}l@{}}\begin{array}{@{}c@{}c@{}c@{}c@{}c@{}c}~\\[-15pt]
\fwboxR{0pt}{\raisebox{2pt}{$\left\{\rule{0pt}{14pt}\right.$}\hspace{-8.5pt}}\hspace{2pt}\cfig{212_#1#2#3#4_1}\hspace{-2pt}\fwboxL{0pt}{\hspace{-4.5pt}\raisebox{2pt}{$\left.\rule{0pt}{14pt}\right\}$}}\\[-10pt]
\phantom{{}_{1}}\hspace{2pt}\,\Omega_1\\[-7pt]
\end{array}
\end{array}\vspace{-0pt}\label{omega212|#1#2#3#4}}}

\newcommand{\dataTableQB}[4]{\ifthenelse{\equal{#2}{1}}{\renewcommand{\topLabelM}{\g{\smash{\hspace{0.4pt}\bar{\hspace{-0.4pt}1\hspace{-0.4pt}}\hspace{0.4pt}}}}}{\renewcommand{\topLabelM}{\g{#2}}}
\ifthenelse{\equal{#4}{1}}{\renewcommand{\topLabelE}{\g{\smash{\hspace{0.4pt}\bar{\hspace{-0.4pt}1\hspace{-0.4pt}}\hspace{0.4pt}}}}}{\renewcommand{\topLabelE}{\g{#4}}}\renewcommand{\topLabel}{\fwbox{25pt}{\rule{0pt}{6.5pt}\r{#1}\topLabelM\b{#3}\topLabelE}}
\renewcommand{\theequation}{$\hspace{-0pt}I^{\fwbox{25pt}{\!\![\r{3}\g{1}\b{2}]}}_{\topLabel}\hspace{-1pt}$}
\vspace{-12pt}\eq{\hspace{-120pt}\hspace{-10pt}\raisebox{-0pt}{$\cfig{312_#1#2#3#4}$}\hspace{5pt}\raisebox{-10pt}{$\cfig{312_#1#2#3#4_table}$}\hspace{20pt}\hspace{-100pt}\vspace{-22pt}\label{int312|#1#2#3#4}}
\renewcommand{\theequation}{$\Omega^{\fwbox{25pt}{\!\![\r{3}\g{1}\b{2}]}}_{\topLabel}\hspace{-1pt}$}
~\\[-16pt]\noindent{\textbf{\b{Defining Contours}:}}\\[-12pt]
\vspace{-10pt}\eq{\fwbox{00pt}{\hspace{05pt}\begin{array}{@{}l@{}}\begin{array}{@{}c@{}c@{}c@{}c@{}c@{}c}~\\[-15pt]
\fwboxR{0pt}{\raisebox{2pt}{$\left\{\rule{0pt}{24pt}\right.$}\hspace{-8.5pt}}\hspace{5pt}\cfig{312_#1#2#3#4_1}\hspace{5pt}\fwboxL{0pt}{,}&\hspace{5pt}\cfig{312_#1#2#3#4_2}\hspace{5pt}\fwboxL{0pt}{,}&\hspace{5pt}\cfig{312_#1#2#3#4_3}\hspace{2pt}\fwboxL{0pt}{\hspace{-4.5pt}\raisebox{2pt}{$\left.\rule{0pt}{24pt}\right\}$}}\\[-10pt]
\phantom{{}_{1}}\Omega_1&\phantom{{}_{2}}\Omega_2&\phantom{{}_{3}}\Omega_3\\[-7pt]
\end{array}
\end{array}}\label{omega312|#1#2#3#4}\vspace{-0pt}
}}

\newcommand{\dataTableQC}[4]{\ifthenelse{\equal{#2}{1}}{\renewcommand{\topLabelM}{\g{\smash{\hspace{0.4pt}\bar{\hspace{-0.4pt}1\hspace{-0.4pt}}\hspace{0.4pt}}}}}{\renewcommand{\topLabelM}{\g{#2}}}
\ifthenelse{\equal{#4}{1}}{\renewcommand{\topLabelE}{\g{\smash{\hspace{0.4pt}\bar{\hspace{-0.4pt}1\hspace{-0.4pt}}\hspace{0.4pt}}}}}{\renewcommand{\topLabelE}{\g{#4}}}\renewcommand{\topLabel}{\fwbox{25pt}{\rule{0pt}{6.5pt}\r{#1}\topLabelM\b{#3}\topLabelE}}
\renewcommand{\theequation}{$\hspace{-0pt}I^{\fwbox{25pt}{[\r{3}\g{1}\b{3}]}}_{\topLabel}\hspace{-1pt}$}
\vspace{-5pt}\eq{\hspace{-120pt}\hspace{-10pt}\raisebox{40pt}{$\cfig{313_#1#2#3#4}$}\hspace{5pt}\raisebox{-2pt}{$\cfig{313_#1#2#3#4_table}$}\hspace{20pt}\hspace{-100pt}\vspace{-22pt}\label{int313|#1#2#3#4}}
\renewcommand{\theequation}{$\Omega^{\fwbox{25pt}{[\r{3}\g{1}\b{3}]}}_{\topLabel}\hspace{-1pt}$}
~\\[-5pt]\noindent{\textbf{\b{Defining Contours}:}}\\[-10pt]
\eq{\hspace{-100pt}\fwboxL{280pt}{\hspace{-80pt}\begin{array}{@{}l@{}}\begin{array}{@{}c@{}c@{}c@{}c@{}c@{}c}
\fwboxR{0pt}{\raisebox{2pt}{$\left\{\rule{0pt}{20pt}\right.$}\hspace{-8.5pt}}\cfig{313_#1#2#3#4_1}\fwboxL{0pt}{,}&\cfig{313_#1#2#3#4_2}\fwboxL{0pt}{,}&\cfig{313_#1#2#3#4_3}\fwboxL{0pt}{,}&\cfig{313_#1#2#3#4_4}\fwboxL{0pt}{,}&\cfig{313_#1#2#3#4_5}\fwboxL{0pt}{,}&\cfig{313_#1#2#3#4_6}\fwboxL{0pt}{,}\\[-10pt]
\phantom{{}_{1}}\Omega_1&\phantom{{}_{2}}\Omega_2&\phantom{{}_{3}}\Omega_3&\phantom{{}_{4}}\Omega_4&\phantom{{}_{5}}\Omega_5&\phantom{{}_{6}}\Omega_6
\end{array}\\[-5pt]
\begin{array}{@{}c@{}c@{}c@{}c@{}c@{}c}
\cfig{313_#1#2#3#4_7}\fwboxL{0pt}{\hspace{0pt},}&\cfig{313_#1#2#3#4_8}\fwboxL{0pt}{\hspace{-8.5pt}\raisebox{2pt}{$\left.\rule{0pt}{20pt}\right\}$}}\\[-10pt]
\phantom{{}_{1}}\Omega_7&\phantom{{}_{8}}\Omega_8
\end{array}\\[45pt]~
\end{array}}\label{omega313|#1#2#3#4}\vspace{-45pt}
\hspace{-100pt}}}

\newcommand{\dataTableQCB}[4]{\ifthenelse{\equal{#2}{1}}{\renewcommand{\topLabelM}{\g{\smash{\hspace{0.4pt}\bar{\hspace{-0.4pt}1\hspace{-0.4pt}}\hspace{0.4pt}}}}}{\renewcommand{\topLabelM}{\g{#2}}}
\ifthenelse{\equal{#4}{1}}{\renewcommand{\topLabelE}{\g{\smash{\hspace{0.4pt}\bar{\hspace{-0.4pt}1\hspace{-0.4pt}}\hspace{0.4pt}}}}}{\renewcommand{\topLabelE}{\g{#4}}}\renewcommand{\topLabel}{\fwbox{25pt}{\rule{0pt}{6.5pt}\r{#1}\topLabelM\b{#3}\topLabelE}}
\renewcommand{\theequation}{$\hspace{-0pt}I^{\fwbox{25pt}{\!\![\r{3}\g{1}\b{3}]}}_{\topLabel}\hspace{-1pt}$}
\vspace{-5pt}\eq{\hspace{-120pt}\hspace{-10pt}\raisebox{40pt}{$\cfig{313_#1#2#3#4}$}\hspace{5pt}\raisebox{-2pt}{$\cfig{313_#1#2#3#4_table}$}\hspace{20pt}\hspace{-100pt}\vspace{-22pt}\label{int313|#1#2#3#4}}
\renewcommand{\theequation}{$\Omega^{\fwbox{25pt}{\!\![\r{3}\g{1}\b{3}]}}_{\topLabel}\hspace{-1pt}$}
~\\[-5pt]\noindent{\textbf{\b{Defining Contours}:}}\\[-10pt]
\eq{\hspace{-100pt}\fwboxL{280pt}{\hspace{-0pt}\hspace{-80pt}\begin{array}{@{}l@{}}\begin{array}{@{}c@{}c@{}c@{}c@{}c@{}c}
\fwboxR{0pt}{\raisebox{2pt}{$\left\{\rule{0pt}{20pt}\right.$}\hspace{-8.5pt}}\cfig{313_#1#2#3#4_1}\fwboxL{0pt}{,}&\cfig{313_#1#2#3#4_2}\fwboxL{0pt}{,}&\cfig{313_#1#2#3#4_3}\fwboxL{0pt}{,}&\cfig{313_#1#2#3#4_4}\fwboxL{0pt}{,}&\cfig{313_#1#2#3#4_5}\fwboxL{0pt}{,}\\[-10pt]
\phantom{{}_{1}}\Omega_1&\phantom{{}_{2}}\Omega_2&\phantom{{}_{3}}\Omega_3&\phantom{{}_{4}}\Omega_4&\phantom{{}_{5}}\Omega_5
\end{array}\\[-5pt]
\begin{array}{@{}c@{}c@{}c@{}c@{}c@{}c}
\cfig{313_#1#2#3#4_6}\fwboxL{0pt}{\hspace{0pt},}&\cfig{313_#1#2#3#4_7}\fwboxL{0pt}{\hspace{0pt},}&\cfig{313_#1#2#3#4_8}\fwboxL{0pt}{\hspace{-8.5pt}\raisebox{2pt}{$\left.\rule{0pt}{20pt}\right\}$}}\\[-10pt]
\phantom{{}_{6}}\Omega_6&\phantom{{}_{1}}\Omega_7&\phantom{{}_{8}}\Omega_8
\end{array}\\[45pt]~
\end{array}\hspace{-100pt}}\hspace{-100pt}\label{omega313|#1#2#3#4}\vspace{-45pt}
}}

\newcommand{\dataTableQCC}[4]{\ifthenelse{\equal{#2}{1}}{\renewcommand{\topLabelM}{\g{\smash{\hspace{0.4pt}\bar{\hspace{-0.4pt}1\hspace{-0.4pt}}\hspace{0.4pt}}}}}{\renewcommand{\topLabelM}{\g{#2}}}
\ifthenelse{\equal{#4}{1}}{\renewcommand{\topLabelE}{\g{\smash{\hspace{0.4pt}\bar{\hspace{-0.4pt}1\hspace{-0.4pt}}\hspace{0.4pt}}}}}{\renewcommand{\topLabelE}{\g{#4}}}\renewcommand{\topLabel}{\fwbox{25pt}{\rule{0pt}{6.5pt}\r{#1}\topLabelM\b{#3}\topLabelE}}
\renewcommand{\theequation}{$\hspace{-0pt}I^{\fwbox{25pt}{\!\![\r{3}\g{1}\b{3}]}}_{\topLabel}\hspace{-1pt}$}
\vspace{-5pt}\eq{\hspace{-120pt}\hspace{-10pt}\raisebox{40pt}{$\cfig{313_#1#2#3#4}$}\hspace{5pt}\raisebox{-2pt}{$\cfig{313_#1#2#3#4_table}$}\hspace{20pt}\hspace{-100pt}\vspace{-22pt}\label{int313|#1#2#3#4}}
\renewcommand{\theequation}{$\Omega^{\fwbox{25pt}{\!\![\r{3}\g{1}\b{3}]}}_{\topLabel}\hspace{-1pt}$}
~\\[-5pt]\noindent{\textbf{\b{Defining Contours}:}}\\[-10pt]
\eq{\hspace{-100pt}\fwboxL{280pt}{\hspace{-80pt}\begin{array}{@{}l@{}}\begin{array}{@{}c@{}c@{}c@{}c@{}c@{}c}
\fwboxR{0pt}{\raisebox{2pt}{$\left\{\rule{0pt}{20pt}\right.$}\hspace{-8.5pt}}\cfig{313_#1#2#3#4_1}\fwboxL{0pt}{,}&\cfig{313_#1#2#3#4_2}\fwboxL{0pt}{,}&\cfig{313_#1#2#3#4_3}\fwboxL{0pt}{,}&\cfig{313_#1#2#3#4_4}\fwboxL{0pt}{,}&\cfig{313_#1#2#3#4_5}\fwboxL{0pt}{,}\\[-10pt]
\phantom{{}_{1}}\Omega_1&\phantom{{}_{2}}\Omega_2&\phantom{{}_{3}}\Omega_3&\phantom{{}_{4}}\Omega_4&\phantom{{}_{5}}\Omega_5
\end{array}\\[-5pt]
\begin{array}{@{}c@{}c@{}c@{}c@{}c@{}c}
\cfig{313_#1#2#3#4_6}\fwboxL{0pt}{\hspace{0pt},}&\cfig{313_#1#2#3#4_7}\fwboxL{0pt}{\hspace{0pt},}&\cfig{313_#1#2#3#4_8}\fwboxL{0pt}{\hspace{-8.5pt}\raisebox{2pt}{$\left.\rule{0pt}{20pt}\right\}$}}\\[-10pt]
\phantom{{}_{6}}\Omega_6&\phantom{{}_{1}}\Omega_7&\phantom{{}_{8}}\Omega_8
\end{array}\\[100pt]~
\end{array}}\label{omega313|#1#2#3#4}\hspace{-100pt}\vspace{-100pt}
}}

\newcommand{\dataTableQCD}[4]{\ifthenelse{\equal{#2}{1}}{\renewcommand{\topLabelM}{\g{\smash{\hspace{0.4pt}\bar{\hspace{-0.4pt}1\hspace{-0.4pt}}\hspace{0.4pt}}}}}{\renewcommand{\topLabelM}{\g{#2}}}
\ifthenelse{\equal{#4}{1}}{\renewcommand{\topLabelE}{\g{\smash{\hspace{0.4pt}\bar{\hspace{-0.4pt}1\hspace{-0.4pt}}\hspace{0.4pt}}}}}{\renewcommand{\topLabelE}{\g{#4}}}\renewcommand{\topLabel}{\fwbox{25pt}{\rule{0pt}{6.5pt}\r{#1}\topLabelM\b{#3}\topLabelE}}
\renewcommand{\theequation}{$\hspace{-0pt}I^{\fwbox{25pt}{\!\![\r{3}\g{1}\b{3}]}}_{\topLabel}\hspace{-1pt}$}
\vspace{-10pt}\eq{\hspace{-120pt}\hspace{-10pt}\raisebox{40pt}{$\cfig{313_#1#2#3#4}$}\hspace{5pt}\raisebox{-2pt}{$\cfig{313_#1#2#3#4_table}$}\hspace{20pt}\hspace{-100pt}\vspace{-22pt}\label{int313|#1#2#3#4}}
\renewcommand{\theequation}{$\Omega^{\fwbox{25pt}{\!\![\r{3}\g{1}\b{3}]}}_{\topLabel}\hspace{-1pt}$}
~\\[-16pt]\noindent{\textbf{\b{Defining Contours}:}}\\[-12pt]
\eq{\hspace{-100pt}\fwboxL{280pt}{\hspace{-80pt}\begin{array}{@{}l@{}}\begin{array}{@{}c@{}c@{}c@{}c@{}c@{}c}
\fwboxR{0pt}{\raisebox{2pt}{$\left\{\rule{0pt}{20pt}\right.$}\hspace{-8.5pt}}\cfig{313_#1#2#3#4_1}\fwboxL{0pt}{,}&\cfig{313_#1#2#3#4_2}\fwboxL{0pt}{,}&\cfig{313_#1#2#3#4_3}\fwboxL{0pt}{,}&\cfig{313_#1#2#3#4_4}\fwboxL{0pt}{,}\\[-10pt]
\phantom{{}_{1}}\Omega_1&\phantom{{}_{2}}\Omega_2&\phantom{{}_{3}}\Omega_3&\phantom{{}_{4}}\Omega_4
\end{array}\\[-5pt]
\begin{array}{@{}c@{}c@{}c@{}c@{}c@{}c}
\cfig{313_#1#2#3#4_5}\fwboxL{0pt}{,}&\cfig{313_#1#2#3#4_6}\fwboxL{0pt}{\hspace{0pt},}&\cfig{313_#1#2#3#4_7}\fwboxL{0pt}{\hspace{0pt},}\\[-10pt]
\phantom{{}_{5}}\Omega_5&\phantom{{}_{6}}\Omega_6&\phantom{{}_{1}}\Omega_7
\end{array}\\[-5pt]
\begin{array}{@{}c@{}c@{}c@{}c@{}c@{}c}
\cfig{313_#1#2#3#4_8}\fwboxL{0pt}{\hspace{-8.5pt}\raisebox{2pt}{$\left.\rule{0pt}{45pt}\right\}$}}\\[-10pt]
\phantom{{}_{8}}\Omega_8
\end{array}\\[160pt]~
\end{array}}\vspace{-140pt}\label{omega313|#1#2#3#4}\hspace{-100pt}
}}

\newcommand{\dataTableQCE}[4]{\ifthenelse{\equal{#2}{1}}{\renewcommand{\topLabelM}{\g{\smash{\hspace{0.4pt}\bar{\hspace{-0.4pt}1\hspace{-0.4pt}}\hspace{0.4pt}}}}}{\renewcommand{\topLabelM}{\g{#2}}}
\ifthenelse{\equal{#4}{1}}{\renewcommand{\topLabelE}{\g{\smash{\hspace{0.4pt}\bar{\hspace{-0.4pt}1\hspace{-0.4pt}}\hspace{0.4pt}}}}}{\renewcommand{\topLabelE}{\g{#4}}}\renewcommand{\topLabel}{\fwbox{24pt}{\rule{0pt}{6.5pt}\r{#1}\topLabelM\b{#3}\topLabelE}}
\renewcommand{\theequation}{$\hspace{-0pt}I^{\fwbox{25pt}{\!\![\r{3}\g{1}\b{3}]}}_{\topLabel}\hspace{-1pt}$}
\vspace{-5pt}\eq{\hspace{-120pt}\hspace{-10pt}\raisebox{40pt}{$\cfig{313_#1#2#3#4}$}\hspace{5pt}\raisebox{-2pt}{$\cfig{313_#1#2#3#4_table}$}\hspace{20pt}\hspace{-100pt}\vspace{-22pt}\label{int313|#1#2#3#4}}
\renewcommand{\theequation}{$\Omega^{\fwbox{25pt}{\!\![\r{3}\g{1}\b{3}]}}_{\topLabel}\hspace{-1pt}$}
~\\[-5pt]\noindent{\textbf{\b{Defining Contours}:}}\\[-10pt]
\eq{\hspace{-100pt}\fwboxL{280pt}{\hspace{-80pt}\begin{array}{@{}l@{}}\begin{array}{@{}c@{}c@{}c@{}c@{}c@{}c}
\fwboxR{0pt}{\raisebox{2pt}{$\left\{\rule{0pt}{20pt}\right.$}\hspace{-8.5pt}}\cfig{313_#1#2#3#4_1}\fwboxL{0pt}{,}&\cfig{313_#1#2#3#4_2}\fwboxL{0pt}{,}&\cfig{313_#1#2#3#4_3}\fwboxL{0pt}{,}\\[-10pt]
\phantom{{}_{1}}\Omega_1&\phantom{{}_{2}}\Omega_2&\phantom{{}_{3}}\Omega_3
\end{array}\\[-5pt]
\begin{array}{@{}c@{}c@{}c@{}c@{}c@{}c}
\cfig{313_#1#2#3#4_4}\fwboxL{0pt}{,}&\cfig{313_#1#2#3#4_5}\fwboxL{0pt}{,}\\[-10pt]
\phantom{{}_{4}}\Omega_4&\phantom{{}_{5}}\Omega_5
\end{array}\\[-5pt]
\begin{array}{@{}c@{}c@{}c@{}c@{}c@{}c}
\cfig{313_#1#2#3#4_6}\fwboxL{0pt}{\hspace{0pt},}&\cfig{313_#1#2#3#4_7}\fwboxL{0pt}{\hspace{0pt},}&\cfig{313_#1#2#3#4_8}\fwboxL{0pt}{\hspace{-8.5pt}\raisebox{2pt}{$\left.\rule{0pt}{45pt}\right\}$}}\\[-10pt]
\phantom{{}_{6}}\Omega_6&\phantom{{}_{1}}\Omega_7&\phantom{{}_{8}}\Omega_8
\end{array}\\[-65pt]~
\end{array}}\vspace{55pt}\label{omega313|#1#2#3#4}\hspace{-100pt}
}}

\newcommand{\dataTableQD}[4]{\ifthenelse{\equal{#2}{1}}{\renewcommand{\topLabelM}{\g{\smash{\hspace{0.4pt}\bar{\hspace{-0.4pt}1\hspace{-0.4pt}}\hspace{0.4pt}}}}}{\renewcommand{\topLabelM}{\g{#2}}}
\ifthenelse{\equal{#4}{1}}{\renewcommand{\topLabelE}{\g{\smash{\hspace{0.4pt}\bar{\hspace{-0.4pt}1\hspace{-0.4pt}}\hspace{0.4pt}}}}}{\renewcommand{\topLabelE}{\g{#4}}}\renewcommand{\topLabel}{\fwbox{26pt}{\rule{0pt}{6.5pt}\r{\smash{\bar{\hspace{-0.4pt}1\hspace{-0.4pt}}\hspace{0.8pt}\bar{\hspace{-0.4pt}1\hspace{-0.4pt}}\hspace{0.8pt}\bar{\hspace{-0.4pt}1\hspace{-0.4pt}}\hspace{0.4pt}}}\topLabelM\b{#3}\topLabelE}}
\renewcommand{\theequation}{$\hspace{-0pt}I^{\fwbox{26pt}{\!\![\r{4}\g{1}\b{2}]}}_{\topLabel}\hspace{-1pt}$}
\vspace{-8pt}\eq{\hspace{-120pt}\hspace{-10pt}\raisebox{-0pt}{$\cfig{412_#1#2#3#4}$}\hspace{5pt}\raisebox{-14pt}{$\cfig{412_#1#2#3#4_table}$}\hspace{20pt}\hspace{-100pt}\vspace{-22pt}\label{int412|#1#2#3#4}}
\renewcommand{\theequation}{$\hspace{-0pt}\Omega^{\fwbox{26pt}{\!\![\r{4}\g{1}\b{2}]}}_{\topLabel}\hspace{-1pt}$}
~\\[-15pt]\noindent{\textbf{\b{Defining Contours}:}}\\[-20pt]
\vspace{-14pt}\eq{\fwbox{00pt}{\hspace{05pt}\begin{array}{@{}l@{}}\begin{array}{@{}c@{}c@{}c@{}c@{}c@{}c}~\\[-15pt]
\fwboxR{0pt}{\raisebox{2pt}{$\left\{\rule{0pt}{24pt}\right.$}\hspace{-8.5pt}}\hspace{5pt}\cfig{412_#1#2#3#4_1}\hspace{5pt}\fwboxL{0pt}{,}&\hspace{5pt}\cfig{412_#1#2#3#4_2}\fwboxL{0pt}{\hspace{-4.5pt}\raisebox{2pt}{$\left.\rule{0pt}{24pt}\right\}$}}\\[-14pt]
\phantom{{}_{1}}\Omega_1&\phantom{{}_{2}}\Omega_2\\[-7pt]
\end{array}
\end{array}}\label{omega412|#1#2#3#4}\vspace{1pt}
}}

\newcommand{\dataTableQKBT}[4]{\ifthenelse{\equal{#2}{1}}{\renewcommand{\topLabelM}{\g{\smash{\hspace{0.4pt}\bar{\hspace{-0.4pt}1\hspace{-0.4pt}}\hspace{0.4pt}}}}}{\renewcommand{\topLabelM}{\g{#2}}}
\ifthenelse{\equal{#4}{1}}{\renewcommand{\topLabelE}{\g{\smash{\hspace{0.4pt}\bar{\hspace{-0.4pt}1\hspace{-0.4pt}}\hspace{0.4pt}}}}}{\renewcommand{\topLabelE}{\g{#4}}}\renewcommand{\topLabel}{\fwbox{30pt}{\rule{0pt}{6.5pt}\r{\smash{\bar{\hspace{-0.4pt}1\hspace{-0.4pt}}\hspace{0.8pt}\bar{\hspace{-0.4pt}1\hspace{-0.4pt}}\hspace{0.8pt}\bar{\hspace{-0.4pt}1\hspace{-0.4pt}}\hspace{0.4pt}}}\smash{\g{\hspace{0.4pt}\bar{\hspace{-0.4pt}0\hspace{-0.4pt}}\hspace{0.4pt}}}\b{#3}\smash{\g{\hspace{0.4pt}\bar{\hspace{-0.4pt}0\hspace{-0.4pt}}\hspace{0.4pt}}}}}
\renewcommand{\theequation}{$\hspace{-0pt}I^{\fwbox{30pt}{\!\![\r{4}\g{0}\b{3}]}}_{\topLabel}\hspace{-1pt}$}
\vspace{-10pt}\eq{\hspace{-120pt}\hspace{-10pt}\raisebox{-0pt}{$\cfig{403_#1#2#3#4}$}\hspace{5pt}\raisebox{-13.5pt}{$\cfig{403_#1#2#3#4_table}$}\hspace{20pt}\hspace{-100pt}\vspace{-22pt}\label{int403|#1#2#3#4}}
\renewcommand{\theequation}{$\hspace{-0pt}\Omega^{\fwbox{30pt}{\!\![\r{4}\g{0}\b{3}]}}_{\topLabel}\hspace{-1pt}$}
~\\[-14pt]\noindent{\textbf{\b{Defining Contours}:}}\\[-30pt]
\vspace{-12pt}\eq{\fwbox{00pt}{\hspace{05pt}\begin{array}{@{}l@{}}\begin{array}{@{}c@{}c@{}c@{}c@{}c@{}c}~\\[-15pt]
\fwboxR{0pt}{\raisebox{2pt}{$\left\{\rule{0pt}{24pt}\right.$}\hspace{-8.5pt}}\hspace{5pt}\cfig{403_#1#2#3#4_1}\hspace{5pt}\fwboxL{0pt}{,}&\hspace{5pt}\cfig{403_#1#2#3#4_2}\fwboxL{0pt}{\hspace{-4.5pt}\raisebox{2pt}{$\left.\rule{0pt}{24pt}\right\}$}}\\[-13pt]
\phantom{{}_{1}}\Omega_1&\phantom{{}_{2}}\Omega_2\\[-5pt]
\end{array}
\end{array}}\label{omega403|#1#2#3#4}\vspace{-0pt}
}}

\newcommand{\dataTableQKT}[4]{\ifthenelse{\equal{#2}{1}}{\renewcommand{\topLabelM}{\g{\smash{\hspace{0.4pt}\bar{\hspace{-0.4pt}1\hspace{-0.4pt}}\hspace{0.4pt}}}}}{\renewcommand{\topLabelM}{\g{#2}}}
\ifthenelse{\equal{#4}{1}}{\renewcommand{\topLabelE}{\g{\smash{\hspace{0.4pt}\bar{\hspace{-0.4pt}1\hspace{-0.4pt}}\hspace{0.4pt}}}}}{\renewcommand{\topLabelE}{\g{#4}}}\renewcommand{\topLabel}{\fwbox{26pt}{\rule{0pt}{6.5pt}\r{#1}\smash{\g{\hspace{0.4pt}\bar{\hspace{-0.4pt}0\hspace{-0.4pt}}\hspace{0.4pt}}}\b{#3}\smash{\g{\hspace{0.4pt}\bar{\hspace{-0.4pt}0\hspace{-0.4pt}}\hspace{0.4pt}}}}}
\renewcommand{\theequation}{$\hspace{-0pt}I^{\fwbox{26pt}{\!\![\r{3}\g{0}\b{3}]}}_{\topLabel}\hspace{-1pt}$}
\vspace{-12pt}\eq{\hspace{-80pt}\hspace{-36pt}\raisebox{-2pt}{$\cfig{303_#1#2#3#4}$}\hspace{5pt}\raisebox{-3.pt}{$\cfig{303_#1#2#3#4_table}$}\hspace{20pt}\hspace{-80pt}\vspace{-28pt}\label{int303|#1#2#3#4}}
\renewcommand{\theequation}{$\hspace{-0pt}\Omega^{\fwbox{26pt}{\!\![\r{3}\g{0}\b{3}]}}_{\topLabel}\hspace{-1pt}$}
~\\[4pt]\noindent{\textbf{\b{Defining Contours}:}}\vspace{-30pt}\eq{\hspace{5pt}\begin{array}{@{}l@{}}\begin{array}{@{}c@{}c@{}c@{}c@{}c@{}c}~\\[-15pt]
\fwboxR{0pt}{\raisebox{2pt}{$\left\{\rule{0pt}{22pt}\right.$}\hspace{-8.5pt}}\hspace{2pt}\cfig{303_#1#2#3#4_1}\hspace{-2pt}\fwboxL{0pt}{\hspace{-4.5pt}\raisebox{2pt}{$\left.\rule{0pt}{22pt}\right\}$}}\\[-18pt]
\phantom{{}_{1}}\hspace{2pt}\,\Omega_1\\[-7pt]
\end{array}
\end{array}\vspace{-0pt}\label{omega303|#1#2#3#4}}}

\makeindex
\title{\texorpdfstring{\mbox{{\huge Prescriptive Master Integrals}{\LARGE~of}}\\[-5pt]\mbox{{\huge Maximal Weight at Two Loops}}}{Prescriptive Master Integrals with Maximal Weight at Two Loops}}
\author[\hspace{-0pt}\dagger,\ast]{\vspace{-24pt} Jacob~L.~Bourjaily}\emailAdd{jbourj@ias.edu}

\affiliation[\dagger]{School of Natural Sciences, Institute for Advanced Study, Princeton, NJ, 08540, USA} 
\affiliation[\ast]{Institute for Gravitation and the Cosmos, Department of Physics,\\Pennsylvania State University, University Park, PA 16802, USA}

\abstract{%
We describe the construction of a spanning set of individually pure, planar master integrals involving massless particles in four dimensions which include all maximal-weight contributions at two loops. In this basis, every independent region associated with infrared divergence is individually matched by specific masters, with all other masters being manifestly finite in four dimensions. 
}

\preprint{}

\begin{document}

\maketitle\thispagestyle{empty}
\pagenumbering{roman}\clearpage

\setcounter{section}{0}

\newpage
\pagenumbering{arabic}
\vspace{0pt}%
\section{Introduction and Overview}\label{sec:introduction}\vspace{-4pt}

We construct a set of individually-pure master Feynman integrals for involving massless propagators at two loops which span {all} maximal-weight contributions to four-dimensional planar theories. These integrals are defined and constructed \emph{prescriptively}, as the cohomological duals \mbox{\cite{Mastrolia:2018uzb,Frellesvig:2019kgj,Frellesvig:2019uqt,Frellesvig:2020qot,Weinzierl:2020gda,Caron-Huot:2021iev,Caron-Huot:2021xqj,lu2025intersection}} of a spanning set of compact contour integrals  \cite{Gluza:2010ws,Caron-Huot:2012awx} $\{\Omega^{\r{j}}\}$,
\eq{\oint\limits_{\Omega^\r{j}}\hspace{-2pt}\dbar^4\hspace{-1pt}\ell_1\dbar^4\hspace{-1pt}\ell_2\;\mathcal{I}_{\b{i}}=\delta\indices{\r{j}}{\b{i}}\,.\label{homological_duality}}
This construction ensures that each master satisfies \emph{canonical} differential equations \cite{Henn:2013pwa,Henn:2014qga}, hopefully making them well-suited for analytic or numeric integration as in the integrals described in \mbox{\cite{Drummond:2010mb,Drummond:2010cz,Henn:2013nsa,Gehrmann:2018yef,Herrmann:2019upk}}.

The contours which define our masters are particular cut configurations involving on-shell propagators; as such, the scheme-independent coefficient of the master $\mathcal{I}^{\r{j}}$ in the expansion of \emph{any} amplitude,
\eq{\mathcal{A}=\sum_{\r{j}}\mathfrak{a}^{\r{j}}\mathcal{I}_{\r{j}}{+}\left(\text{strictly lower-weight contributions}\right)\vspace{-5pt}\label{partial_match}}
can be determined by an \emph{on-shell function} (see \emph{e.g.}~\cite{ArkaniHamed:book,Elvang:2013cua,Henn:2014yza,Arkani-Hamed:2014bca,Bourjaily:2016mnp})
\eq{\mathfrak{a}^{\r{j}}\equivR\!\!\oint\limits_{\Omega^{\r{j}}}\hspace{-2pt}\dbar^4\hspace{-1pt}\ell_1\dbar^4\hspace{-1pt}\ell_2\;\mathcal{A}\vspace{-4pt}}
---computable without reference to gauge redundancy in terms of tree amplitudes. Importantly, this is true even for theories such as QCD for which the lower-weight terms in (\ref{partial_match}) are non-vanishing. In theories such as maximally supersymmetric Yang-Mills theory \cite{Brink:1976bc,Gliozzi:1976qd,Arkani-Hamed:2008owk}, this basis spans all contributions to amplitudes \mbox{\cite{Bern:2006kd,BjerrumBohr:2006yw,BjerrumBohr:2008ji}}.

The contours which define our basis have been chosen to span \emph{all} independent collinear and soft-collinear regions responsible for infrared divergence  \cite{Bourjaily:2021ewt}. This not only ensures that all (leading-weight) sources of infrared divergence appear \emph{individually} and isolated in (\ref{partial_match}), but also that any integral \emph{not} defined to be dual to a contour associated with such a region is necessarily free of all infrared divergences and therefore requires no regularization for its integration in four dimensions. (All ultraviolet divergences have less than maximal weight.)\\

This work describes the construction of this master integral basis, with details described in \mbox{appendix~\ref{prescriptive_definition_of_masters}}. Precise expressions for all master integrals in terms of momentum twistor variables \cite{Hodges:2009hk}, reviewed in \mbox{appendix~\ref{kinematics_appendix}}, are included as ancillary files to this work's submission to the \texttt{arXiv}, which we describe in \mbox{appendix~\ref{ancillary_appendix}}. 

\newpage
\subsection{Organization and Outline}\label{organization}\vspace{-4pt}

This work is organized as follows. In section \ref{masters_and_integral_reduction} we review the construction, structure, and use of master integrals for the representation of perturbative quantities in quantum field theory. In particular, we review the familiar choices of masters in \mbox{section~\ref{constructing_bases_of_masters_subsection}} and describe how coefficients may be determined using integration-by-parts `projections' or generalized unitarity in \mbox{section~\ref{constructing_bases_of_masters_subsection}}. 

We define the notion of \emph{purity} of transcendental functions in \mbox{section~\ref{weight_and_purity_general_section}} and discuss degrees of impurity in \mbox{section~\ref{different_degrees_of_impurity_subsection}}, and the impurities of \emph{scalar} Feynman integrals in \mbox{section~\ref{impurities_of_feynman_ints}}. 

In \mbox{section~\ref{power_counting_masters}} we define masters constructed by \emph{power-counting} and define the space of masters constructed here in \mbox{section~\ref{four_dimensional_masters_with_triangle_power_counting}}. These masters are chosen \emph{prescriptively} as the duals of particular choices of (linear combinations of) compact cycles of integration; these choices are elucidated in \mbox{section~\ref{prescriptively_defining_master_basis_ints}} and \mbox{section~\ref{periods_of_feynman_integrals}}. 

In \mbox{section~\ref{conclusions}} we conclude with a discussion of future work and applications.\\[-8pt]

Finally, we have included a brief discussion of kinematics relevant for our concrete results in \mbox{appendix~\ref{kinematics_appendix}}, which defines the invariants that appear in the ancillary files included with this work's submission to the \texttt{arXiv}. Concrete expressions for our master integrals, together with explicit details for their defining contours of integration are provided in the \textsc{Mathematica} package \texttt{two\uscore loop\uscore masters.m}, for which we have prepared a companion \texttt{Notebook} \texttt{two\uscore loop\uscore master\uscore integrals.nb} which demonstrates the package's main usage and documents its principal functionality.

In \mbox{appendix~\ref{prescriptive_definition_of_masters}} we provide a complete description of the contours used to define all the master integrals in our basis. These contours are represented graphically, in notation that is described in \mbox{appendix~\ref{organizing_principles}} and \mbox{appendix~\ref{graphical_notation}}. 

\newpage

\section{Masters Integrals, Integral Reduction, Purity and Rigidity}\label{masters_and_integral_reduction}\vspace{-4pt}

The Feynman rules for perturbative quantum field theory encode scattering amplitudes as sums over differential forms on a space of internal `loop' momenta, which must be integrated. The \emph{particular} differential forms that arise from the Feynman expansion are never unique: each loop momentum enjoys translation invariance, and the assignment of $L$ \emph{particular} loop momentum variables to an $L$-loop Feynman graph depends on a choice for how momentum conservation should be imposed throughout the graph.\footnote{This can be done by choosing $L$ propagators in the complement of any spanning tree and solving for the momenta in every other propagator in terms of these. More generally, any choice of \emph{cycle basis} for the graph can be used \cite{Bourjaily:2020qca}.}

The equivalence between Feynman integrals whose \emph{integrands} differ by translated or transformed assignments of loop momenta is an example of a \emph{cohomological} equivalence: they differ by something \emph{cohomologically trivial}, a `total derivative'. Colloquially, we say they differ by an \emph{integration by parts} (`IBP') \emph{identity}.

Even when there exists a natural or otherwise canonical assignment of loop momenta across a graph (such as for planar (or more precisely, plane-\emph{embedded}) Feynman integrals), it is useful to think of integrals more abstractly as representing elements of equivalence classes of cohomology. One immediate advantage of this way of thinking is the recognition that such differential forms form a vector space of \emph{finite} dimension, for which we can choose any \emph{basis}. 

Any basis of cohomologically independent differential forms on loop momenta---regardless of whether these arise via the Feynman rules for some theory---is called a set of \emph{master integrals}. Conveniently, the size of this basis is typically \emph{much} smaller than the number of Feynman \emph{diagrams} encountered in perturbation theory.\\[-8pt]

Consider any $L$-loop observable in some quantum field theory; the Feynman rules provide us with an expansion
\eq{\mathcal{O}^{L}({p_{[n]}},\epsilon_{[n]},\ldots)=\sum_{\Gamma}\int\limits_{\hspace{-4pt}\ell_i\in\mathbb{R}^{d}\hspace{-16pt}}\hspace{-1pt}f_{\Gamma}(p_{[n]},\epsilon_{[n]},\ldots,\ell_{[L]})\,.\label{feynman_sum}}
Suppose that $\mathfrak{B}\equivR\{\mathcal{I}_i(p_{[n]},\ell_{[L]})\}$ is some master integral basis large enough to \emph{span} the integrals appearing in the Feynman expansion (\ref{feynman_sum}); that is, for any Feynman diagram $f_\Gamma$, there exist a set of loop-momentum-independent coefficients $c_\Gamma^i$ such that\\[-10pt]
\eq{f_{\Gamma}(p_{[n]},\epsilon_{[n]},\ldots,\ell_{[L]})\equivR\sum_i c_\Gamma^i(p_{[n]},\epsilon_{[n]},\ldots)\,\mathcal{I}_i(p_{[n]},\ell_{[L]}).\vspace{-5pt}}
(We review how such coefficients can be determined via `IBP reduction' or `generalized unitarity' in \mbox{section~\ref{ibp_projection_and_unitarity}}.) 

In terms of these masters then, the observable $\mathcal{O}^L$ (\ref{feynman_sum}) may be expressed 
\eq{\begin{split}\mathcal{O}^{L}=\sum_{\Gamma,i}\int\limits_{\hspace{-4pt}\ell_a\in\mathbb{R}^{d}\hspace{-16pt}}\hspace{-1pt}c_\Gamma^i\,\mathcal{I}_i(p_{[n]},\ell_{[L]})&=\sum_{\Gamma,i}c_\Gamma^i\,\int\limits_{\hspace{-4pt}\ell_a\in\mathbb{R}^{d}\hspace{-16pt}}\hspace{-1pt}\mathcal{I}_i(p_{[n]},\ell_{[L]})\\
&\equivL\sum_{\Gamma,i}c_\Gamma^i(p_{[n]},\epsilon_{[n]},\ldots)\,I_i(p_{[n]})\,.\\[-9pt]\end{split}}
Writing quantities in this way is known as as \emph{integral reduction}. For useful reviews and particular applications, see \emph{e.g.}~\cite{Melrose:1965kb,Passarino:1978jh,vanNeerven:1983vr,Denner:2002ii,Denner:2005nn,Cachazo:2008vp,Mastrolia:2010nb,Ellis:2011cr,Mastrolia:2011pr,Badger:2012dv,Mastrolia:2012an,Mastrolia:2013kca,Badger:2013sta,Mastrolia:2016dhn,Ita:2015tya,Zhang:2016kfo}.

Importantly, for any renormalizable quantum field theory, the number of master integrals required to compute various observables \emph{saturates} at some multiplicity \cite{Bourjaily:2020qca} (see also \mbox{\cite{Ossola:2006us,Mastrolia:2012wf,Kleiss:2012yv,Feng:2012bm}}); that is, there exists a \emph{finite} collection of masters which span \emph{all} observables---at \emph{arbitrary multiplicity} and involving \emph{arbitrary external states}. Moreover, the size of this basis grows only \emph{exponentially} with loop order (although the number of Feynman graphs---and hence the work required to determine coefficients---still grows factorially). 

Thus, from a purely practical point of view, a key advantage of integral reduction is that a finite list of integrals are required to represent an infinitude of observables; the `master' integrals chosen for $\mathfrak{B}$ can be integrated and tabulated once and for all, and are immediately reusable for all observables at that loop order in that theory (and often many other theories as well).\\[-8pt]

\vspace{-6pt}\subsection{Constructing Bases of Master Integrals}\label{constructing_bases_of_masters_subsection}\vspace{-4pt}

A master integral \emph{basis} is any set of cohomologically independent differential forms on loop momenta; typically, we require these integrals consist of \emph{Feynman-like integrals}: integrals with denominators defined by products of (scalar) propagators encoding some (momentum-conserving) Feynman graph. These integrals need not correspond to any particular Feynman \emph{diagrams} of any particular quantum field theory, but because they involve only propagators constructed from Feynman graphs, they can (in principle) be integrated using ordinary methods (such as Feynman/Schwinger parameterization \cite{Feynman:1949zx}, \cite{Schwinger:1949ra}, Mellin-Barnes \cite{barnes1908new,mellin1910abriss}, etc.).

To be clear, our notion of a `basis' of master integrals is broader than how this term is used in most of the literature, as we do not require a basis to be \emph{complete}. That is, we do not require that bases span the entire cohomology of differential forms on Feynman integrals; nor even do we require that a basis spans the space of external differentials (with respect to kinematics) acting on the basis. Our definition would permit even a single (not cohomologically trivial) Feynman integral to form a one-dimensional basis of masters. Of course, a basis consisting of a single Feynman integral would not be especially \emph{useful} for the representation of amplitudes.

Given an \emph{incomplete} basis, it is natural to wonder whether or not it is large enough to span the amplitudes (if not individual Feynman diagrams) of some theory. Regardless of the answer to this question, however, it should be clear that even an incomplete basis can be used to represent \emph{some part} of the contributions to any amplitude in any particular field theory. 

In this work, we describe the construction of such a basis which spans all maximal-weight Feynman integrals at two loops; this excludes ultraviolet structure arising from bubbles, for example, but includes {all} leading sources of infrared divergence. Thus, this basis may be used to faithfully capture all such contributions in any four-dimensional quantum field theory involving massless particles---even if this subset is incomplete for many theories.\\[-8pt]

It is worthwhile to briefly discuss the class of master integrals most commonly used in physics. Readers interested in a more thorough discussion may consult the excellent texts \cite{Smirnov:2004ym,Smirnov:2006ry,Smirnov:2012gma} (see also \cite{Kosower:2011ty,Badger:2013gxa,Johansson:2013sda}) or explore their implementation in various public computational tools (see \emph{e.g.}~\mbox{\cite{Nogueira:1991ex,Laporta:2000dsw,Binoth:2000ps,Gluza:2007rt,Smirnov:2008iw,Smirnov:2008py,Peraro:2014cba,Ochman:2015fho}}).

A particularly familiar class of \emph{master integrals} may be defined by considering integrands constructed from \emph{scalar} propagators $1/P(\vec{\ell})$ given by some graph
\eq{\mathcal{I}[d;\nu_a,\nu_b,\ldots,\nu_c]\equivR\,\,d^{(d{-}2\epsilon)L}\vec{\ell}\frac{1}{P_a(\vec{\ell})^{\nu_a}P_b(\vec{\ell})^{\nu_b}\cdots P_c(\vec{\ell})^{\nu_c}}\,.\label{standard_ibp_integrals}}
Integrals of this form represent classes in the (twisted) cohomology on the space $\mathbb{C}^{d\,L}\backslash\langle P_a(\vec{\ell})\!\cdots\!P_{c}(\vec{\ell}){=}0\rangle$ \cite{Caron-Huot:2021iev,Caron-Huot:2021xqj}. Upon integration against the Feynman contour of real momenta, they are known to be \emph{meromorphic} in $\epsilon$ and the exponents $\nu_i$, with simple poles appearing only at integer values. Varying over the integer lattice of possible values of the $\nu_i$ provides a collection of integrals which span the finite-dimensional cohomology; a mater integral basis, therefore, may be defined as a finite collection of integrals (independent under integration by parts identities) of the form (\ref{standard_ibp_integrals}) which span this cohomology.

Differentiation acts naturally on the space of these integrals: whether acting on the external kinematics or the internal loop momenta, it is clear that derivatives result in linear combinations of integrands of the same form as (\ref{standard_ibp_integrals}) but with powers of propagators shifted by integers. Derivatives with respect to the loop momenta generate `integration by parts' identities which allow one to test cohomological independence and to \emph{project} integrands into this space of master integrals. These have operations may been effectively implemented using, for example, Laporta's algorithm~\cite{Laporta:2000dsw}.\\[-8pt]

Unfortunately, as we will describe below, master integrals of the form (\ref{standard_ibp_integrals}) are among the most difficult to actually integrate and the most complex as analytic functions. Moreover, these integrals widely share redundant (and hence unnecessary) overlapping sources of infrared and ultraviolet divergence, requiring regularization and complicating the computation of ultimately finite observables.\\[-8pt]

\subsection{Integral Reduction: Integration by Parts Projections and Unitarity}\label{ibp_projection_and_unitarity}\vspace{-4pt}

Given a complete basis of master integrals, any individual Feynman diagram may be represented as a sum with loop-independent coefficients. For masters of the form (\ref{standard_ibp_integrals}), this is achieved by first tabulating the projection of all loop-dependent monomials into this basis for cohomology, and then applying these projectors to any Feynman diagram one encounters in the expansion of an observable. This results in coefficients that are nearly as complicated (algebraically) as the Feynman expansion itself, and great efforts and considerable cleverness has been employed to `reconstruct' these coefficients analytically (see \emph{e.g.}~\cite{vonManteuffel:2014ixa,Peraro:2016wsq,Abreu:2019odu,Peraro:2019okx,vonHippel:2025okr}). 

While this procedure is well known, well implemented, and widely used (see \emph{e.g.}~\cite{Peraro:2019svx,Peraro:2019cjj}), it requires a great deal of computational complexity and overhead to employ. And it is not at all obvious how one would employ these tools in the case of an \emph{incomplete} basis of masters. Is there any other way to determine such coefficients? 

Suppose that $\mathfrak{B}\equivR\{\mathcal{I}_{\b{1}},\ldots,\mathcal{I}_{\b{p}}\}$ is any (possibly incomplete) set of cohomologically independent set of Feynman integrands. Their cohomological independence is equivalent to the existence of a set of $p$ compact contours of integration $\{\Omega^i\}$ such that the \emph{period} \cite{kontsevich2001periods} matrix (or `pairing')  \mbox{\cite{Mastrolia:2018uzb,Frellesvig:2019kgj,Frellesvig:2019uqt,Frellesvig:2020qot,Weinzierl:2020gda,Caron-Huot:2021iev,Caron-Huot:2021xqj,lu2025intersection}}
\eq{\mathbf{M}\indices{\r{i}}{\b{j}}\equivR\oint\limits_{\Omega^{\r{i}}}\hspace{-2pt}\dbar^{d\,L}\hspace{-1pt}\vec{\ell}\hspace{0pt}\,\,\mathcal{I}_{\b{j}}\,}
must be full rank---and hence invertible. 

Given \emph{any} differential form $f(\vec{\ell})$---including any Feynman integral---one can compute the periods 
\eq{\mathfrak{f}^{\r{j}}\equivR\oint\limits_{\Omega^{\r{i}}}\hspace{-2pt}\dbar^{d\,L}\hspace{-1pt}\vec{\ell}\hspace{0pt}\,\,f(\vec{\ell}),}
from which it is easy to see that the (loop-independent) coefficients 
\eq{\mathfrak{c}^{\r{i}}\hspace{-1pt}[f]\equivR\sum_{\t{j}}\mathfrak{f}^{\t{j}}(\mathbf{M}^{\text{-}1})^{\phantom{j}\r{i}}_{\t{j}}}
encode the projection of $f(\vec{\ell})$ onto the subspace of cohomology within the \emph{span} of the basis $\mathfrak{B}$:
\eq{f(\ell)\simeq\sum_{\r{i}}\mathfrak{c}^{\r{i}}\hspace{-1pt}[f]\,\mathcal{I}_{\r{i}}\;{+}\;\text{(terms cohomologically independent of $\mathrm{span}(\mathfrak{B})$)}}

This procedure is truly general: it can be employed to represent any individual Feynman diagram in terms of any choice of basis; when the basis is incomplete, it is detectable by the fact that the difference
\eq{f(\ell)-\sum_{\r{i}}\mathfrak{c}^{\r{i}}\![f]\,\mathcal{I}_{\r{i}}}
will be cohomologically non-trivial---implying there exists some new contour, $\Omega_f$, on which the difference integrates to something non-vanishing. 

But something magical happens for this expansion when applied not to individual Feynman diagrams but to entire amplitudes. Provided that every contour $\Omega_{\r{j}}$ `cuts' at least $(L{+}1)$ propagators (by `encircling' their vanishing)---which is always possible \cite{Caron-Huot:2021iev,Caron-Huot:2021xqj}---then an amplitude's periods 
\eq{\mathfrak{a}^{\r{j}}\equivR\oint\limits_{\Omega^{\r{i}}}\hspace{-2pt}\dbar^{d\,L}\hspace{-1pt}\vec{\ell}\hspace{0pt}\,\,\mathcal{A}(\vec{\ell})\label{amplitude_periods}}
can be computed \emph{directly} in terms of products of gauge-invariant tree amplitudes. This is the method known as \emph{generalized} unitarity  \cite{Bern:1994zx,Bern:1994cg,Britto:2004nc} (see also \cite{Drummond:2008bq,Anastasiou:2006jv}). 

The periods of amplitudes (\ref{amplitude_periods}) are called \emph{leading singularities} or more generally `on-shell functions', a class of functions which about which we have learned a great deal in recent decades. In the case of four-dimensional theories of massless particles, the kinematic dependence of these functions can be represented in terms of volume-forms on a subspace of an Grassmannian manifold with a rich geometric, combinatoric, and analytic structure \cite{ArkaniHamed:2009dn,ArkaniHamed:2009dg,ArkaniHamed:2010gg,Bourjaily:2012gy,Arkani-Hamed:2014via,Arkani-Hamed:2014bca}.\\[-8pt]

\subsection{Avoiding \emph{Unnecessary} Complexity when Choosing Masters}\label{avoiding_unnecessary_complexity}\vspace{-4pt}

Although the construction of master integral bases of the form (\ref{standard_ibp_integrals}) and the determination coefficients (via `IBP projection') of observables in terms of these masters \cite{Chetyrkin:1981qh,Tkachov:1981wb} have been well implemented in a number of public codes, it is worth noting that beyond one loop order, only a small subset of the integrals required for four dimensional quantum field theories have ever been \emph{integrated}: most are not known in any analytic form involving familiar (or even exotic) special functions, and they can prove extremely difficult to evaluate numerically via Monte Carlo.

The difficulty of actually integrating Feynman integrals---masters or otherwise---beyond one loop is partly due to the appearance of genuinely novel and exotic functional structures: starting at two loops, Feynman integrals are known to require elliptic integrals or those involving more `rigid' (higher-dimension) components such as Calabi-Yau geometries \cite{Bourjaily:2017bsb,Bourjaily:2018ycu,Bourjaily:2018yfy,Bourjaily:2019hmc}. These functional novelties have been the focus of a great deal of research in recent years (see \emph{e.g.}~\cite{Walden:2020odh,Kristensson:2021ani,Bourjaily:2022bwx,Wilhelm:2022wow,Morales:2022csr,Frellesvig:2023bbf}), but we are still far from as complete an understanding as now exists for polylogarithmic functions, say (see \emph{e.g.}~\cite{Goncharov:2010jf,Duhr:2011zq,Duhr:2012fh,Golden:2013lha,Golden:2013xva,Golden:2014pua,Dixon:2020cnr}).

But these genuine novelties only explain a small part of the difficulties of integration beyond one loop. We suspect that a more systematic (if not greater) difficulty arises from the simple fact that the bases of master integrals in (\ref{standard_ibp_integrals}) are simply \emph{unnecessarily} complicated functions that are \emph{unnecessarily} difficult to integrate or express analytically. 

We see at least two distinct sources of such complication that we view as unnecessary (and which we have avoided in our choice of masters here). First, common choices of master integrals have a high degree of \emph{impurity}. That is, they do not satisfy canonical differential equations, and it can require enormous efforts (beyond one loop) to construct from them new masters which do. We describe this in more detail with examples in \mbox{section~\ref{impurities_of_feynman_ints}} below.

But even for masters which \emph{do} satisfy canonical differential equations, there can be enormous redundancies in their \emph{support}, including an abundance of \emph{unnecessary} and overlapping divergences. One example of this would be the infrared divergences of scalar box integrals involving massless momenta, for which \emph{all} divergences are associated with contact term sub-integrals, and therefore can be removed; another example are the ultraviolet divergences of bubble integrals, which are shared by all, rendering the difference between any pair of bubbles perfectly finite in four dimensions.\\[-8pt]

To better appreciate these complexities, it is worthwhile to more sharply define what it means for a function (or Feynman integral) to be \emph{pure}.

\subsection{The \emph{Weight} and \emph{Purity} of (`Transcendental') Functions}\label{weight_and_purity_general_section}\vspace{-4pt}

One measure of the complexity of a Feynman-like integral is its \emph{purity} or \emph{degree} of \emph{impurity}. The purity or impurity of a function is defined relative to some notion of `transcendental' \emph{weight}; these concepts require a more technical digression (which we postpone for now), but are easy enough to understand and rigorously describe in the case of (classical and generalized) \emph{polylogarithms}.\\[-8pt]

Let us start with a discussion of the classical polylogarithms. To the $\b{n}$th polylogarithm $\text{Li}_{\b{n}}(\r{z})$ we assign a weight of $\b{n}$. These functions satisfy simple (canonical) differential equations, graded by their weight:
\vspace{-4pt}\eq{\frac{\partial}{\partial\r{z}}\text{Li}_{\b{n}}(\r{z})=\frac{1}{\r{z}}\,\text{Li}_{\b{n\text{-}1}}(\r{z})\;\qquad\bigger{\Leftrightarrow}\qquad\text{Li}_{\b{n}}(\r{z})\equivR\int\limits_0^{\r{z}}\frac{d\g{t}}{\g{t}}\,\text{Li}_{\b{n\text{-}1}}(\g{t})\vspace{-4pt}\label{classical_polylogs_defined}}
where $\text{Li}_{\b{1}}(\r{z})\equivR\!\log(1/(1{-}\r{z}))$. We assign any rational or algebraic function weight zero, and the weight of any product of functions is the sum of their weights.  The notion of weights can be extended to their valuations on algebraic input so that, for example, we would assign the weight 1 to $\pi$ and $\log(2)$, weight $2$ to $\mathrm{Li}_2(1)$, weight $7$ to $\mathrm{Li}_{7}(\sqrt{5})$, and so-on. Sums of functions with mixed weights may be assigned a weight of the maximal-weight term of the sum.

Notice from the definition (\ref{classical_polylogs_defined}) that the classical polylogarithms satisfy simple differential equations. In particular, because differentiation strictly lowers a function's weight, we call $\text{Li}_{\b{n}}(\r{z})$ a \emph{pure} function; equivalently, we say that these functions \emph{satisfy canonical differential equations}.\\[-8pt]

Any function whose differentiation does not \emph{strictly} lower its weight is called \emph{impure}. Consider for example the product of $\text{Li}_{\b{n}}(\r{z})$ by any rational or algebraic function $f(\r{z})$:
\vspace{-4pt}\eq{f(\r{z})\,\text{Li}_{\b{n}}(\r{z})\,;\label{impure_exempli}\vspace{-4pt}}
this function would not be pure: differentiation with respect to $\r{z}$ results in the homogeneous (in weight) term from the product rule:
\vspace{-4pt}\eq{\frac{\partial}{\partial\r{z}}\left[f(\r{z})\,\text{Li}_{\b{n}}(\r{z})\right]=\left(\frac{\partial}{\partial\r{z}}f(\r{z})\right)\text{Li}_{\b{n}}(\r{z}){+}f(\r{z})\left(\frac{\partial}{\partial\r{z}}\text{Li}_{\b{n}}(\r{z})\right).\vspace{-4pt}}
The two terms above have different weights, with the leading term having the same weight as the original function. Thus, because differentiation does not \emph{strictly} lower the weight of $f(\r{z})\,\text{Li}_{\b{n}}(\r{z})$, this function would be said to be \emph{impure}.\\[-8pt]

The classical polylogarithms may be generalized recursively to a broader class of `weight-$\b{w}$' functions of many variables as follows \cite{goncharov:hyperlogs}. Let 
\vspace{-4pt}\eq{G(\b{r_1},\b{\ldots}\,,\b{r_w};\r{z})\equivR\int\limits_0^{\r{z}}\!\!\frac{d\g{t}}{\g{t}{-}\b{r_1}}G(\b{r_2},\b{\ldots},\b{r_w};\g{t})\,\quad\text{with}\quad G(;\r{z})\equivR{-}1\,.\label{generalized_polylogarithm_defined}\vspace{-4pt}}
In terms of these, the {classical} polylogarithm $\text{Li}_{\b{n}}(\r{z})$ would be denoted
\vspace{-4pt}\eq{\text{Li}_{\b{n}}(\r{z})\equivR G(\underbrace{\b{0},\b{\ldots}\,,\b{0}}_{\b{n\text{-}1}\text{ times}},\b{1};\r{z})\,.\vspace{-4pt}}
Other generalizations of the classical polylogarithms naturally also fall into the class defined by (\ref{generalized_polylogarithm_defined}); for example, a `harmonic' polylogarithm \cite{Remiddi:1999ew} (see also \cite{Maitre:2005uu}) is one for which all of the arguments $\b{r_i}$ in (\ref{generalized_polylogarithm_defined}) take values $\b{r_i}\!\in\!\{\text{-}1,0,1\}$. Other generalizations include those of \mbox{\cite{Chen:1977oja,Brown:2004ugm,Goncharov:2009kx,Brown:2013gia,Brown:2015ztw}}.

These generalized (or `multiple') polylogarithms in (\ref{generalized_polylogarithm_defined}) are defined to satisfy the same kind of \emph{canonical} differential equations as the classical polylogarithms. As such, they are \emph{pure} functions of weight defined by the number of their \emph{letters} $\b{r_i}$.\\[-8pt]

It is worth pausing to note that, of course,  although many Feynman-like integrals are polylogarithmic, this is not universally true. More generally, a set of functions $\{I\}$ may be said to be \emph{pure} if they can be graded by \emph{some} notion of weights,\\[-12pt]
\eq{\{I(\r{\vec{z}})\}\equivL\{I_{\b{0}}(\r{\vec{z}})\}\oplus\!\{I_{\b{1}}(\r{\vec{z}})\}\!\oplus\!\ldots\!\oplus\{I_\b{w}(\r{\vec{z}})\}\oplus\ldots}
such that there exists a set of differential operators\footnote{Functions of higher `rigidity' require higher-order differential operators.} $D_{\r{\vec{z}}}$ which strictly lower this weight: $D_{\r{\vec{z}}}I_{\b{w}}(\r{\vec{z}})\in\mathrm{span}[\ldots\oplus\!\{I_{\b{w\text{-}1}}(\r{\vec{z}})\}]$, with coefficients taking values in the space of weight-zero functions. But this degree of generality in our discussion is not required so long as we confine our interests (for now) to functions built as linear combinations of generalized polylogarithms.\\[-8pt]

\subsubsection{Differentiating Different \emph{Degrees} of Impurity}\label{different_degrees_of_impurity_subsection}\vspace{-4pt}
\epigraph{\fwboxR{0pt}{\text{``}}\emph{There are no differences but differences of degree between different degrees of difference and no difference}'' \cite{James:Hegelisms}}

~\\[-44pt]

Notice that our definition of purity implies that if $\{I^{i}_{\b{w}}(\r{\vec{z}})\}$ is any set of linearly independent\footnote{We leave aside for now the question of how one may test  the linear independence of some set of generalized polylogarithmic functions.} pure functions of a given weight, then so would be
\eq{J_{\b{w}}^{j}(\r{\vec{z}})\equivR\sum_{i}\mathbf{c}\indices{j}{i}\,I^{i}_{\b{w}}(\r{\vec{z}})\vspace{-6pt}\label{constant_linear_combination_of_pure_functions}}
for any choice of \emph{constant} coefficients $\mathbf{c}\indices{j}{i}$. Thus, purity is not by itself an extremely strong criterion even when choosing masters. 

Multiplication of any pure function of a given weight by any weight-zero function does not change its weight; and so it may be said that a \emph{typical} function of some weight will take the form\\[-12pt]
\vspace{-0pt}\eq{f_{\g{1}}(\r{\vec{z}})\big(\text{some pure function}\big){+}\ldots{+}f_{\g{q}}(\r{\vec{z}})\big(\text{some \emph{other} pure function}\big).\label{impure_sum}\vspace{-2pt}}
For such a function, we say that it has a \emph{degree of impurity} given by the linear \emph{rank} of the coefficients $f_{\g{q}}(\r{\vec{z}})$---and not merely the number of weight-zero coefficients $f_{\g{q}}(\r{\vec{z}})$ that appear, as this can be changed by taking linear combinations of pure functions with constant coefficients as in (\ref{constant_linear_combination_of_pure_functions}).\\[-8pt] 

Notice that the linear rank of any set of \emph{constants} is always zero; as such pure functions have a \emph{vanishing} degree of impurity.

\subsubsection{The Purity and Impurities of Feynman(-Like) Integrals}\label{impurities_of_feynman_ints}\vspace{-4pt}

One loop Feynman integrals involving $p\!\leq\!d$ scalar propagators in $d$ spacetime dimensions are always polylogarithmic with at most a single degree of impurity \cite{Arkani-Hamed:2017ahv,Bourjaily:2019exo}. For example, a `\emph{scalar box}' integral in four (or more) dimensions is \emph{proportional} to a pure (weight-$\b{2}$) function; as such, it can be rendered pure if normalized appropriately:
\vspace{-10pt}\eq{\hspace{-5pt}\fig{-2.25pt}{1}{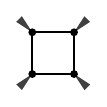}\hspace{-5pt}=f(\vec{p})\big(\text{some \emph{pure} function of weight 2}\big).\vspace{-12pt}\label{purity_of_boxes_in_4d}}
Here, the external legs simply denote \emph{any} inflowing external momentum. If none of these momenta are massless (or if the internal propagators are massive), then the integral is in general finite and needs no regularization in four dimensions. Our notion of purity and degrees of impurity can easily be extended to (\emph{e.g.} dimensionally) regularized integrals, but we believe such a diversion to be unnecessary: the differential equations relating Feynman integrals and defining their weights can be discussed at the level of differential forms without any reference to the divergence-generating Feynman contour of integration.\\[-8pt]

The \emph{near}-purity of scalar $p\!\leq\!d$-gon Feynman integrals is a very convenient thing, as these form a basis for (the non-vanishing part of) \emph{all} one loop Feynman integrals in $d$ spacetime dimensions.

But this near-purity is not in fact a general feature of scalar Feynman integrals even at one loop. Any integral with more propagators than the number of spacetime dimensions will always have a higher degree of impurity.  

Consider for example the \emph{scalar} pentagon integral (with no loop dependence in its numerator) in four dimensions; upon loop integration, it takes the form 
\begin{align}\nonumber\\[-30pt]\hspace{-5pt}\fig{-2.25pt}{1}{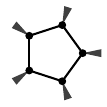}\hspace{-2pt}=&\hspace{-8pt}~\phantom{+}f_1\hspace{-4pt}\fig{-2.25pt}{1}{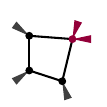}\hspace{-5pt}{+}f_2\hspace{-4pt}\fig{-2.25pt}{1}{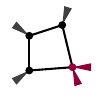}\hspace{-5pt}{+}f_3\hspace{-7pt}\fig{-2.25pt}{1}{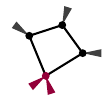}{+}f_4\hspace{-4pt}\fig{-2.25pt}{1}{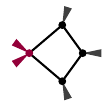}{+}f_5\hspace{-7pt}\fig{-2.25pt}{1}{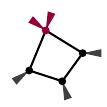}\label{pentagon_impurity_example}\\[-10pt]
&\hspace{-8pt}~{+}f_6\hspace{-4pt}\fig{-2.25pt}{1}{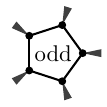}\nonumber\\[-34pt]\nonumber\end{align}
These pictures represent \emph{integrated} expressions. The `(parity-)odd' integral above is in fact zero when integrated against the parity-even contour of loop integration, and so it may be said that only the first 5 `box' contributions are relevant. Because each scalar box integral is proportional to something pure (\ref{purity_of_boxes_in_4d}) and because each of these integrals is linearly independent, the expansion (\ref{pentagon_impurity_example}) may be viewed as exactly in the form (\ref{impure_sum}); as each of the 5 (or 6) coefficients are linearly independent, we conclude that the scalar pentagon integral has an impurity of degree 5 (or 6)\footnote{We may want to discuss the purity of differential forms on loop momenta irrespective of the contours of loop integration; as a differential form, the scalar pentagon would then have an impurity of degree 6.}. Notice that this assessment is independent of the normalization of the pentagon integral, or even how the box integrals are normalized.\\[-8pt]

While at one loop (in any spacetime dimension), a \emph{basis} of master integrals may always be chosen to consist solely of \emph{scalar} $p\!\leq\!d$-gons (with loop independent numerators) which are individually pure upon appropriate normalization, the same is not true at two or more loops: integrals constructed using only scalar propagators (without loop dependence in their numerators) generally have higher degrees of impurity, which cannot be spanned by this class of Feynman integrals. 

At two loops in four spacetime dimensions, it turns out that all planar two-loop integrals with fewer than 8 propagators have impurities of degree one; but consider the scalar `penta-box' integral, which, upon loop integration takes the form\footnote{As with the scalar pentagon in four dimensions (\ref{pentagon_impurity_example}), there is also one further independent differential form that appears, but vanishes upon loop integration.}
\begin{align}\nonumber\\[-30pt]\hspace{-20pt}\fig{-2.25pt}{1}{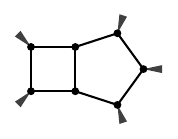}\hspace{-4pt}=&\hspace{-5pt}\phantom{{+}}f_1\hspace{-5pt}\fig{-2.25pt}{1}{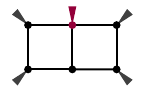}\hspace{-5pt}{+}f_2\hspace{-5pt}\fig{-2.25pt}{1}{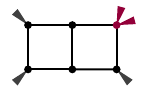}\hspace{-5pt}{+}f_3\hspace{-5pt}\fig{-2.25pt}{1}{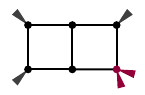}\hspace{-5pt}{+}f_4\hspace{-5pt}\fig{-2.25pt}{1}{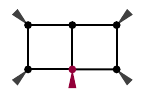}\hspace{-20pt}\label{pentabox_impurity_example}\\[-18pt]
&{+}f_5\big(\text{some \emph{other} {pure} function of weight 4}\big)\nonumber\\
&{+}f_6\big(\text{\hspace{0pt}\emph{yet another} \hspace{-1.75pt}{pure} function of weight 4}\big)\nonumber\\[-24pt]\nonumber\end{align}
Here, the figures represent \emph{scalar} Feynman integrals---those constructed with only scalar propagators according to the graph; the four `double-boxes' appearing in the first line of (\ref{pentabox_impurity_example}) are all proportional to (independent) pure functions of weight 4. It is natural to consider the the last terms in (\ref{pentabox_impurity_example}) as encoding the `(independent) pure parts' of the pentabox integral; but as they are non-vanishing, and cohomologically independent of the other terms in (\ref{pentabox_impurity_example}), they can only be described as \emph{linear combinations of integrals} of the form (\ref{standard_ibp_integrals})---or, equivalently, as integrals \emph{with non-trivial loop-dependence in their numerators}.\\[-8pt]

This example makes clear that in order to construct bases of master Feynman-like integrals that are individually pure---that is, that individually satisfy canonical differential equations---it is \emph{necessary} to allow for integrals with loop dependence in their numerators. This conclusion is independent of any preferences for infrared or ultraviolet structure (which also suggest that certain linear combinations of otherwise pure Feynman integrals be chosen for elements of a master integral basis).\\[-8pt]

\subsection[Complexities Beyond Polylogarithms: Ellipticity and Higher \emph{Rigidities}]{\mbox{Complexities Beyond Polylogarithms: Ellipticity \& Higher \emph{Rigidities}}}\label{rigidity_subsection}\vspace{-4pt}

While it is generally true that any one-loop Feynman integral in any quantum field theory (independent of its dimension or spectrum) is polylogarithmic \cite{Bourjaily:2019exo,Arkani-Hamed:2017ahv}, this is not true in general beyond one loop. 

Starting at two loops, four dimensional Feynman integrals can involve (any number of) elliptic curves \cite{Bourjaily:2017bsb} or even $K3$ surfaces \cite{Bourjaily:2018ycu,Bourjaily:2018yfy}, requiring a broader class of functions \emph{beyond} the generalized polylogarithms defined in (\ref{generalized_polylogarithm_defined}). Such generalizations to `elliptic polylogarithms' now exist and have been widely studied in recent years (see \emph{e.g.}~\cite{Walden:2020odh,Kristensson:2021ani,Bourjaily:2022bwx,Wilhelm:2022wow,Morales:2022csr,Frellesvig:2023bbf}). This is in part because their complexities pose a real bottleneck to many concrete computations---lacking, as we currently do, a level of understanding of these functions analogous to what has been developed for generalized polylogarithms. 

In \cite{Bourjaily:2018yfy}, the term \emph{rigidity} was introduced to encode the degree of complexity \emph{beyond} polylogarithms. Any meromorphic differential form is holomorphic on the complement of its divisors, and rigidity is best understood in terms of the geometry of $\mathbb{C}^{d\,L}\backslash\langle P_a(\vec{\ell})\cdots P_c(\vec{\ell}){=}0\rangle$---integrals representing elements of the cohomology of this space. Polylogarithms correspond to geometries which are everywhere locally the products of doubly-punctured Riemann spheres---the space on which $d\log(z){=}dz/z$ is an element of cohomology. 

Rigidity can be defined by the largest dimension of any connected component \emph{not} involving a doubly-punctured sphere. Thus, an elliptic integral would have rigidity one (as would any integral involving a higher-genus Riemann surface); a `$K3$' integral would involve the geometry of a K3 surface and have rigidity two, and so-on. 

It is not known what variety of algebraic varieties appear in Feynman integrals, but an upper bound on rigidity follows from the maximal weight ($\lfloor d\,L/2\rfloor$) of any Feynman integral and the fact that all one loop integrals are polylogarithmic. In four dimensional theories at two loops, rigidity is bounded by 2, and so we expect to encounter such Feynman integrals with this degree of novelty at two loops. 

Conveniently, integrals involving massless propagators are much less rigid---much more polylogarithmic---than those with massive propagators. And planar integrals appear to be less rigid than non-planar ones. At two loops, the only integral with rigidity two is the non-planar \emph{tardigrade} \cite{Bourjaily:2018yfy}
\vspace{-5pt}\eq{\fig{-2pt}{1}{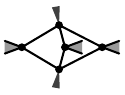}\label{tardigrade}\vspace{-10pt}}
while the planar integrals 
\vspace{-3pt}\eq{\fig{-2pt}{1}{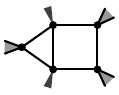}\qquad\text{and}\qquad\fig{-2pt}{1}{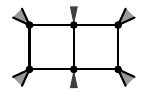}\label{elliptic_bt_and_double_box}\vspace{-3pt}}
are elliptic. In these figures, an external wedge braced by solid lines denotes a \emph{necessarily massive} inflowing momentum, while a wedge without this bracing denotes any \emph{non-vanishing} (but possibly massless) momentum. (These conventions are reviewed in \mbox{appendix~\ref{graphical_notation}}.) For amplitudes involving massless external particles, a massive inflowing momentum requires at least two particles involved; and so, the tardigrade (\ref{tardigrade}) is first encountered for integrals involving 8 external massless particles, and the the examples in (\ref{elliptic_bt_and_double_box}) require 8 and 10 external particles, respectively.\\[-8pt]

These novelties unquestionably cause complexity even at the level of building master integrals, and frustrate (if not impede) our ability to understand these functions analytically. But they pose no fundamental obstruction to integral reduction or computing periods: more rigid integrals simply require a specification of contours $\{\Omega_{\r{j}}\}$ and periods that extends beyond those required for polylogarithms.

For elliptic integrals, leading singularities defined with respect to their possible cycles are equally well defined and computable in terms of trees via unitarity \cite{Bourjaily:2021vyj,Bourjaily:2022tep}. Moreover, in the case of maximally supersymmetric Yang-Mills theory, these `novel' leading singularities are in fact Yangian invariant \cite{Bourjaily:2020hjv}.\\[-8pt]

\newpage
\section{Master Integral Bases Defined via \emph{Power-Counting}}\label{power_counting_masters}

\subsection[Organizing the Loop Dependence of Numerators of Feynman Integrals]{The Loop Dependence of \emph{Numerators} of Feynman Integrals}\vspace{-4pt}

Any one-loop Feynman integral can be expressed as a product of \emph{ordered} propagators with some loop-dependent numerator
\eq{I=\!\int\limits_{\ell\in\mathbb{R}^{d}}\!\!\!\!d^d\ell\,\,\,\mathcal{I}(\ell)\,,\qquad\text{with}\qquad\mathcal{I}(\ell)\equivL\frac{N(\ell)}{P_a(\ell)P_b(\ell)\cdots P_c(\ell)}\label{generic_integral_one_loop}}
where $N(\ell)$ is a \emph{polynomial} in $\ell$ (with coefficients that are rational functions of external momenta) and each (inverse) propagator may be written
\eq{P_i(\ell)\equivL(\ell{-}x_i)^2{-}m_i^2\;\;\bigger{\Leftrightarrow}\;\;\,\text{`}\x{\ell}{i}\text{'}\label{inverse_prop_general}}
for some choices of $x_i\!\!\in\!\mathbb{R}^{d}$ for each $i\!\in\!\{a,b,\ldots,c\}$. The notation `$\x{\ell}{i}$' introduced above in (\ref{inverse_prop_general}) may be understood as symbolic at this stage; but it takes on a sharper meaning when interpreted in embedding space \cite{Weinberg:2010fx,SimmonsDuffin:2012uy,Bourjaily:2019exo} or (a realization thereof,) momentum-twistor space \cite{Hodges:2009hk}. These formalisms, and the translations between them and momentum space, are briefly reviewed in \mbox{appendices~\ref{embedding_space_appendix} and \ref{momentum_twistors_appendix}}.

If we define $\ell_i\equivR\ell{-}x_i$ to be the momentum flowing through the `$i$th' propagator, then momentum conservation requires that the difference between consecutive pairs, $\ell_a{-}\ell_{b}\equivL\,p_a$, must equal the external momentum flowing into the diagram at the vertex connecting the edges with momenta $\ell_a$ and $\ell_{b}$. We may solve momentum conservation easily enough; using translational invariance of the $\ell$ integral, we are free to take $x_a\!\mapsto\!0\!\in\!\mathbb{R}^d$ and identify $\ell_a\equivR\ell$; upon making this choice, the momenta flowing through all other edges will be determined 
\eq{\ell_c=\ell{-}(p_a{+}\ldots{+}p_b)\equivL\, \ell{-}p_{a\cdots b}{=}\,\ell{-}x_c\,.}

Now, it is a general statement (easy to prove) that any polynomial $N(\ell)$ can be expressed in the space of polynomials constructed as products of inverse propagators. Let's define
\eq{[\ell]^k\equivR\mathrm{span}_{x^{[k]}\in\mathbb{R}^d}\prod_{\alpha\in[k]}(\ell{-}x^{\alpha})^2}
to be the vector space of Lorentz-invariant polynomials in $\ell$ constructed as $k$-fold products of inverse (massless) propagators, differing by their arbitrary displacements $x^{\alpha}\in\mathbb{R}^d$. The space $[\ell]^0$ consists of the single, \emph{loop-independent} monomial `$1$'\footnote{While we are typically interested in polynomials $N(\ell)$ with fixed scaling dimensions, these can always be viewed as arising from loop-independent \emph{coefficients}, and do not affect questions of independence or rank.}; and it is easy to see that $[\ell]^1\!\equivL[\ell]$ can be spanned by the $(d{+}2)$\footnote{In dimensional regularization, where $\ell\!\in\!\mathbb{R}^{d{-}2\epsilon}$, this number should be $(d{+}3)$---with the extra component relative to (\ref{naive_ell1_basis}) spanned by $\ell^{-2\epsilon}\equivR\ell\!\cdot\!e^{{-}2\epsilon}$.} (Lorentz-invariant inner products) monomials
\eq{[\ell]\simeq\mathrm{span}\{1,\underbrace{\ell\!\cdot\!e^1,\ldots,\ell\!\cdot\!e^d}_{\ell^\mu},\ell^2\}\,\label{naive_ell1_basis}\vspace{-4pt}}
and so-on. It is not hard to see (especially in the embedding space formalism) that these vector spaces of polynomials have rank
\eq{\mathrm{rank}([\ell]^k)=\binom{d{+}k}{d}{+}\binom{d{+}k{-}1}{d}\,\label{rank_of_ellk}}
for integer dimensions $d$. Importantly, \emph{any} polynomial $N(\ell)$ of degree $2k$ (or degree $2k{-}1$) in the components of $\ell$ appearing in a Feynman integral (\ref{generic_integral_one_loop}) must be expressible in terms of those of $[\ell]^k$ with loop-independent coefficients. 

One reason for a preference for the consideration of `$[\ell]$' over the irreducible scalar products \cite{Kosower:2011ty,Badger:2013gxa,Johansson:2013sda} appearing on the RHS of (\ref{naive_ell1_basis}), say, is that the former is both translationally and Lorentz invariant, while the latter requires a choice of origin for $\ell$ and a basis for its components. Moreover, even for massive propagators, $P_i(\ell)\in[\ell]$, and any pair of propagators whose momenta differ only by external legs will live in the same space $[\ell]$.

Notice that there is redundancy in these vector spaces; for example $[\ell]^0\subset[\ell]^1$; more generally, multiplication by $1$ induces an injective map $[\ell]^k\!\hookrightarrow\![\ell]^{k{+}1}$, which shows that $[\ell]^a\!\!\subset\![\ell]^b$ for all $a\!\leq\!b$. This is a triviality: polynomials of any degree can always be viewed as subspaces of those with higher degree; but it suggests that we define
\vspace{-6pt}\eq{\hat{[\ell]^k}\equivR[\ell]^k\backslash[\ell]^{k{-}1}\,,}
to consist of all polynomials constructed as $2k$ and $2k{-}1$-degree powers of the components of $\ell$ \emph{but not smaller}. From (\ref{rank_of_ellk}), it is easy to see that
\eq{\mathrm{rank}(\hat{[\ell]^k})=\binom{d{+}k}{k}{-}\binom{d{+}k{-}2}{k{-}2}\,.}

In terms of these, we can meaningfully discuss the vector space of Feynman integrals of the form 
\vspace{-4pt}\eq{\fig{-2pt}{1}{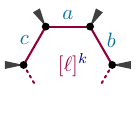}\;\bigger{\Leftrightarrow}\;\frac{\phantom{{}^k}\r{[\ell]}^{\b{k}}}{P_{\t{a}}(\r{\ell})P_{\t{b}}(\r{\ell})\cdots P_{\t{c}}(\r{\ell})}\,.\vspace{-10pt}}
These integrals may or may not be independent under cohomology, but their linear span can be used to define a set of one-loop master integrals.\\[-8pt]

This discussion generalizes to higher loops in a natural way \cite{Bourjaily:2020qca}. For example, we may consider the space of integrals of the form
\vspace{-4pt}\eq{\begin{split}\fig{-2pt}{1}{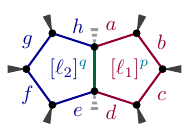}\;&\bigger{\Leftrightarrow}\\[-20pt]
&\hspace{-50pt}\frac{\phantom{{}^p}\r{[\ell_1]}^{\t{p}}\b{[\ell_2]}^{\t{q}}}{\r{P}_{\r{a}}(\r{\ell_1})\r{P}_{\r{b}}(\r{\ell_1})\r{P}_{\r{b}}(\r{\ell_1})\r{P}_{\r{c}}(\r{\ell_1})\r{P}_{\r{d}}(\r{\ell_1})\g{P(}\r{\ell_1}{-}\b{\ell_2}\g{)}\b{P}_{\b{e}}(\b{\ell_2})\b{P}_{\b{f}}(\b{\ell_2})\b{P}_{\b{g}}(\b{\ell_2})\b{P}_{\b{h}}(\b{\ell_2})}\,.\\[-2pt]\end{split}\label{example_two_loop_master_set}}
The master integrals we construct here are defined as integrands of this form. 

Because for any $\t{p},\t{q}\!>\!0$ these sets of numerators naturally span the space of their contact terms, it is natural to divide the space of such masters into `{\color{contactCount}contact terms}' spanned by their sub-topologies and those at the `{\color{topCount}top-level}', which are outside the space spanned by their contact terms. In the example (\ref{example_two_loop_master_set}), it turns out that for four-dimensional theories, for any choice of $\t{p},\t{q}\!>1$, these spaces of integrals are entirely spanned by their contact terms; this is a reflection of the Passarino-Veltman reduction into $\t{p}\!\leq\!d$-gons at one loop \cite{Melrose:1965kb,Passarino:1978jh}.\\[-8pt]

One advantage of discussing sets of masters in such a way is that their definition is \emph{manifestly} translationally invariant. This is in contrast to discussions of sets of integrals defined by numerators with fixed polynomial degrees of components of the loop momenta---the spaces of which depend (strongly) on choices of origin for the loop momenta.\\[-8pt]

\subsection{Building Master Integrals with Fixed Power-Counting}\label{building_masters_with_fixed_power_counting}\vspace{-4pt}

Consider the set of Feynman integrals defined a graph with $\r{p}$ propagators and numerators chosen from the set $[\ell]^{\b{k}}$:
\vspace{-4pt}\eq{\mathfrak{B}_{\r{p}}^{(\b{k})}\equivR\frac{[\ell]^{\b{k}}}{P_{\r{1}}(\ell)\cdots P_{\r{p}}(\ell)}\vspace{-6pt}}
Not only does $P_{\r{i}}(\ell)\!\in\![\ell]^{\r{k}}$ but so do all products of any subset of $\g{q}\!\leq\!\r{k}$ of the propagators. Thus, the space $\mathfrak{B}_{\r{p}}^{(\b{k})}$ naturally includes Feynman-like integrals involving any number of propagators from $\r{p}{-}\b{k}$ to as many as $\r{p}$. These subspaces are called \emph{contact terms} of the integral encoded by the original $\r{p}$ propagators. 

Moreover, it is easy to see that every element of $\mathfrak{B}_{\r{p}}^{(\b{k})}$ scales like the product of $(\r{p}{-}\b{k})$ propagators as $\ell\!\to\!\infty$, and thus we'd say that it lives within the space of masters with `$\g{q}\equivR(\r{p}{-}\b{k})$-gon power-counting':
\vspace{-4pt}\eq{\mathfrak{B}^{\g{q}}\equivR\bigoplus_{\b{r}{=}0}^{\infty}\mathfrak{B}^{\b{r}}_{\g{q}{+}\b{r}}\,.\vspace{-6pt}}
(Here, we consider these spaces to consist of all integrals with some number of propagators, not merely those of some fixed graph.) Interestingly, not all these spaces are independent: it is not hard to see that $\mathfrak{B}^{\b{r{+}1}}_{\g{q}{+}\b{r{+}1}}\subset\mathfrak{B}^{\b{r}}_{\g{q}{+}\b{r}}$ for all $\b{r}\!>\!(d{-}\g{q})$, and also for $\b{r}{=}(d{-}\g{q})$ if $(d{-}\g{q})\!>\!0$. This is a famous result of Passarino and Veltman \cite{Melrose:1965kb,Passarino:1978jh} (see also \cite{Bourjaily:2020qca} for a more thorough discussion in this notation).\\[-8pt]

When extended to higher loops, we define the space of $\g{p}$-gon power-counting integrals to be those which start from some set of Feynman integrals taken to be `\emph{scalar}' $\g{p}$-gons.\footnote{The precise definition of this set of integrals is up to the user, and need not have trivial loop dependence in their numerators.} At two loops, we define scalar $\g{p}$-gons to be defined by the Feynman integrals with \emph{girth} $\g{p}$ and for which any edge contraction would result in a graph with lower girth. Thus, when the scalar $\g{p}$-gon integrals for $\g{p}{=}2,3,4,5,\ldots$ would be given by the Feynman integrals
\vspace{-10pt}\eq{\begin{array}{@{}c@{}c@{}c@{}}
\text{scalar $\g{2}$-gons:}&\fig{-2pt}{1}{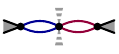}&\fig{-2pt}{1}{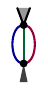}\\[-10pt]
\text{scalar $\g{3}$-gons:}&\fig{-2pt}{1}{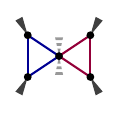}&\fig{-2pt}{1}{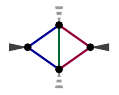}\\[-12pt]
\text{scalar $\g{4}$-gons:}&\fig{-2pt}{1}{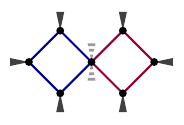}&\fig{-2pt}{1}{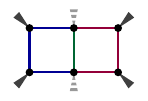}\\[-12pt]
\text{scalar $\g{5}$-gons:}&\fig{-2pt}{1}{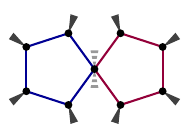}&\fig{-2pt}{1}{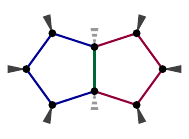}\\[-12pt]
\vdots&\vdots&\vdots\\[-14pt]
\end{array}\label{scalar_p_gons_defined}}
(For $\g{p}{=}d$, we may actually exclude the `kissing' topology for the scalars.) Importantly, each of these integrals is has an impurity of degree at most 1 in $d$ dimensions, and therefore can be normalized in order to define a \emph{pure} set of master integrals. 

The $\g{p}$-gon power counting basis is then defined as follows: we assign loop-independent numerators to the scalar $\g{p}$-gons defined above, and define the rest of the basis to consist of all Feynman-like integrals which scale like one or both of these scalars as $\ell\!\to\!\infty$; that is, it consists of the scalar graphs, combined with any additional propagators compensated by factors of $\r{[\ell_1]}$, $\b{[\ell_2]}$ or $\g{[\ell_1{-}\ell_2]}$ accordingly. In all cases, this defines a finite-dimensional subspace of cohomology and therefore a meaningful set of master integrals.\\[-8pt]

\begin{table}[b!]$$\fwbox{0pt}{\begin{array}{@{}l@{}l@{$\hspace{10pt}$}l@{}l@{}}\raisebox{14pt}{$\mathfrak{B}_{\r{4}\g{1}\b{4}}^4$}\fwboxR{24pt}{\fig{-18pt}{0.8}{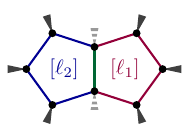}\hspace{-20pt}}&\rankBreakdownTable{\r{[\ell_1]}\b{[\ell_2]}}{2}{24}{2}{8}&\raisebox{14pt}{$\mathfrak{B}_{\r{4}\g{1}\b{3}}^4$}\fwboxR{19pt}{\fig{-18pt}{0.8}{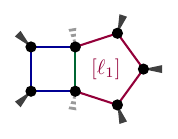}\hspace{-20pt}}&\rankBreakdownTable{\r{[\ell_1]}}{1}{4}{1}{0}\\[-18pt]
\multicolumn{4}{c}{\raisebox{14pt}{$\mathfrak{B}_{\r{3}\g{1}\b{3}}^4$}\fwboxR{24pt}{\fig{-12pt}{0.8}{313_box}\hspace{-12.05pt}}\rankBreakdownTable{1}{1}{0}{0}{0}}
\end{array}}$$\vspace{-24pt}\caption{Master integrals with box (`$\g{4}$-gon') power-counting in four dimensions.}\label{box_power_counting_basis}\vspace{-40pt}\end{table}

For example, the set of all planar two-loop integrals with `box' (or `$\g{4}$-gon' power-counting) in four dimensions are enumerated in \mbox{Table~\ref{box_power_counting_basis}}. This basis may simply be defined by the set of all integrals which scale asymptotically like the `double box' at infinite loop momentum. 
Notice that we have excluded the `kissing' topology, because
\vspace{-10pt}\eq{\fig{-2.5pt}{1}{414_box}\bigger{\;\supset\;\;}\fig{-2.5pt}{1}{404_box}\,.\vspace{-10pt}}

These integrals span the space of \emph{dual-}conformal integrals in four dimensions, and therefore span amplitudes in planar maximally supersymmetric Yang-Mills theory (`sYM'). Explicit forms of these integrals were defined in \cite{Bourjaily:2015jna} (see also \cite{ArkaniHamed:2010gh,Bourjaily:2013mma,Bourjaily:2017wjl}).\\[-8pt]

At one loop, all integrals with maximal weight are spanned by those with $(d{-}1)$-gon power counting, and we (strongly) suspect the same to be true at two loops. This is not to say that all integrals with triangle power counting in four dimensions will be maximal weight. Indeed, the scalar integral
\vspace{-10pt}\eq{\fig{-2pt}{1}{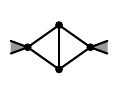}\vspace{-15pt}}
has triangle power counting by our definition, but  (with proper normalization) it is known to integrate to $20\zeta(3)=20\,\mathrm{Li}_3(1)$---a `number' of sub-leading weight.\\[-8pt]

\vspace{\fill}

\subsection[Two Loop Masters with Triangle Power-Counting in Four Dimensions]{\mbox{Two Loop Masters with Triangle Power-Counting in Four Dimensions}}\label{four_dimensional_masters_with_triangle_power_counting}\vspace{-1pt}

The space of master integrals with triangle power-counting in four dimensions should span all integrals with maximal transcendental weight. These integrals were enumerated in \cite{Bourjaily:2020qca}, and we summarize the counting of their degrees of freedom in \mbox{Table~\ref{triangle_pc_basis}}. Notice that we have been careful to separate degrees of freedom between those that are even under parity from those that are odd. Although parity-odd integrals are cohomologically non-trivial, they contribute nothing to integrated expressions, and many readers may find the parity-even counting of primary interest.

\begin{table}[b!]
$$\fwbox{0pt}{\begin{array}{@{}l@{}l@{$\hspace{10pt}$}l@{}l@{}}\raisebox{14pt}{$\mathfrak{B}_{\r{4}\g{1}\b{3}}^3$}\fwboxR{24pt}{\fig{-18pt}{0.8}{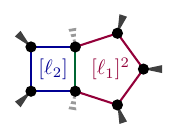}\hspace{-20pt}}&\rankBreakdownTable{\r{[\ell_1]^{2}}\b{[\ell_2]}}{2}{78}{2}{38}&\raisebox{14pt}{$\mathfrak{B}_{\r{4}\g{1}\b{2}}^3$}\fwboxR{19pt}{\fig{-18pt}{0.8}{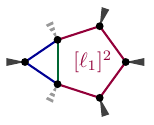}\hspace{-20pt}}&\rankBreakdownTable{\r{[\ell_1]^{2}}}{1}{14}{1}{4}\\[-16pt]
\raisebox{14pt}{$\mathfrak{B}_{\r{3}\g{1}\b{3}}^3$}\fwboxR{24pt}{\fig{-12pt}{0.8}{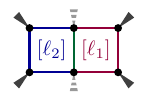}\hspace{-9.2pt}}&\rankBreakdownTable{\r{[\ell_1]}\b{[\ell_2]}}{4}{22}{4}{6}&\raisebox{14pt}{$\mathfrak{B}_{\r{3}\g{1}\b{2}}^3$}\fwboxR{19pt}{\fig{-12pt}{0.8}{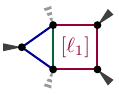}\hspace{-8pt}}&\rankBreakdownTable{\r{[\ell_1]}}{2}{3}{1}{0}\\[-10pt]
\raisebox{14pt}{$\mathfrak{B}_{\r{4}\g{0}\b{4}}^3$}\fwboxR{24pt}{\fig{-15pt}{0.8}{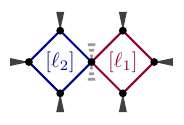}\hspace{-16pt}}&\rankBreakdownTable{\r{[\ell_1]}\b{[\ell_2]}}{2}{24}{2}{8}&\raisebox{14pt}{$\mathfrak{B}_{\r{4}\g{0}\b{3}}^3$}\fwboxR{19pt}{\fig{-15pt}{0.8}{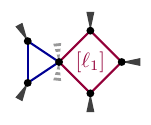}\hspace{-18pt}}&\rankBreakdownTable{\r{[\ell_1]}}{1}{4}{1}{0}\\[-10pt]
\raisebox{14pt}{$\mathfrak{B}_{\r{2}\g{1}\b{2}}^3$}\fwboxR{24pt}{\fig{-12pt}{0.8}{212_tri}\hspace{-3pt}}&\rankBreakdownTable{1}{1}{0}{0}{0}&\raisebox{14pt}{$\mathfrak{B}_{\r{3}\g{0}\b{3}}^3$}\fwboxR{19pt}{\fig{-12pt}{0.8}{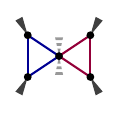}\hspace{-5.5pt}}&\rankBreakdownTable{1}{1}{0}{0}{0}
\end{array}}$$\vspace{-24pt}\caption{Master integrals with triangle (`$\g{3}$-gon') power-counting in four dimensions.}\label{triangle_pc_basis}\vspace{-40pt}\end{table}

\newpage
\subsection{\emph{Prescriptively} Choosing Basis Elements for Master Integrals}\label{prescriptively_defining_master_basis_ints}\vspace{-4pt}

So far, we have only described an vector space of master integrals defined by their power-counting; we have said nothing at all about the criteria used to choose \emph{particular} master integrals to form a basis. 

As we have seen, there are many reasons to prefer integrals which are \emph{pure}---equivalently, that satisfy canonical (or nipotent) differential equations. When purity was first introduced in \cite{ArkaniHamed:2010gh}, it was defined not in terms of differential equations, but by the periods or residues of Feynman integrals. 

Notice that we have been using `$\dbar^4\hspace{-1pt}\ell\text{'}\equivR\frac{d^4\hspace{-1pt}\ell}{(2\pi)^4}$. This is motivated in part so that 
\eq{\mathrm{Residue}\Big[\frac{d^4\hspace{-1pt}\r{\ell}}{f_1(\r{\ell})f_2(\r{\ell})f_3(\r{\ell})f_4(\r{\ell})},f_i(\r{\ell}){=}0\Big]=\hspace{-5pt}\oint\limits_{|f_i(\r{\ell})|=\epsilon}\hspace{-5pt}\frac{\dbar^4\hspace{-1pt}\r{\r{\ell}}}{f_1(\r{\ell})f_2(\r{\ell})f_3(\r{\ell})f_4(\r{\ell})}.}
In \cite{ArkaniHamed:2010gh}, an integral was defined as \emph{pure} if it had the property that \emph{all} its maximal co-dimension residues (equivalently, its polylogarithmic periods) were kinematic constants. We have avoided the use of `residue' here because it is known that compact contours (\emph{e.g.} elliptic) \emph{beyond} those that are \emph{polylogarithmic} are required to match amplitudes and define leading singularities \cite{Bourjaily:2020hjv}. The more general term of `periods'---defined as algebraic differential forms integrated over algebraically-prescribed cycles of integration \cite{kontsevich2001periods}---is more general, and necessary for our discussion of master integrals beyond one loop. 

Thus, to ensure that our masters are individually pure, it suffices to define them by the criterion that they are the cohomological duals \mbox{\cite{Mastrolia:2018uzb,Frellesvig:2019kgj,Frellesvig:2019uqt,Frellesvig:2020qot,Weinzierl:2020gda,Caron-Huot:2021iev,Caron-Huot:2021xqj}} of certain compact cycles.\\[-8pt]

Consider for example the double-box integral
\eq{\fig{-2pt}{1}{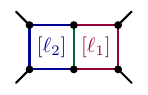}\bigger{\Leftrightarrow\;}\frac{\r{[\ell_1]}\b{[\ell_2]}}{\x{\r{\ell_1}}{{a}}\x{\r{\ell_1}}{{b}}\x{\r{\ell_1}}{c}\x{\r{\ell_1}}{\inf}\g{\x{\r{\ell_1}}{\b{\ell_2}}}\x{\b{\ell_2}}{c}\x{\b{\ell_2}}{d}\x{\b{\ell_2}}{a}\x{\b{\ell_2}}{\inf}}\,.\label{313_numerator_general}}
Here we have explicitly included the propagators involving the point at `infinity' as they are needed for embedding space (or, equivalently, \emph{line} at `infinity' in momentum-twistor space), to make it clear that some contours exist where one or both of the momenta are taken to infinity. This set of integrals defines a ${\color{totalCount}36}{=}\r{6}\!\times\!\b{6}{=}\mathrm{rank}(\r{[\ell_1]})\!\times\!\mathrm{rank}(\b{[\ell_2]})$-dimensional vector space of masters. 

We may start with \emph{any} choice of basis for these integrals (which need not be pure or in any way nice): $\{\hat{\mathcal{I}}_{\b{1}},\ldots,\hat{\mathcal{I}}_{\b{36}}\}$. For the set of propagators in (\ref{313_numerator_general}), there turn out to be \textbf{\r{156}} distinct compact contours of integration $\{\hat{\Omega}^{\r{1}},\ldots\hat{\Omega}^{\r{156}}\}$; all of these are polylogarithmic (and three of involve double-poles, signaling weight-drops). Using these contours and the initial (arbitrary) choice of basis, we construct the $\r{156}\!\times\!{\color{totalCount}36}$ period matrix
\eq{\hat{\mathbf{M}}\indices{\r{[156]}}{{\color{totalCount}[36]}}\equivR\oint\limits_{\hat{\Omega}^{\r{[156]}}}\hspace{-5pt}\dbar^4\hspace{-1pt}\ell_1\,\dbar^4\hspace{-1pt}\ell_2\,\,\hat{\mathcal{I}}_{{\color{totalCount}[36]}}\,.\label{initial_periods}}
This period matrix has rank 36, signaling that all masters initially chosen are cohomologically independent. From this long list of contours, we define 36 independent linear combinations to comprise our \emph{master contours}
\eq{\Omega^{\r{[36]}}\equivR\sum_{\t{j}\in\t{[156]}}\mathbf{P}\indices{\r{[36]}}{\t{j}}\hat{\Omega}^{\t{j}}\,,}
with constant coefficients encoded by the \emph{master contour projectors} $\mathbf{P}\indices{\r{[36]}}{\t{[156]}}$ such that $\mathbf{M}\indices{\r{[36]}}{{\color{totalCount}[36]}}\equivR\sum_{\t{j}\in\t{[156]}}\mathbf{P}\indices{\r{[36]}}{\t{j}}\hat{\mathbf{M}}\indices{\t{j}}{{\color{totalCount}[36]}}$ is full rank, and hence invertible. From these, our \emph{prescriptive} masters may be defined
\eq{\mathcal{I}_{{\color{totalCount}[36]}}\equivR\sum_{\t{j}\in\t{[36]}}(\mathbf{M}^{\text{-}1})_{{\color{totalCount}[36]}}^{\phantom{{\color{totalCount}[36]}}\t{j}}\,\hat{\mathcal{I}}_{\t{j}}\,,}
from which it is easy to see that 
\eq{\oint\limits_{\mathbf{P}\indices{\r{[36]}}{\t{j}}\hat{\Omega}^{\t{j}}}\hspace{-5pt}\dbar^4\hspace{-1pt}\ell_1\,\dbar^4\hspace{-1pt}\ell_2\,\,\mathcal{I}_{{\color{totalCount}[36]}}=\delta\indices{\r{[36]}}{{\color{totalCount}[36]}}}

It turns out that once we have diagonalized our masters against (any) spanning set of 36 {master contours}, \emph{all} the other periods are constant. This would not be true if we had only diagonalized against a subset of the 36 independent master contours---suggesting that there is a sense in which these masters form a meaningfully `complete' set.\\[-10pt]

\subsubsection{The Many Periods of Feynman Integrals}\label{periods_of_feynman_integrals}\vspace{-4pt}

There are of course many choices of master contours against which we may choose to define our master integrals, but all choices can be organized in a systematic way according to which subsets of propagators involving external momenta are cut. 

In the case of (\ref{313_numerator_general}), there are seven propagators involving the external kinematics; of the 156 compact contours of integration, there are 14 which put all these on-shell: 
\eq{\fwbox{0pt}{\begin{array}{@{}c@{}c@{}c@{}c@{}c@{}c@{}c@{}}\fwbox{60pt}{\fig{-2pt}{1}{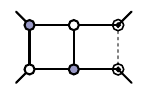}}&\fwbox{60pt}{\fig{-2pt}{1}{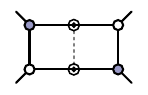}}&\fwbox{60pt}{\fig{-2pt}{1}{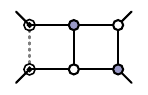}}&\fwbox{60pt}{\fig{-2pt}{1}{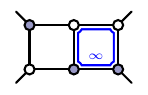}}&\fwbox{60pt}{\fig{-2pt}{1}{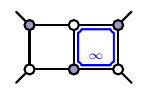}}&\fwbox{60pt}{\fig{-2pt}{1}{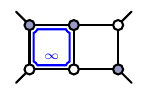}}&\fwbox{60pt}{\fig{-2pt}{1}{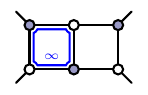}}\\[-10pt]
\fwbox{60pt}{\fig{-2pt}{1}{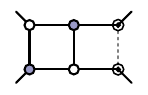}}&\fwbox{60pt}{\fig{-2pt}{1}{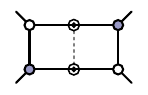}}&\fwbox{60pt}{\fig{-2pt}{1}{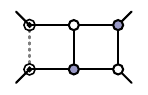}}&\fwbox{60pt}{\fig{-2pt}{1}{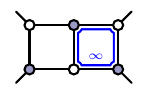}}&\fwbox{60pt}{\fig{-2pt}{1}{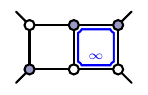}}&\fwbox{60pt}{\fig{-2pt}{1}{313_110110_i5}}&\fwbox{60pt}{\fig{-2pt}{1}{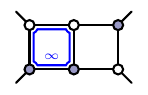}}\end{array}}\vspace{-10pt}\nonumber}
(The conventions for these figures are described in \mbox{appendix~\ref{graphical_notation}}.) The periods of any choice of the {\color{totalCount}36} initial masters (\ref{initial_periods}), when restricted to these 14 contours, span a space of rank {\color{topCount}8}---reflecting the number of `{\color{topCount}top-level}' degrees of freedom associated with the graph (see \mbox{Table~\ref{triangle_pc_basis}}). The rest of the space ({\color{contactCount}28} of the {\color{totalCount}36} masters) can be accounted for by \emph{{\color{contactCount}contact-terms}}:
\vspace{-35pt}$$\raisebox{-55pt}{$\fwbox{0pt}{\begin{array}{@{}c@{$\hspace{-8pt}$}c@{$\hspace{-8pt}$}c@{}}\\[00pt]\fig{-2pt}{1}{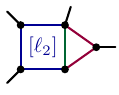}&\fig{-2pt}{1}{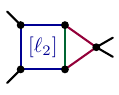}&\fig{-2pt}{1}{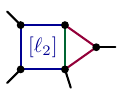}\\[-15pt]
\hspace{11pt}\fig{-2pt}{1}{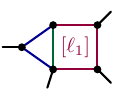}&\hspace{11pt}\fig{-2pt}{1}{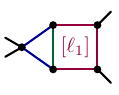}&\hspace{11pt}\fig{-2pt}{1}{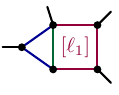}\\[-14pt]\end{array}\hspace{-9pt}\begin{array}{@{}c@{}}\\[0pt]\fig{0pt}{1}{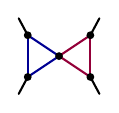}\\[-14pt]\end{array}\hspace{-9pt}\begin{array}{@{}c@{$\hspace{-3pt}$}c@{$\hspace{-3pt}$}c@{$\hspace{-3pt}$}c@{}}\\[0pt]\fig{-2pt}{1}{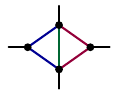}&\fig{-2pt}{1}{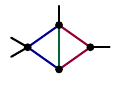}&\fig{-2pt}{1}{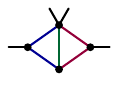}&\fig{-2pt}{1}{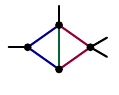}\\[-14pt]
\fig{-2pt}{1}{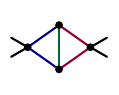}&\fig{-2pt}{1}{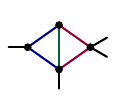}&\fig{-2pt}{1}{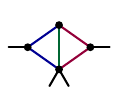}&\fig{-2pt}{1}{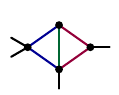}\\[-15pt]\end{array}}$}\vspace{-80pt}$$
\newpage

\noindent(Here, the first double-triangle integral on the upper-left should be counted twice, as there are two images with distinct leg distributions.) These contact terms include 6 box-triangles (of types $[\r{3}\g{1}\b{2}]$ and $[\r{2}\g{1}\b{3}]$, each representing a {\color{topCount}3}-dimensional vector space of masters (independent of their contact terms) together with 10 scalar integrals. Each scalar integral clearly can be used to span one element of the vector space of masters in (\ref{313_numerator_general}). 

Using the contours we have chosen to define our masters (see \mbox{appendix~\ref{prescriptive_definition_of_masters}}), the final set of individually pure master integrals for the topology (\ref{313_numerator_general}) consists of the following set of masters:
\eq{\fwboxR{0pt}{\hspace{20pt}\fig{40pt}{1}{313_110110_num}\hspace{-55pt}}\hspace{30pt}\fig{-2pt}{1}{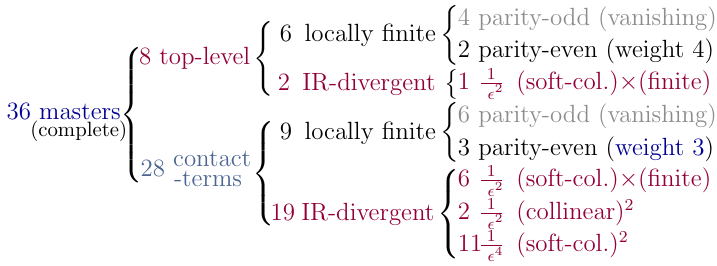}}

Depending on the distribution of massless momenta flowing into a given graph, the number and variety of contours change considerably. This requires that different choices be made for each case.

\subsubsection{On Choosing a Spanning Set of Cycles}\label{choosing_cycles_wisely}\vspace{-4pt}

Although any spanning set of master contours would be sufficient to prescriptively define a concrete set of individually pure master integrals, the structure of the basis and the form of amplitudes represented in terms thereof changes considerably depending on which choices are made. 

We have chosen our master contours to span all independent regions associated with soft or collinear infrared divergence. Once this has been done, all \emph{other} master integrals must manifestly vanish on these contours (by the orthogonality of their master periods). To be clear: this does not mean that every distinct region associated with such a divergence can be associated with a particular master contour choice; nevertheless, we have checked that (at least for our choices of master contours), orthogonality against these subsets of regions associated with infrared divergence \emph{suffice} to ensure the local finiteness \cite{Bourjaily:2021ewt} of all integrals \emph{not} defined by a contour associated with such a divergence. (We do not know the degree to which this was guaranteed or merely fortunate.)

To illustrate this point, consider an integral of type `\ref{int312|11111}' (in notation described in \mbox{appendix~\ref{organizing_principles}}):
\eq{\fig{-2pt}{1}{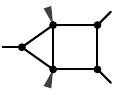}}
This integral supports 11 distinct polylogarithmic contours which cut all the 6 propagators involving external momenta; among these, the 7 contours 
\vspace{-5pt}\eq{\hspace{-20pt}\fig{-2pt}{1}{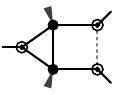}\fig{-2pt}{1}{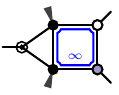}\fig{-2pt}{1}{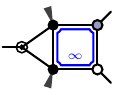}\fig{-2pt}{1}{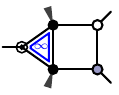}\fig{-2pt}{1}{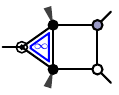}\fig{-2pt}{1}{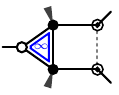}\fig{-2pt}{1}{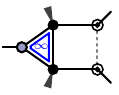}\hspace{-15pt}\label{312_divergent_contours}\vspace{-5pt}}
are all associated with {some} region of infrared divergence. Only three linear combinations of these periods could be chosen to define our masters with this topology; we have chosen the first three of those above (see \ref{omega312|11111}). Thus, it at least \emph{seems} highly non-trivial that integrals such as that defined dual to the contour
\vspace{-10pt}\eq{\fig{-2pt}{1}{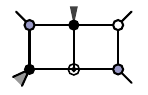}\vspace{-10pt}}
which is only \emph{ensured} to manifestly vanish on the first three contours of (\ref{312_divergent_contours}), also happens to vanish on all of these. (This does \emph{not} happen for masters of topology \ref{int313|110211} defined by contours which themselves are dual to contours associated with infrared divergence---see \ref{omega313|110211}).\\[-8pt]

Beyond a preference for master contours involving regions associated with infrared divergence, there are a number of general principles that should guide any choice of master contours. (These lessons were described in \cite{Bourjaily:2021hcp}, after they were learned the hard way through the works of \cite{Bourjaily:2019iqr,Bourjaily:2019gqu,Bourjaily:2021iyq}.) Chief among these principles is that the \emph{set} of master contours should map to itself (up to a signed permutation of the identity) under \emph{parity} or any discrete graph symmetries. 

To illustrate this point, while the contours 
\vspace{-10pt}\eq{\left\{\rule{0pt}{20pt}\right.\fig{-2pt}{1}{312_11111_1},\fig{-2pt}{1}{312_11111_2},\fig{-2pt}{1}{312_11111_inf1}\left.\rule{0pt}{20pt}\right\}\vspace{-14pt}}
\emph{could have been} chosen to prescriptively define the masters of type \ref{int312|11111}, these contours do not form a closed set under parity; as such, the corresponding masters would not form a closed set under parity, and all integrals (`parents') which contain these integrals as contact terms would themselves carry chiral remnants of this choice. 

Similarly, for the case of \ref{int313|110111}, whose master contours \ref{omega313|110111} are defined by
\vspace{-6pt}\eq{\hspace{-100pt}\fwboxL{280pt}{\hspace{-80pt}\begin{array}{@{}l@{}}\begin{array}{@{}c@{}c@{}c@{}c@{}c@{}c}
\fwboxR{0pt}{\raisebox{2pt}{$\left\{\rule{0pt}{20pt}\right.$}\hspace{-8.5pt}}\cfig{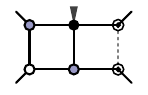}\fwboxL{0pt}{,}&\cfig{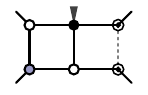}\fwboxL{0pt}{,}&\cfig{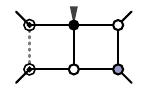}\fwboxL{0pt}{,}&\cfig{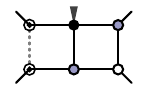}\fwboxL{0pt}{,}&\cfig{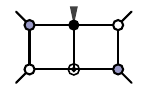}\fwboxL{0pt}{,}&\cfig{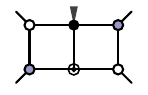}\fwboxL{0pt}{,}\\[-10pt]
\phantom{{}_{1}}\Omega_1&\phantom{{}_{2}}\Omega_2&\phantom{{}_{3}}\Omega_3&\phantom{{}_{4}}\Omega_4&\phantom{{}_{5}}\Omega_5&\phantom{{}_{6}}\Omega_6
\end{array}\\[-5pt]
\begin{array}{@{}c@{}c@{}c@{}c@{}c@{}c}
\cfig{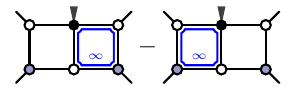}\fwboxL{0pt}{\hspace{0pt},}&\cfig{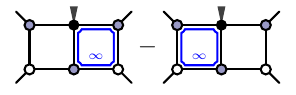}\fwboxL{0pt}{\hspace{-8.5pt}\raisebox{2pt}{$\left.\rule{0pt}{20pt}\right\},$}}\\[-10pt]
\phantom{{}_{1}}\Omega_7&\phantom{{}_{8}}\Omega_8
\end{array}\\[45pt]~
\end{array}}\vspace{-45pt}
\hspace{-100pt}}%
if we had traded $\Omega_7$ and $\Omega_8$ for 
\vspace{-8pt}\eq{\left\{\rule{0pt}{20pt}\right.\fig{-2pt}{1}{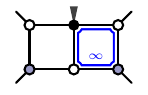},\fig{-2pt}{1}{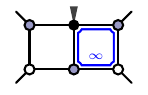}\left.\rule{0pt}{20pt}\right\}\vspace{-8pt}}
the resulting integrals would not map onto themselves under the reflection symmetry of the graph, resulting in asymmetric prescriptions for the masters associated with parent integrals
\vspace{-4pt}\eq{\fig{-2pt}{1}{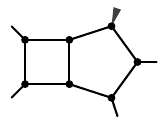}\quad\text{and}\quad\fig{-2pt}{1}{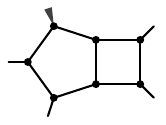}\vspace{-4pt}.}

\section{Conclusions and Future Directions}\label{conclusions}\vspace{-3pt}

A key advantage of a \emph{prescriptive} basis of master integrands is that the coefficients of any amplitude can be directly calculated in terms of on-shell functions. In the case of maximally supersymmetric Yang-Mills theory in the planar limit, the basis we have described here necessarily includes all contributions to all scattering amplitudes at two loops; it would be interesting to compare how amplitudes would be represented in this basis relative to those used in \cite{Bourjaily:2015jna}, or to what results from on-shell recursion \cite{BCF,BCFW} extended to loop-level \cite{CaronHuot:2010zt,ArkaniHamed:2010kv,Bourjaily:2023apy}. This has in fact already been implemented, the details of which will be described in the forthcoming \cite{bourjaily:amplitudesInNewBasis}.

Besides the fact that leading singularities are \emph{individually} matched in this basis, the fact that the infrared structure is so carefully refined provides a source of optimism that observables such as the ratio function may be rendered \emph{manifestly} locally finite in terms of these masters \cite{Bourjaily:2021ewt}. This is indeed the case, but involves interesting novelties: it turns out that infrared divergences must and do cancel in a non-trivial way between two-loop masters and the products of one-loop masters, leading to a new class of infrared-finite integrals whose denominators cannot be interpreted as arising from any single Feynman diagram. This will be described in more detail in the forthcoming \cite{bourjaily:IRstructureOfamplitudes}. 

But as emphasized above, even for theories whose amplitudes are \emph{not} maximally transcendental (such as pure Yang-Mills), these masters unambiguously capture all leading-weight contributions to any observable. These would of course a non-trivial and important subset of the contributions to amplitudes. Using the generalized superfunctions described in \cite{Bourjaily:2025tpr}, it should be possible to encode all such contributions using supersymmemtric coefficients (an improvement relative to \cite{Bourjaily:2021ujs}). Similarly, the tools recently developed for describing Yang-Mills coupled to charged matter \cite{Bourjaily:2026adf} should allow these masters to be used for more general QCD amplitudes. We hope to explore these applications in the near future.\\[-8pt]

Although we have complained about the difficulty of integrating more familiar master integrals such as those of (\ref{standard_ibp_integrals}), it is fair to observe that the masters presented here have yet to be integrated. As they are individually pure, and specifically tailored to particular regions of loop-momentum support, we strongly suspect these masters to be integrable using methods of differential equations or direct integration (see \emph{e.g.}~\cite{Kotikov:1990kg,Bern:1992em,Gehrmann:1999as,Bourjaily:2018aeq,Bourjaily:2019jrk,Bourjaily:2019vby,Bourjaily:2021lnz,Spiering:2024sea}).

At present, there appear to be a number of related sources of difficulty in directly applying standard methods of integration to the masters we have described. One source is the lack of existing computational tools to compute differential equations \emph{directly} in terms of masters expressed in terms of momentum twistor variables (but see \emph{e.g.}~\cite{Drummond:2010cz}). Another source is the fact our basis is \emph{incomplete}: because we have only described masters which span integrals of maximal weight, their derivatives with respect to external kinematics lie outside their span. (Of course, differentiation itself could be used to \emph{generate} the span of these lower-weight sets of masters, but this may involve considerable work.) 

Arguably the most straightforward way to apply the existing tools of using differential equations to these masters would require that they be re-expressed in terms of ordinary momentum variables. (This would also have the advantage that a much wider subset of researchers would know what to do with them.) In principle, this is should be straightforward (if laborious) exercise---perhaps one assigned to some large language model. The primary challenge we see in doing this is the algebraic complexity of representing and computing periods analytically in momentum-space variables. 

Translating our masters into more familiar momentum-space variables would also have the advantage that it would be easier to generalize them to include all \emph{non-planar} masters of maximal weight. This would not only provide an interesting comparison with the results described in refs.~\cite{Bourjaily:2019iqr,Bourjaily:2019gqu,Bourjaily:2021iyq}, say, but increase the applicability of these masters to a broader class of quantum field theories in four dimensions.\\[-8pt]

One aspect about our masters which has been (intentionally) de-emphasized in our present work is the fact that we have chosen \emph{not} to use elliptic cycles of integration to define elliptic masters, as promoted in \cite{Bourjaily:2022tep}. Rather, we have used evaluation of integrals along their maximal cuts to define these masters. This choice was motivated primarily by a desire to minimize the `algebraic' complexity of masters, but the result is that many of our masters are likely to involve functions of \emph{mixed rigidity}: combinations of polylogarithmic and elliptic-polylogarithmic functions; moreover, masters of mixed rigidity are likely to involve \emph{multiple} elliptic curves. 

While it would be justifiable to declare these as `impurities' which can (and perhaps arguably \emph{should}) be removed in the way described in \cite{Bourjaily:2022tep}, this is not \emph{necessary} for the use of unitarity in the representation of amplitudes. Choosing elliptic contours to define elliptic masters would have the benefit that all coefficients of amplitudes in sYM (at least) would be Yangian-invariant,  \cite{Drummond:2009fd,Bourjaily:2020hjv}; however, this comes at an (enormous) cost in algebraic and structural complexity in their definition. As of this writing, it seems preferable to \emph{not} require that elliptic integrals be chosen pure.\\[-8pt]

Finally, there are a number of clear avenues of further exploration and extension. It would be interesting, for example, to explore \emph{other} choices of master contours; even requiring that all independent regions associated with infrared divergence are chosen as master contours, there is a considerable range of possible choices that one can make. Our choices have largely been influenced by a desire to separate masters according to their support in maximally supersymmetric Yang-Mills theory, but we have not been nearly as observant regarding the possible support of contours in pure Yang-Mills, say. In pure Yang-Mills, there are a number of contours involving infinite loop momenta with support, and it may be advantageous to choose a set of masters more judiciously in order to most directly capture such contributions. We leave such investigations to future work. 

It would also be natural to extend these ideas to higher loop-orders. Although it was shown in \cite{Bourjaily:2017wjl} and that all amplitudes in planar sYM can be expressed in terms of masters with box power-counting at three loops,\footnote{This is also to be true at least for MHV amplitudes in \emph{non-planar} maximally supersymmetric Yang-Mills \cite{Bourjaily:2018omh}.} the only amplitudes for which closed formulae are known today to arbitrary multiplicity are MHV amplitudes \cite{ArkaniHamed:2010gh}, and these involve individual integrals with spurious, overlapping support and also cancelling contributions of sub-leading weight. It would be nice to improve these expressions and extend them to all amplitudes with arbitrary N${}^k$MHV-degree.

\vspace{\fill}
\subsection*{Acknowledgements}
This work was supported in part by a grant from the US Department of Energy (No.\ DE-SC00019066) and the J.~Robert Oppenheimer Visiting Professor Endowment at the Institute for Advanced Study, Princeton. This work was performed in part at Aspen Center for Physics, which is supported by National Science Foundation grant PHY-2210452.

\newpage
\appendix

\section{Dual Momenta, Embedding Space \& Momentum Twistors}\label{kinematics_appendix}\vspace{-4pt}

In the description of our constructions here, we have tried (for the most part) to be agnostic about the choice of variables used to define master integrals or their contours, both because our results do not depend on such choices and because not all readers may be equally familiar with the peculiarities of notation required for various choices. 

That being said, the \emph{explicit} numerators provided in the ancillary files included as this work's submission the the \texttt{arXiv} are specifically encoded in terms of invariants involving momentum twistor variables. Because we anticipate some readers who may want to \emph{use} these results are less familiar with these variables and constructions built from them, we review some of the most relevant material here. Interested readers may wish to consult \mbox{\emph{e.g.}~\cite{Weinberg:2010fx,SimmonsDuffin:2012uy,Bourjaily:2019exo}} for a more detailed discussion.

\subsection{Dual Momenta for Planar Feynman Integrals}\label{dual_momenta_appendix}\vspace{-4pt}

As described in \mbox{section~\ref{building_masters_with_fixed_power_counting}} above, it is very natural to describe propagators of plane Feynman diagrams in terms of the \emph{differences} of moment flowing through various edges. This corresponds to the choice that we encode the momentum flowing through the `$a$th' propagator as involving the squared momentum $P_a(\ell)\equivL(x_{\ell}{-}x_{a})^2$. When two adjacent propagators differ by some inflowing momentum, the inflowing momentum is encoded by the differences between these `dual-momentum' coordinates: 
\eq{\hspace{-30pt}\fig{-2pt}{1.25}{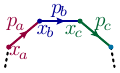}\quad \dbar^4\hspace{-1pt}\ell\,\frac{1}{P_{\r{a}}(\ell)P_{\b{b}}(\ell)P_{\g{c}}(\ell)\cdots}=\dbar^4\hspace{-1pt}x_{\ell}\,\frac{1}{(x_\ell{-}\r{x_a})^2(x_{\ell}{-}\b{x_b})^2(x_\ell{-}\g{x_c})^2\cdots}\hspace{-20pt}\label{momentum_differences_general}}
with the momentum flowing into each vertex defined as $p_{\r{a}}\equivL \b{x}_{\b{b}}{-}\r{x}_{\r{a}}$, and so-on. 

Without loss of generality, the momentum flowing into any planar Feynman integral can be viewed as arising from a set of massless momenta, where two `legs' encode the momentum of any non-null momentum. Thus, we may restrict our attention to the situation in which all the momentum involved in a Feynman graph are encoded by some number $n$ (greater than or equal to the number legs of the graph) of momentum-conserving, massless momenta, conventionally ordered clockwise around the graph. 

For a set of $[n]$ ordered momenta $\{p_{a}\}$ with $a\!\in\![n]\equivR\{1,\ldots,n\}$, we associate $p_a\equivL x_{a{+}1}{-}x_{a}$ with cyclic indexing understood.

\subsection{Embedding Space: General Notation and Invariants}\label{embedding_space_appendix}

The space of inverse propagators---which are of polynomial degree two in the components of loop momenta---can be embedded naturally into the linear span of inverse propagators in the embedding space formalism (see \emph{e.g.}~\cite{Weinberg:2010fx,SimmonsDuffin:2012uy,Bourjaily:2019exo}). Here, each $x^\mu$ with $\mu\!\in\![d]$ gets mapped to a point described by $M\in[d{+}2]$ homogeneous coordinates on $\mathbb{P}^{d{+}1}$ according to 
\eq{x_a^\mu\mapsto|a)^{M}\equivR\left(\begin{array}{@{}c@{}}x_a^{\mu}\\x_a^\mu x_a^\nu\,g_{\mu\nu}\\1\end{array}\right)}
with a metric defined by 
\eq{g^{M\,N}\equivR\left(\begin{array}{@{$\hspace{-2pt}$}c@{$\;$}c@{$\;\;\;\;$}c@{$\;$}}-2g^{\mu\nu}&0&0\\0&0&1\\0&1&0\end{array}\right)}
so that
\eq{\x{a}{b}\equivR |a)^M|b)^N\,g_{MN}=(x_a{-}x_b)^2\,\x{a}{\inf}\x{b}{\inf}}
where we have made explicit the reference to a point `at infinity' 
\eq{|\inf)\equivR\left(\begin{array}{@{$\hspace{-2pt}$}c@{}}\phantom{{}^\mu}0^\mu\\1\\0\end{array}\right)\,.}
In terms of these the integration measure becomes
\eq{\dbar^4\ell\hspace{-1pt}=\dbar^4\hspace{-1pt}x_{\ell}=\dbar^6\hspace{-1pt}|\ell)\frac{(2\pi\,i)^2\delta(\x{\ell}{\inf}{-}1)\delta(\x{\ell}{\ell})}{\x{\ell}{\inf}^4}\equivL\big[\dbar^5\hspace{-1pt}|\ell)\big](2\pi\,i)\delta(\x{\ell}{\ell})\equivL\big[\dbar^4\hspace{-1pt}|\ell)\big]\,,}
so that, for example,
\eq{\dbar^4\hspace{-1pt}x_{\ell}\frac{1}{(x_\ell{-}x_{\r{a}})^2(x_{\ell}{-}x_{\b{b}})^2(x_{\ell}{-}x_{\g{c}})^2(x_{\ell}{-}x_{\t{d}})^2}=\big[\dbar^4\hspace{-1pt}|\ell)\big]\frac{\x{\r{a}}{\inf}\x{\b{b}}{\inf}\x{\g{c}}{\inf}\x{\t{d}}{\inf}}{\x{\ell}{\r{a}}\x{\ell}{\b{b}}\x{\ell}{\g{c}}\x{\ell}{\t{d}}}}
while
\eq{\dbar^4\hspace{-1pt}x_{\ell}\frac{(x_{\r{a}}-x_{\g{c}})^2(x_{\b{b}}{-}x_{\t{d}})^2}{(x_\ell{-}x_{\r{a}})^2(x_{\ell}{-}x_{\b{b}})^2(x_{\ell}{-}x_{\g{c}})^2(x_{\ell}{-}x_{\t{d}})^2}=\big[\dbar^4\hspace{-1pt}|\ell)\big]\frac{\x{\r{a}}{\g{c}}\x{\b{b}}{\t{d}}}{\x{\ell}{\r{a}}\x{\ell}{\b{b}}\x{\ell}{\g{c}}\x{\ell}{\t{d}}}}
would be `dual-(momentum-)conformal'. (Note: this integral is \emph{not} pure.)

Notice that embedding space makes it obvious that there are \emph{two} solutions to the cut-equations of any box integral---as the only non-linear constraint appears via the `simplicity' part of the measure $\delta(\x{\ell}{\ell})$.

\subsection{A \emph{Chiral} Realization of Embedding Space: Momentum Twistors}\label{momentum_twistors_appendix}

Momentum twistor space may be viewed as a \emph{chiral} realization of embedding space, motivated by (ordered) sets of massless momenta. The idea here is to associate with each point $x_a$, not merely an abstract point $|\ell)$ in some projective space, but as a \emph{line} `$(a)$' in momentum twistor space, $\mathbb{P}^3$,  \cite{Hodges:2009hk}. By this we mean the linear \emph{span} of a pair of distinct twistors $\bar{z}_i\!\in\!\mathbb{P}^3$, represented homogeneously in terms of four homogeneous coordinates $z_i^a$. That is 
%
%
\eq{\r{x_a}\mapsto\mathrm{span}\{z_{\r{a{-}1}},z_{\r{a}}\}\equivL\text{`}\r{(a)}\text{'}\qquad\hspace{-0pt}\fig{-2pt}{1.25}{x_and_ps_glyph}\quad\bigger{\Rightarrow}\quad\fig{-2pt}{1.25}{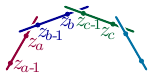}\qquad\,}

In (momentum) twistor space, the only fundamental invariant is the \emph{four-bracket},\\[-8pt]
\eq{\ab{\r{a}\,\b{b}\,\g{c}\,\t{d}}\equivR\mathrm{det}(z_{\r{a}},z_{\b{b}},z_{\g{c}},z_{\t{d}})\,,}
and two lines $\r{(a)}$ and $\b{(b)}$ intersect iff their corresponding points $\r{x_{a}}$ and $\b{x_{b}}$ are light-like separated. In momentum twistor space, this is tested by whether or not the four-bracket $\ab{\r{(a)}\b{(b)}}\equivR\ab{\r{a\hspace{1pt}\text{-}1\,a}\,\b{b\hspace{1pt}\text{-}1\,b}}=0$.

A null four-momentum $p_a\equivL x_{a{+}1}{-}x_a$ is therefore encoded by the pair of lines $(a{+}1)\equivR\mathrm{span}\{z_{a{+}1},z_a\}$ and $(a)\equivR\mathrm{span}\{z_a,z_{a{-}1}\}$, which automatically intersect at a point---namely, $z_a$. Thus, an arbitrary configuration of $n$ twistors $z_a$ automatically encodes a momentum-conserving set of massless momenta defined by the cyclic pairs of consecutive lines. 

The principal distinction between momentum twistors and normal embedding space is that rather than having a simple point at infinity such that $\x{a}{\inf}{=}1$, we have a bi-twistor $(\inf)\equivR\mathrm{span}\{z_{\inf_1},z_{\inf_2}\}$ defined so that $\ab{\r{(a)}(\inf)}\equivR\ab{\r{a\hspace{1pt}\text{-}1\,a}}\equivR\mathrm{det}\{\lambda_{\r{a{-}1}},\lambda_{\r{a}}\}$ where $\lambda$'s are the ordinary spinor-helicity variables $p_{\r{a}}\equivR\lambda_{\r{a}}\tilde\lambda_{\r{a}}$. 

A loop momentum is then represented by a line `$(\ell)$' which is itself encoded by the span of a pair of momentum twistors. By long tradition, we associate the lines $(\ell_1)$ and $(\ell_2)$ with the pairs of twistors $\{A,B\}$ and $\{C,D\}$. The measure on loop momentum space is then
\eq{\dbar^4\hspace{-1pt}x_{\ell}=\frac{\big[\ab{A\,B\,\dbar^2\hspace{-1pt}z_A}\big]\big[\ab{A\,B\,\dbar^2\hspace{-1pt}z_B}\big]}{\ab{(\ell)(\inf)}^4}\equivL\big[\dbar^4\hspace{-1pt}(\ell)\big]\,.}
So that, in terms of these, 
\eq{\frac{\dbar^4\hspace{-1pt}x_{\ell}}{(x_\ell{-}x_{\r{a}})^2(x_{\ell}{-}x_{\b{b}})^2(x_{\ell}{-}x_{\g{c}})^2(x_{\ell}{-}x_{\t{d}})^2}=\big[\dbar^4\hspace{-1pt}(\ell)\big]\frac{\ab{\r{(a)}(\inf)}\ab{\b{(b)}(\inf)}\ab{\g{(c)}(\inf)}\ab{\t{(d)}(\inf)}}{\ab{(\ell)\r{(a)}}\ab{(\ell)\b{(b)}}\ab{(\ell)\g{(c)}}\ab{(\ell)\t{(d)}}}.}

Due to the sharp equivalence with embedding space, we feel empowered to liberally confuse these notation, writing $\x{\r{a}}{\b{b}}\Leftrightarrow\ab{\r{(a)}\b{(b)}}$---provided that there is understanding that the meaning of `infinity' changes slightly between the two systems of coordinates.

\subsubsection{Geometrically Defined Points in Momentum-Twistor Space}\vspace{-4pt}

One key advantage to the momentum twistor formalism (and the reason our expressions have been written in this way) is that it is easier to describe \emph{chiral} solutions to cut equations. For example, the two solutions to cutting four propagators\\[-8pt]
\vspace{-4pt}\eq{\fig{-2pt}{1}{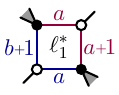}\quad\text{and}\quad\fig{-2pt}{1}{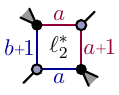}\vspace{-4pt}}
would correspond to the pair of lines $(AB)_i^*$ in momentum twistor space that simultaneously intersect the four lines $\r{(a\hspace{1pt}\text{-}1\,a)},\r{(a\,a\text{+}1)},\b{(b\hspace{1pt}\text{-}1\,b)},$ and $\b{(b\,b\text{+}1)}$. One solution would be $(AB)_1^*\equivR\text{`}(\r{a}\,\b{b})\text{'}$, and another would be the intersection of the two planes $(AB)_2^*\equivR\r{(a\hspace{1pt}\text{-}1\,a\,a\text{+}1)}\tcap\b{(b\hspace{1pt}\text{-}1\,b\,b\text{+}1)}$. In the former, the $\lambda$'s of the momenta will be parallel (so the contour is labelled white), while in the later, the $\tilde\lambda$'s would be parallel (so the massless vertices of the contour are coloured blue). 

Such \emph{geometrically-defined} points in momentum-twistor space may be defined concretely by the simple fact that any twistor may be expanded into a basis of any (linearly independent) others. In terms of homogeneous coordinates, this fact is encoded by \emph{Cramer's rule} \cite{Cramer1750}:
\vspace{-4pt}\eq{\r{z_{\r{a}}}\ab{\b{b}\,\g{c}\,\t{d}\,e}{+}\b{z}_{\b{b}}\ab{\g{c}\,\t{d}\,e\,\r{a}}{+}\g{z}_\g{c}\ab{\t{d}\,e\,\r{a}\,\b{b}}{+}\t{z}_{\t{d}}\ab{e\,\r{a}\,\b{b}\,\g{c}}{+}z_e\ab{\r{a}\,\b{b}\,\g{c}\,\t{d}}{=}0\,.\vspace{-4pt}}
This identity allows us to define the intersection between any line and plane as the rank-one space
\eq{\begin{split}\\[-24pt](\r{a\,b})\tcap(\b{c\,d\,e})&\equivR \r{z}_{\r{a}}\ab{\r{b}\,\b{c\,d\,e}}{+}\r{z}_{\r{b}}\ab{\b{c\,d\,e}\,\r{a}}\\
&\equivL{-}(\b{c\,d\,e})\tcap(\r{a\,b})\hspace{40pt}={-}\big(\hspace{-1pt}\b{z}_{\b{c}}\ab{\b{d\,e}\,\r{a\,b}}{+}\b{z}_{\b{d}}\ab{\b{e\,c}\,\r{a\,b}}{+}\b{z}_{\b{e}}\ab{\b{c\,d}\,\r{a\,b}}\hspace{-1pt}\big)\\[-26pt]
\label{line_plane_inter_in_mt}\end{split}}
or the intersection of two planes as
\eq{\begin{split}\\[-24pt](\r{a\,b\,c})\tcap(\b{d\,e\,f})&\equivR (\r{a\,b})\ab{\r{c}\,\b{d\,e\,f}}{+}(\r{b\,c})\ab{\r{a}\,\b{d\,e\,f}}{+}(\r{c\,a})\ab{\r{b}\,\b{c\,d\,e}}\\
&\equivL{-}(\b{d\,e\,f})\tcap(\r{a\,b\,c})\\
&=-\big(\hspace{-1pt}(\b{d\,e})\ab{\b{f}\,\r{a\,b\,c}}{+}(\b{e\,f})\ab{\b{d}\,\r{a\,b\,c}}{+}(\b{f\,d})\ab{\b{e}\,\r{a\,b\,c}}\hspace{-1pt}\big)\,.\\[-26pt]
\label{plane_cap_mt}\end{split}}

\subsubsection{Translations Between Spinors and Momentum Twistors}\vspace{-4pt}

The translation between momentum-twistor expressions and ordinary spinor expressions is achieved through the relation:
\eq{\ab{\r{a}\,\b{b}\,\g{c}\,\t{d}}=\ab{\r{a}\,\b{b}}\ab{\g{c}|(x_{\g{c}\,\b{b}})(x_{\b{b}\,\t{d}})|\t{d}}{+}\ab{\b{b}\,\g{c}}\ab{\r{a}|(x_{\r{a}\,\b{b}})(x_{\b{b}\,\t{d}})|\t{d}}{-}\ab{\b{b}\,\t{d}}\ab{\r{a}|(x_{\r{a}\,\b{b}})(x_{\b{b}\,\g{c}})|\g{c}}\,\label{general_four_bracket}}
where $x_{\r{a}\,\b{b}}\equivR x_{\b{b}}{-}x_{\r{a}}{=}(p_{\r{a}}{+}\ldots{+}p_{\b{b{-}1}})$. Special instances of (\ref{general_four_bracket}) of note are
\begin{align}\fwboxL{63pt}{\ab{\r{a\hspace{1pt}\text{-}1}\,\r{a}\,\b{b}\,\g{c}}}&\equivL\fwbox{40pt}{\ab{\hspace{-1pt}(\hspace{-1pt}\r{a}\hspace{-1pt})\hspace{1pt}\b{b}\,\g{c}}}=\ab{\r{a\hspace{1pt}\text{-}1\,a}}\ab{\b{b}|x_{\b{b}\,\r{a}}|x_{\r{a}\,\g{c}}|\g{c}};\label{a_with_bc_in_mt}\\
\fwboxL{63pt}{\ab{\r{a\hspace{1pt}\text{-}1}\,\r{a}\,\r{a\text{+}1}\,\b{b}}}&\equivL\fwbox{40pt}{\ab{\r{a}\,\b{\bar{b}}}}=\ab{\r{a\hspace{1pt}\text{-}1}\,\r{a}}\ab{\r{a}\,\r{a\text{+}1}}\langle\b{b}|x_{\r{a}\,\b{b}}|\r{a}];\label{a_bbar_in_mt}\\
\fwboxL{63pt}{\ab{\r{a\hspace{1pt}\text{-}1\,a}\,\b{b\hspace{1pt}\text{-}1\,b}}}&\equivL\fwbox{40pt}{\ab{\hspace{-1pt}(\hspace{-1pt}\r{a}\hspace{-1pt})\hspace{0pt}(\hspace{-1pt}\b{b}\hspace{-1pt})\hspace{-1pt}}}=\ab{\r{a\hspace{1pt}\text{-}1\,a}}\ab{\b{b\hspace{1pt}\text{-}1\,b}}x^2_{\r{a}\,\b{b}}=\ab{\r{a\hspace{1pt}\text{-}1\,a}}\ab{\b{b\hspace{1pt}\text{-}1\,b}}(p_\r{a}{+}\ldots{+}p_{\b{b\hspace{1pt}\text{-}1}})^2.\label{mandelstam_in_mt}
\end{align}

In addition to these ordinary invariants, there are three that arise regularly in the periods of one and two loop Feynman integrals. The first of these is associated with the one-loop box, and encodes the square root of the Gramm matrix \cite{Gram1883} of four points in dual-momentum coordinates:
\eq{\begin{split}
\Delta[\r{a},\!\b{b},\!\g{c},\!\t{d}]^2&\equivR\big(\hspace{-1pt}\x{\r{a}}{\g{c}}\x{\b{b}}{\t{d}}{-}\x{\r{a}}{\b{b}}\x{\g{c}}{\t{d}}{-}\x{\b{b}}{\g{c}}\x{\t{d}}{\r{a}}\hspace{-1pt}\big)^2\hspace{-3pt}{-}4\,\x{\r{a}}{\b{b}}\x{\b{b}}{\g{c}}\x{\g{c}}{\t{d}}\x{\t{d}}{\r{a}}\\
&=\big(\hspace{-1pt}\ab{\hspace{-1pt}\r{a_1\hspace{-1pt}a_2}\g{c_1\hspace{-1pt}c_2}\hspace{-1pt}}\ab{\hspace{-1pt}\b{b_1\hspace{-1pt}b_2}\t{d_1\hspace{-1pt}d_2}\hspace{-1pt}}{-}\ab{\hspace{-1pt}\r{a_1\hspace{-1pt}a_2}\b{b_1\hspace{-1pt}b_2}\hspace{-1pt}}\ab{\hspace{-1pt}\g{c_1\hspace{-1pt}c_2}\t{d_1\hspace{-1pt}d_2}\hspace{-1pt}}{-}\ab{\hspace{-1pt}\b{b_1\hspace{-1pt}b_2}\g{c_1\hspace{-1pt}c_2}\hspace{-1pt}}\ab{\hspace{-1pt}\t{d_1\hspace{-1pt}d_2}\r{a_1\hspace{-1pt}a_2}\hspace{-1pt}}\hspace{-1pt}\big)^2\\
&\phantom{\,=\,}\;\;{-}4\,\ab{\hspace{-1pt}\r{a_1\hspace{-1pt}a_2}\b{b_1\hspace{-1pt}b_2}\hspace{-1pt}}\ab{\hspace{-1pt}\b{b_1\hspace{-1pt}b_2}\g{c_1\hspace{-1pt}c_2}\hspace{-1pt}}\ab{\hspace{-1pt}\g{c_1\hspace{-1pt}c_2}\t{d_1\hspace{-1pt}d_2}\hspace{-1pt}}\ab{\hspace{-1pt}\t{d_1\hspace{-1pt}d_2}\r{a_1\hspace{-1pt}a_2}\hspace{-1pt}}\\
&\equivL\big(\hspace{-1pt}\ab{\hspace{-1pt}(\hspace{-1pt}\r{a}\hspace{-1pt})\hspace{0pt}(\hspace{-1pt}\g{c}\hspace{-1pt})\hspace{-1pt}}\ab{\hspace{-1pt}(\hspace{-1pt}\b{b}\hspace{-1pt})\hspace{0pt}(\hspace{-1pt}\t{d}\hspace{-1pt})\hspace{-1pt}}{-}\ab{\hspace{-1pt}(\hspace{-1pt}\r{a}\hspace{-1pt})\hspace{0pt}(\hspace{-1pt}\b{b}\hspace{-1pt})\hspace{-1pt}}\ab{\hspace{-1pt}(\hspace{-1pt}\g{c}\hspace{-1pt})\hspace{0pt}(\hspace{-1pt}\t{d}\hspace{-1pt})\hspace{-1pt}}{-}\ab{\hspace{-1pt}(\hspace{-1pt}\b{b}\hspace{-1pt})\hspace{0pt}(\hspace{-1pt}\g{c}\hspace{-1pt})\hspace{-1pt}}\ab{\hspace{-1pt}(\hspace{-1pt}\t{d}\hspace{-1pt})\hspace{0pt}(\hspace{-1pt}\r{a}\hspace{-1pt})\hspace{-1pt}}\hspace{-1pt}\big)^2\\
&\phantom{\,=\,}\;\;{-}4\,\ab{\hspace{-1pt}(\hspace{-1pt}\r{a}\hspace{-1pt})\hspace{0pt}(\hspace{-1pt}\b{b}\hspace{-1pt})\hspace{-1pt}}\ab{\hspace{-1pt}(\hspace{-1pt}\b{b}\hspace{-1pt})\hspace{0pt}(\hspace{-1pt}\g{c}\hspace{-1pt})\hspace{-1pt}}\ab{\hspace{-1pt}(\hspace{-1pt}\g{c}\hspace{-1pt})\hspace{0pt}(\hspace{-1pt}\t{d}\hspace{-1pt})\hspace{-1pt}}\ab{\hspace{-1pt}(\hspace{-1pt}\t{d}\hspace{-1pt})\hspace{0pt}(\hspace{-1pt}\r{a}\hspace{-1pt})\hspace{-1pt}}\\
&\propto\big(\hspace{-1pt}x_{\r{a}\,\g{c}}^2x_{\b{b}\,\t{d}}^2{-}x_{\r{a}\,\b{b}}^2x_{\g{c}\,\t{d}}^2{-}x_{\b{b}\,\g{c}}^2x_{\t{d}\,\r{a}}^2\hspace{-1pt}\big)^2\hspace{-3pt}{-}4x_{\r{a}\,\b{b}}^2x_{\b{b}\,\g{c}}^2x_{\g{c}\,\t{d}}^2x_{\t{d}\,\r{a}}^2\\
&=\big(\hspace{-1pt}(p_{\r{a}}{+}p_{\b{b}})^2(p_{\b{b}}{+}p_{\g{c}})^2{-}p^2_{\r{a}}\,p^2_{\g{c}}{-}\,p^2_{\b{b}}\,p^2_{\t{d}}\hspace{-1pt}\big)^2\hspace{-3pt}{-}4\,p^2_{\r{a}}\,p^2_{\b{b}}\,p^2_{\g{c}}\,p^2_{\t{d}}
\end{split}\label{four_point_grammian_defined}}
where the constant of proportionality in the penultimate line above is
\eq{(\x{\r{a}}{\inf}\x{\b{b}}{\inf}\x{\g{c}}{\inf}\x{\t{d}}{\inf})^2\equivR\big(\hspace{-1pt}\ab{\hspace{-1pt}\r{a_1\hspace{-1pt}a_2}\hspace{-1pt}}\ab{\hspace{-1pt}\b{b_1\hspace{-1pt}b_2}\hspace{-1pt}}\ab{\hspace{-1pt}\g{c_1\hspace{-1pt}c_2}\hspace{-1pt}}\ab{\hspace{-1pt}\t{d_1\hspace{-1pt}d_2}\hspace{-1pt}}\hspace{-1pt}\big)^2\,.}
The case where one of the four arguments is the point or line at infinity `$\inf$' is relevant to the three-mass triangle integral at one loop; in this case, $\Delta[\r{a},\!\b{b},\!\g{c},\!\inf]$ is given by 
\eq{\begin{split}\frac{\Delta[\r{a},\!\b{b},\!\g{c},\!\inf]}{\x{\r{a}}{\inf}\x{\b{b}}{\inf}\x{\g{c}}{\inf}}&=\sqrt{\big(\hspace{-1pt}x^2_{\r{a}\,\g{c}}{-}x^2_{\r{a}\,\b{b}}{-}x^2_{\b{b}\,\g{c}}\hspace{-1pt}\big)^2\hspace{-3pt}{-}4\,x^2_{\r{a}\,\b{b}}\,x^2_{\b{b}\,\g{c}}}\\
&=\sqrt{\big(\hspace{-1pt}(p_{\r{a}}{+}p_{\b{b}})^2{-}p_{\r{a}}^2{-}p_{\b{b}}^2\hspace{-1pt}\big)^2\hspace{-3pt}{-}4\,p_{\r{a}}^2\,p^2_{\b{b}}}.
\end{split}}

At two loops one also encounters the algebraic quantity `$\Delta_2^{\hspace{0pt}\r{a}}\!\left[\b{b},\!\b{c};\!\t{d},\!\t{e}\right]$' which is defined by 
\eq{\begin{split}\\[-30pt]\left(\hspace{-1pt}\Delta_2^{\hspace{0pt}\r{a}}\!\left[\b{b},\!\b{c};\!\t{d},\!\t{e}\right]\hspace{-1pt}\right)^2\equivR&\big[\hspace{-1.5pt}\ab{\hspace{-1pt}(\hspace{-0.75pt}\r{a}\hspace{-0.75pt})\tcap(\b{b_1}(\hspace{-0.75pt}\b{c}\hspace{-0.75pt})\hspace{-1pt})(\hspace{-0.75pt}\t{d}\hspace{-0.75pt})(\hspace{-0.75pt}\t{e}\hspace{-0.75pt})\tcap(\hspace{-1pt}(\hspace{-0.75pt}\r{a}\hspace{-0.75pt})\b{b_2}\hspace{-1pt})\hspace{-1pt}}{+}\ab{\hspace{-1pt}(\hspace{-0.75pt}\r{a}\hspace{-0.75pt})\tcap(\b{b_2}(\hspace{-0.75pt}\b{c}\hspace{-0.75pt})\hspace{-1pt})(\hspace{-0.75pt}\t{d}\hspace{-0.75pt})(\hspace{-0.75pt}\t{e}\hspace{-0.75pt})\tcap(\hspace{-1pt}(\hspace{-0.75pt}\r{a}\hspace{-0.75pt})\b{b_1}\hspace{-1pt})\hspace{-1pt}}\hspace{-1.5pt}\big]^2\\
&\hspace{-4pt}{-}4\hspace{1pt}\ab{\hspace{-1pt}(\hspace{-0.75pt}\r{a}\hspace{-0.75pt})\tcap(\b{b_1}(\hspace{-0.75pt}\b{c}\hspace{-0.75pt})\hspace{-1pt})(\hspace{-0.75pt}\t{d}\hspace{-0.75pt})(\hspace{-0.75pt}\t{e}\hspace{-0.75pt})\tcap(\hspace{-1pt}(\hspace{-0.75pt}\r{a}\hspace{-0.75pt})\b{b_1}\hspace{-1pt})\hspace{-1pt}}\ab{\hspace{-1pt}(\hspace{-0.75pt}\r{a}\hspace{-0.75pt})\tcap(\b{b_2}(\hspace{-0.75pt}\b{c}\hspace{-0.75pt})\hspace{-1pt})(\hspace{-0.75pt}\t{d}\hspace{-0.75pt})(\hspace{-0.75pt}\t{e}\hspace{-0.75pt})\tcap(\hspace{-1pt}(\hspace{-0.75pt}\r{a}\hspace{-0.75pt})\b{b_2}\hspace{-1pt})\hspace{-1pt}};
\label{delta2_in_embedding}\\[-6pt]
\end{split}}
we can spell this out more explicitly for momentum twistor arguments as
\begin{align}\nonumber\\[-30pt]&\hspace{-30pt}\left(\hspace{-1pt}\Delta_2^{\hspace{-3pt}(\hspace{-1pt}\r{a_1}\r{a_2}\hspace{-1pt})}\!\left[(\b{b_1b_2}),\!(\b{c_1c_2});\!(\t{d_1d_2}),\!(\t{e_1e_2})\right]\hspace{-1pt}\right)^2=\label{delta2_in_mt_expanded}\\
&\big[\hspace{-1.5pt}\ab{\hspace{-1pt}(\hspace{-1pt}\r{a_1a_2}\hspace{-1pt})\tcap(\b{b_1}\b{c_1c_2}\hspace{-1pt})\t{d_1d_2}(\hspace{-1pt}\t{e_1e_2}\hspace{-1pt})\tcap(\hspace{-1pt}\r{a_1a_2}\b{b_2}\hspace{-1pt})\hspace{-1pt}}{+}\ab{\hspace{-1pt}(\hspace{-1pt}\r{a_1a_2}\hspace{-1pt})\tcap(\b{b_2}\b{c_1c_2}\hspace{-1pt})\t{d_1d_2}(\hspace{-1pt}\t{e_1e_2}\hspace{-1pt})\tcap(\hspace{-1pt}\r{a_1a_2}\b{b_1}\hspace{-1pt})\hspace{-1pt}}\hspace{-1.5pt}\big]^2\nonumber\\
&\hspace{-4pt}{-}4\hspace{1pt}\ab{\hspace{-1pt}(\hspace{-1pt}\r{a_1a_2}\hspace{-1pt})\tcap(\b{b_1}\b{c_1c_2}\hspace{-1pt})\t{d_1d_2}(\hspace{-1pt}\t{e_1e_2}\hspace{-1pt})\tcap(\hspace{-1pt}\r{a_1a_2}\b{b_1}\hspace{-1pt})\hspace{-1pt}}\ab{\hspace{-1pt}(\hspace{-1pt}\r{a_1a_2}\hspace{-1pt})\tcap(\b{b_2}\b{c_1c_2}\hspace{-1pt})\t{d_1d_2}(\hspace{-1pt}\t{e_1e_2}\hspace{-1pt})\tcap(\hspace{-1pt}\r{a_1a_2}\b{b_2}\hspace{-1pt})\hspace{-1pt}}.\nonumber\\[-10pt]\nonumber
\end{align}
This arises as the Jacobian associated with the cut
\vspace{-5pt}\eq{\fig{-2pt}{1}{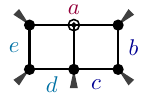}\,.\vspace{-5pt}}
The case $\Delta_2^{\hspace{-1pt}\inf}\hspace{-1pt}[\r{a},\r{b};\b{c},\b{d}]$ arises as the Jacobian of the leading singularity 
\vspace{-5pt}\eq{\fig{-2pt}{1}{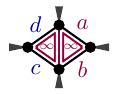}\vspace{-5pt}}
where the loop momenta are taken to be collinear with infinity.\\[-8pt]

Finally, there is the fully antisymmetric `6-bracket' defined on the space of 6 lines in momentum twistor space, 
\eq{\epsilon((\b{a}),(\t{b}),(\t{c}),(\t{d}),(\t{e}),(\r{f_1}\,\r{f_2}))\equivR\frac{1}{12}\sum_{\g{r},\g{s},\g{t},\g{u}\in[4]\fwboxL{0pt}{\;\;\;\hspace{15pt}\text{with}\;\;L_\g{[4]}\!\!\equivR\!\{(\t{b}),(\t{c}),(\t{d}),(\t{e})\}}\hspace{-20pt}}\epsilon^{\g{r\,s\,t\,u}}\ab{(\b{a})(L_{\g{r}})\tcap(L_{\g{s}}\,\r{f_1})(L_{\g{t}})\tcap(L_{\g{u}}\,\r{f_2})}.\label{epsilon_in_mt_defined}}
The translation of this invariant into momentum space is associated with the chiral trace involving Dirac $\gamma$ matrices,
\eq{\frac{\epsilon(\t{a},\t{b},\t{c},\t{d},\t{e},\t{f})}{\x{\t{a}}{\inf}\x{\t{b}}{\inf}\x{\t{c}}{\inf}\x{\t{d}}{\inf}\x{\t{e}}{\inf}\x{\t{f}}{\inf}}={-}\frac{1}{2}x_{\t{ab}}^{\g{\mu}} x_{\t{bc}}^{\g{\nu}} x_{\t{cd}}^{\g{\rho}} x_{\t{de}}^{\g{\sigma}} x_{\t{ef}}^{\g{\tau}} x_{\t{fa}}^{\g{\upsilon}}\mathrm{tr}(\gamma^5\gamma_\g{\mu}\gamma_\g{\nu}\gamma_\g{\rho}\gamma_\g{\sigma}\gamma_\g{\tau}\gamma_\g{\upsilon})}
where, again, $x_{\r{a}\b{b}}\equivR(x_{\b{b}}{-}x_{\r{a}})=(p_{\r{a}}{+}\ldots{+}p_{\b{b\text{-}1}})$.

When one of these arguments is the point or line at infinity, `$\inf$', it becomes
\eq{\begin{split}\frac{\epsilon(\t{a},\t{b},\t{c},\t{d},\t{e},\inf)}{\x{\t{a}}{\inf}\x{\t{b}}{\inf}\x{\t{c}}{\inf}\x{\t{d}}{\inf}\x{\t{e}}{\inf}}&=\frac{1}{2}x_{\t{ab}}^{\g{\mu}} x_{\t{bc}}^{\g{\nu}} x_{\t{cd}}^{\g{\rho}} x_{\t{de}}^{\g{\sigma}}\mathrm{tr}(\gamma^5\gamma_{\g{\mu}}\gamma_{\g{\nu}}\gamma_{\g{\rho}}\gamma_{\g{\sigma}})
\\&=2i\,x_{\t{ab}}^{\g{\mu}} x_{\t{bc}}^{\g{\nu}} x_{\t{cd}}^{\g{\rho}} x_{\t{de}}^{\g{\sigma}}\,\epsilon_{\g{\mu\nu\rho\sigma}}\end{split}\,.}
One final instance of this tensor which is especially useful is when one of the arguments encodes a loop momentum $\r{\ell}$; in this case, it is given by 
\eq{\begin{split}\frac{\epsilon(\r{\ell},\t{a},\t{b},\t{c},\t{d},\inf)}{\x{\r{\ell}}{\inf}\x{\t{a}}{\inf}\x{\t{b}}{\inf}\x{\t{c}}{\inf}\x{\t{d}}{\inf}}&=\phantom{{-}}\frac{1}{2}\,x_{\t{ab}}^{\g{\mu}}\, x_{\t{b}\r{\ell}}^{\g{\nu}}\, x_{\r{\ell}\t{c}}^{\g{\rho}}\,x_{\t{cd}}^{\g{\sigma}}\mathrm{tr}(\gamma^5\gamma_{\g{\mu}}\gamma_{\g{\nu}}\gamma_{\g{\rho}}\gamma_{\g{\sigma}})\\
&={-}\frac{1}{2}\,p_{\t{a}}^{\g{\mu}}\, \r{\ell}_{\t{b}}^{\g{\nu}}\, \r{\ell}_{\t{c}}^{\g{\rho}}\,p_{\t{c}}^{\g{\sigma}}\mathrm{tr}(\gamma^5\gamma_{\g{\mu}}\gamma_{\g{\nu}}\gamma_{\g{\rho}}\gamma_{\g{\sigma}})
\end{split}}
where $\r{\ell}_{\t{v}}\equivR \r{x_{\ell}}{-}\t{x_b}$ is the momentum flowing through the $\t{b}$th propagator.

\newpage
\section{Structure, Usage \& Capabilities of Ancillary Package}\label{ancillary_appendix}\vspace{-4pt}
Although the master integral basis we describe is uniquely defined (irrespective of coordinate choices) by prescriptivity relative to the contours to which they are dual, there are obvious advantages to having access to concrete analytic formulae for these masters and also the contours which define them. This not only disambiguates any lack of clarity in the graphical encoding of contours provided in the following \mbox{appendix~\ref{prescriptive_definition_of_masters}}, but also provides the reader with tools to analytically determine coefficients of amplitudes in this basis by evaluating the field theory amplitudes at explicit points in loop momentum space. 

In the ancillary files attached to this work's submission to the \texttt{arXiv}, we have provided a \textsc{Mathematica} package \texttt{two\uscore loop\uscore masters.m} and an associated \texttt{Notebook} \texttt{two\uscore loop\uscore master\uscore integrals.nb} which demonstrates its main usage and documents its principal functionality.

Chief among these functions are:
\begin{itemize}
\item\fun{masterNumerators}\brace{\var{topology}\pattern} and \fun{masterIntegrals}\brace{\var{topology}\pattern}: provide explicit, analytic formulae for the master numerators and integrals for graphs with topology encoded by \var{topology}, expressed in terms of momentum-twistors;
\item\fun{masterContourGraphs}\brace{\var{topology}\pattern}: provides a graphical representation of the polylogarithmic contours defining each master;
\item \fun{polylogarithmicContours}\brace{\var{topology}\pattern}: provides a detailed description of every individual polylogarithmic contour integral---its pole location in momentum twistor variables, and its Jacobian (which defines its orientation);
\item \fun{polylogarithmicContourGraphs}\brace{\var{topology}\pattern}: provides a graphical representation of each of the contours encoded by \fun{polylogarithmicContours};
\item \fun{masterContourProjectors}\brace{\var{topology}\pattern}: defines the precise linear combination of polylogarithmic contours defining masters;
\item\fun{polylogPeriods}\brace{\var{topology}\patternTwo,\var{numeratorList}}: computes the period matrix of integrands with propagators encoded by \var{topology} and numerators given by \var{numeratorList};
\item\fun{showMasterPeriods} and \fun{showMasterPeriodsOnDaughters}: displays the complete period matrix of masters on their top-level contours or those of their daughters;
\item\texttt{\textbf{epsRule}} and \texttt{\textbf{deltaRule}}: provides an explicit expansion of the invariants `$\epsilon\brace{\var{abcdef}\patternTwo}$' `$\Delta\brace{\var{abcd}\patternTwo}$' and `$\Delta_2\brace{\var{a}\patternTwo}\brace{\var{bcde}\patternTwo}$' in terms of momentum-twistor four-brackets;
\item \fun{evaluate}\brace{\var{expression}\pattern}: tools to convert these analytic expressions to numeric evaluations for kinematics specified by (use generated, reference values of, or randomly generated) momentum twistors.\\[-20pt]~
\end{itemize}
These tools were built from those of other \textsc{Mathematica} packages prepared by the author in other collaborations over the years (see \emph{e.g.}~\cite{Bourjaily:2010wh,Bourjaily:2013mma,Bourjaily:2015jna,Bourjaily:2023uln}).

\newpage
\section[Prescriptive Definition of the Master Integral Basis]{Prescriptive Definition of the Master Integral Basis}\label{prescriptive_definition_of_masters}\vspace{-4pt}

\subsection{Classes of Masters and their Organizing Principles}\label{organizing_principles}\vspace{-4pt}

Although there are many unifying principles across graphs of a given topology, the scope of possible contours varies considerably with which subsets of legs (if any) are taken to be massless or possibly vanishing. This variation not only changes the number of possible contours, but also their association with infrared divergences, elliptic sub-topologies, and double-poles (signifying weight-drops upon integration). As such, we find it necessary to spell out many cases specifically. 

These cases correspond to a choice of allowed momenta flowing into each vertex---which, in terms of massless particles, how many massless legs must be attached at each vertex. For a graph with the topology of type `$[\r{a}\g{b}\b{c}]$', legs at the vertices $\{\r{a_1},\r{\cdots},\r{a_{\text{-}1}},\g{b_1},\b{c_1},\b{\ldots},\b{c_{\text{-}1}},\g{b_{\text{-}1}}\}$ would each be designated by a number indicating the allowed range of inflowing momenta:
\vspace{-5pt}\eq{\fig{-2pt}{1}{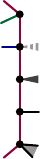}\begin{array}{@{}l@{$\;\;$}l@{}}
\hspace{2pt}\text{`$0$'}&\text{\emph{\r{no}} momentum inflowing}\\[0pt]
\hspace{2pt}\text{`$\hspace{0.4pt}\bar{\hspace{-0.4pt}0\hspace{-0.4pt}}\hspace{0.4pt}$'}&\text{\emph{\r{any}} momentum, including \emph{none}}\\[0pt]
\hspace{2pt}\text{`$\bar{1}$'}&\text{\hspace{0pt}any \emph{\r{non-vanishing}} momentum}\\
\hspace{2pt}\text{`$1$'}&\text{\hspace{0pt}any \emph{\r{massless}} momentum}\\[0pt]
\hspace{2pt}\text{`$2$'}&\text{any \emph{\r{massive}} momentum}
\end{array}}

Examples of how these designations are used are as follows: 
\vspace{-10pt}\eq{\fig{-2pt}{1}{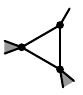}\fig{-2pt}{1}{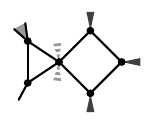}\fig{-2pt}{1}{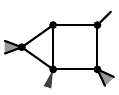}\fig{-2pt}{1}{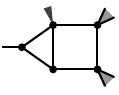}\fig{-2pt}{1}{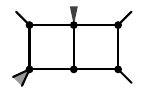}\fig{-2pt}{1}{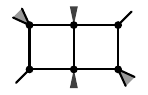}\vspace{-20pt}}
\hspace{30pt}\raisebox{10pt}{\ref{int3|122}\hspace{25pt}\ref{int403|2221121}\hspace{40pt}\ref{int312|12120}\hspace{30pt}\ref{int312|22011}\hspace{28pt}\ref{int313|110211}\hspace{34pt}\ref{int313|121121}}\\[-8pt]
\noindent In addition to the analysis of these cases, there is the question of how the independent master integrals---once defined---should be \emph{ordered} or indexed. Our choices are motivated by a desire to organize masters according to their role in field theory: integrals defined by cuts only using the graph's propagators (and not sending either loop to infinity) are listed first; these integrals span the space of dual-conformal integrals (even if they are not themselves dual-conformal). Among these, those for which field theory periods are typically non-vanishing are listed first.

\subsection{Graphical Notation for the Contours Defining Masters}\label{graphical_notation}\vspace{-4pt}

In the following, explicit contours defining the master integrals in our basis are described graphically. Although fairly dense in notation, we find these figures more helpful than, say, an explicit formulation of the pole locations in some coordinate chart. To better make sense of these pictures, let us briefly define their key ingredients. 

Every edge of a contour diagram denotes a \emph{cut} propagator. Thus, all vertices connect a set of momentum-conserving, \emph{on-shell} momenta. Which particular cut-solution is involved may be indicated by a number (arbitrarily) assigned to the two solutions or otherwise indicated at vertices as follows:
\eq{\fig{-2pt}{1}{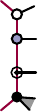}\hspace{2pt}\begin{array}{@{}l@{$\;\;$}l@{}}
\text{massless cut with $\lambda$'s are parallel (`$\bar{\text{MHV}}$')}\\[0pt]
\text{\hspace{2pt}massless cut with $\tilde{\lambda}$'s are parallel (`$\text{MHV}$')}\\[0pt]
\text{\hspace{2pt}massless cut involving \emph{collinear} momenta}\\[0pt]
\text{cut involving massive momenta}
\end{array}}
There is also the case where cut conditions require the momentum flowing through some edge to be simultaneously collinear with two external (massless) momenta; this is only possible if the momentum through that edge is also \emph{soft} (namely, vanishing). This is denoted diagrammatically by \raisebox{-5pt}{\rotatebox{90}{\fig{-2pt}{1}{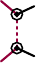}}}\hspace{-4pt}.

Examples of these include
\eq{\fwbox{0pt}{\fig{-2pt}{1}{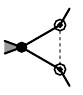}\quad\fig{-2pt}{1}{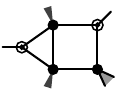}\quad\fig{-2pt}{1}{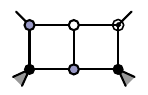}\quad\fig{-2pt}{1}{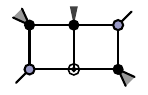}\quad\fig{-2pt}{1}{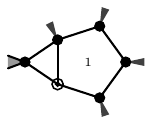}}\vspace{-20pt}}
\hspace{45pt}\raisebox{10pt}{\ref{omega3|112}\hspace{40pt}\ref{omega312|12111}\hspace{42pt}\ref{omega313|120210}\hspace{45pt}\ref{omega313|120121}\hspace{40pt}\ref{omega412|222021}\hspace{34pt}}\\[-8pt]
\noindent In addition, there are cuts involving sending one or more of the loops to infinity denoted with an `$\infty$' mark as in
\eq{\fwbox{0pt}{\fig{-2pt}{1}{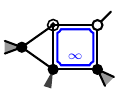}\quad\fig{-2pt}{1}{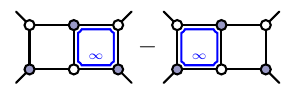}\quad\fig{-2pt}{1}{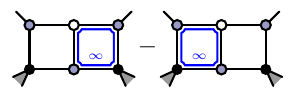}.}\vspace{-15pt}}
\hspace{55pt}\raisebox{10pt}{\ref{omega312|12120}\hspace{78pt}\ref{omega313|110110}\hspace{120pt}\ref{omega313|120210}}\\[-8pt]
\noindent Master contours defined as combinations are either written explicitly as a sum (as above), or implicitly as a the `(parity-)even' or `(parity-)odd' combination.

Because both loops can be sent to infinity simultaneously, this can be done in one of two parities or---the intersection of these two conditions---in a collinear way. Contours defined with both loops taken \emph{collinearly} to infinity are indicated in red, as in the cases
\eq{\fig{-2pt}{1}{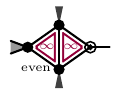}\quad\fig{-2pt}{1}{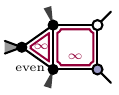}\quad\fig{-2pt}{1}{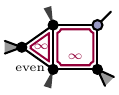}.\vspace{-10pt}}
\hspace{138pt}\raisebox{10pt}{\ref{omega212|1121}\hspace{40pt}\ref{omega312|11121}\hspace{37pt}\ref{omega312|12121}}\\[-16pt]

Finally, when the contour requires involves encircling a double-pole, this is denoted by `dp' as in 
\eq{\fig{-2pt}{1}{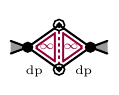}\quad\fig{-2pt}{1}{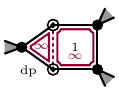}\vspace{-14pt}}
\hspace{168pt}\raisebox{10pt}{\ref{omega212|2020}\hspace{40pt}\ref{omega312|22020}}\\[-8pt]
\noindent (The master integrals dual to such contours are of sub-leading transcendental weight.)

Both because of the difficulty of parsing these contours explicitly and also the fact that these contours require a \emph{conventional} orientation (rendering linear combinations ambiguous), we have provided explicit contour prescriptions for all contours defining master integrals in the ancillary files for this work. 

\newpage
\renewcommand{\thesubsection}{\fwboxR{2pt}{\text{[}}\r{3}]\hspace{-8pt}}
\renewcommand{\theequation}{\thesubsection\hspace{7pt}~}
\subsection{(One-Loop) {Triangles}}\vspace{-4pt}
The \emph{scalar} triangle integrals are defined to be of the form
\eq{\fwboxL{370pt}{\hspace{-10pt}\fig{-2.5pt}{1}{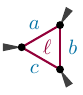}\;\bigger{\Leftrightarrow}\;\frac{\mathfrak{n}_{[\r{3}]}[\t{a},\!\t{b},\!\t{c};\!\inf]}{\x{\r{\ell_1}}{\t{a}}\x{\r{\ell_1}}{\t{b}}\x{\r{\ell_1}}{\t{c}}\x{\r{\ell_1}}{\inf}}.}\label{3_masters_defined}}
\paragraph*{\r{Prescriptive Contours Defining Master Integrals:}}~\\[-12.5pt]
\noindent\rule{430pt}{0.5pt}\\[-14.75pt]
\noindent\rule{430pt}{0.5pt}
\dataTableTri{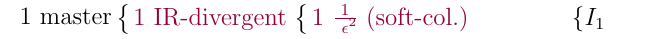}
\noindent\rule{430pt}{0.5pt}
\dataTableTri{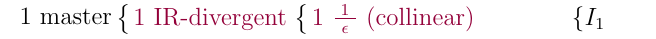}
\noindent\rule{430pt}{0.5pt}
\dataTableTri{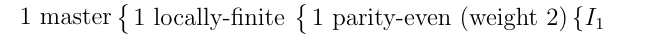}

\renewcommand{\thesubsection}{\fwboxR{2pt}{\text{[}}\r{4}]\hspace{-8pt}}
\renewcommand{\theequation}{\thesubsection\hspace{7pt}~}
\subsection{(One-Loop) {Boxes}}\vspace{-4pt}
The \emph{box} integrals are defined to be of the form
\eq{\fwboxL{370pt}{\hspace{-10pt}\fig{-2.5pt}{1}{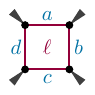}\;\bigger{\Leftrightarrow}\;\frac{\mathfrak{n}^{\hspace{3pt}i}_{[\r{4}]}[\r{\ell}\hspace{-0pt};\!\t{a},\!\t{b},\!\t{c},\!\t{d};\!\inf]}{\x{\r{\ell_1}}{\t{a}}\x{\r{\ell_1}}{\t{b}}\x{\r{\ell_1}}{\t{c}}\x{\r{\ell_1}}{\t{d}}\x{\r{\ell_1}}{\inf}}.}\label{4_masters_defined}}
\paragraph*{\r{Prescriptive Contours Defining Master Integrals:}}~\\[-12.5pt]
\noindent\rule{430pt}{0.5pt}\\[-14.75pt]
\noindent\rule{430pt}{0.5pt}
\setcounter{topology}{0}
\renewcommand{\theequation}{$\hspace{-0pt}I^{[\r{4}]}_{\fwbox{12pt}{}}\hspace{-2pt}$}
\vspace{-15pt}\eq{\hspace{-80pt}\hspace{-36pt}\raisebox{-2pt}{$\cfig{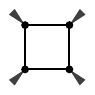}$}\hspace{5pt}\raisebox{-15.pt}{$\cfig{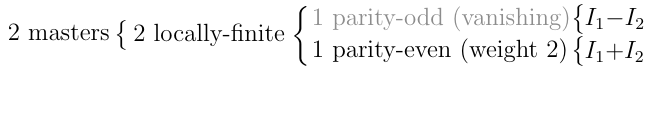}$}\hspace{20pt}\hspace{-80pt}\vspace{-32pt}\label{int4}}
\renewcommand{\theequation}{$\Omega^{[\r{4}]}_{\fwbox{12pt}{}}\hspace{-2pt}$}
~\\[-4pt]\noindent{\textbf{\b{Defining Contours}:}}\vspace{-20pt}\eq{\hspace{5pt}\begin{array}{@{}l@{}}\begin{array}{@{}c@{}c@{}c@{}c@{}c@{}c}~\\[-15pt]
\fwboxR{0pt}{\raisebox{2pt}{$\left\{\rule{0pt}{22pt}\right.$}\hspace{-8.5pt}}\hspace{5pt}\cfig{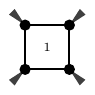}\hspace{5pt}\fwboxL{0pt}{,}&\hspace{5pt}\cfig{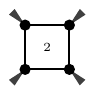}\hspace{2pt}\fwboxL{0pt}{\hspace{-4.5pt}\raisebox{2pt}{$\left.\rule{0pt}{22pt}\right\}$}}\\[-10pt]
\phantom{{}_{1}}\hspace{2pt}\,\Omega_1&\phantom{{}_{2}}\Omega_2\\[-7pt]
\end{array}
\end{array}\vspace{-0pt}\label{omega4}}

\newpage

\renewcommand{\thesubsection}{\fwboxR{2pt}{\text{[}}\r{3}\g{0}\b{3}]\hspace{-8pt}}
\renewcommand{\theequation}{\thesubsection\hspace{7pt}~}
\subsection{{Kissing Triangles}}\vspace{-4pt}
The \emph{kissing triangle} integrals are defined to be of the form
\vspace{-8pt}\eq{\fwboxL{370pt}{\hspace{-10pt}\fig{-2.5pt}{1}{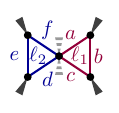}\;\bigger{\Leftrightarrow}\;\frac{\mathfrak{n}_{[\r{3}\g{0}\b{3}]}[\r{a},\!\r{b},\!\r{c};\!\b{d},\!\b{e},\!\b{f};\!\inf]}{\x{\r{\ell_1}}{\r{a}}\x{\r{\ell_1}}{\r{b}}\x{\r{\ell_1}}{\r{c}}\x{\r{\ell_1}}{\inf}\x{\b{\ell_2}}{\b{d}}\x{\b{\ell_2}}{\b{e}}\x{\b{\ell_2}}{\b{f}}\x{\b{\ell_2}}{\inf}}.}\label{303_masters_defined}\vspace{-10pt}}
\paragraph*{\r{Prescriptive Contours Defining Master Integrals:}}~\\[-12.5pt]
\noindent\rule{430pt}{0.5pt}\\[-14.75pt]
\noindent\rule{430pt}{0.5pt}
\dataTableQKT{11}{1}{11}{1}
\noindent\rule{430pt}{0.5pt}
\dataTableQKT{11}{1}{12}{1}
\noindent\rule{430pt}{0.5pt}
\dataTableQKT{11}{1}{22}{1}
\noindent\rule{430pt}{0.5pt}
\dataTableQKT{12}{1}{12}{1}
\noindent\rule{430pt}{0.5pt}
\dataTableQKT{12}{1}{21}{1}
\noindent\rule{430pt}{0.5pt}
\dataTableQKT{12}{1}{22}{1}
\noindent\rule{430pt}{0.5pt}
\dataTableQKT{22}{1}{22}{1}
\noindent\rule{430pt}{0.5pt}

\newpage
\renewcommand{\thesubsection}{\fwboxR{2pt}{\text{[}}\r{4}\g{0}\b{3}]\hspace{-8pt}}
\renewcommand{\theequation}{\thesubsection\hspace{7pt}~}
\subsection{{Kissing Box-Triangles}}\vspace{-4pt}

The \emph{kissing box-triangle} integrals are defined to be of the form
\vspace{-4pt}\eq{\fwboxL{370pt}{\hspace{-10pt}\fig{-2.5pt}{1}{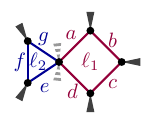}\;\bigger{\Leftrightarrow}\;\frac{\mathfrak{n}^{\hspace{3pt}\r{i}}_{[\r{4}\g{0}\b{3}]}[\r{\ell_1}\hspace{-1pt};\!\r{a},\!\r{b},\!\r{c},\!\r{d};\b{e},\!\b{f},\!\b{g};\!\inf]}{\x{\r{\ell_1}}{\r{a}}\x{\r{\ell_1}}{\r{b}}\x{\r{\ell_1}}{\r{c}}\x{\r{\ell_1}}{\r{d}}\x{\r{\ell_1}}{\inf}\x{\b{\ell_2}}{\b{e}}\x{\b{\ell_2}}{\b{f}}\x{\b{\ell_2}}{\b{g}}\x{\b{\ell_2}}{\inf}}.}\label{403_masters_defined}\vspace{-10pt}}
\paragraph*{\r{Prescriptive Contours Defining Master Integrals:}}~\\[-12.5pt]
\noindent\rule{430pt}{0.5pt}\\[-14.75pt]
\noindent\rule{430pt}{0.5pt}
\dataTableQKBT{222}{1}{11}{1}
\noindent\rule{430pt}{0.5pt}
\dataTableQKBT{222}{1}{12}{1}
\noindent\rule{430pt}{0.5pt}
\dataTableQKBT{222}{1}{22}{1}
\noindent\rule{430pt}{0.5pt}
%

\renewcommand{\thesubsection}{\fwboxR{2pt}{\text{[}}\r{4}\g{0}\b{4}]\hspace{-8pt}}
\renewcommand{\theequation}{\thesubsection\hspace{7pt}~}
\subsection{{Kissing Boxes}}\vspace{-4pt}

The \emph{kissing box} integrals are defined to be of the form
\eq{\fwboxL{370pt}{\hspace{-10pt}\fig{-2.5pt}{1}{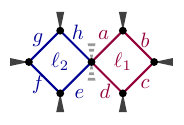}\;\bigger{\Leftrightarrow}\;\frac{\mathfrak{n}^{\hspace{3pt}\r{i}\b{j}}_{[\r{4}\g{0}\b{4}]}[\r{\ell_1}\hspace{-1pt},\!\b{\ell_2}\hspace{-1pt};\!\r{a},\!\r{b},\!\r{c},\!\r{d};\!\b{e},\!\b{f},\!\b{g},\!\b{h};\!\inf]}{\x{\r{\ell_1}}{\r{a}}\x{\r{\ell_1}}{\r{b}}\x{\r{\ell_1}}{\r{c}}\x{\r{\ell_1}}{\r{d}}\x{\r{\ell_1}}{\inf}\x{\b{\ell_2}}{\b{e}}\x{\b{\ell_2}}{\b{f}}\x{\b{\ell_2}}{\b{g}}\x{\b{\ell_2}}{\b{h}}\x{\b{\ell_2}}{\inf}}.}\label{404_masters_defined}\vspace{-10pt}}
\paragraph*{\r{Prescriptive Contours Defining Master Integrals:}}~\\[-12.5pt]
\noindent\rule{430pt}{0.5pt}\\[-14.75pt]
\noindent\rule{430pt}{0.5pt}
\renewcommand{\thesubsection}{\fwboxR{0pt}{\text{[}}\r{4}\g{0}\b{4}]\hspace{-10pt}}
\renewcommand{\theequation}{\roman{subsection}.\arabic{equation}}\setcounter{topologyIndex}{22212221}
\stepcounter{topologyIndex}\setcounter{topology}{404}\renewcommand{\theequation}{$\hspace{-0pt}I^{\fwbox{18pt}{[\r{4}\g{0}\b{4}]}}_{\fwbox{0pt}{}}\hspace{0pt}$}
\vspace{-06.5pt}\eq{\hspace{-105pt}\hspace{-10pt}\raisebox{-0pt}{$\cfig{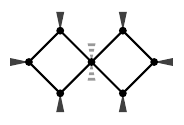}$}\hspace{-0pt}\raisebox{-2pt}{$\cfig{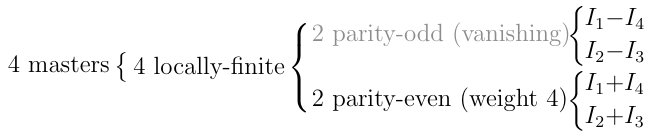}$}\hspace{20pt}\hspace{-100pt}\vspace{-22pt}\label{int414}}
\renewcommand{\theequation}{$\Omega^{\fwbox{18pt}{[\r{4}\g{0}\b{4}]}}_{\fwbox{0pt}{}}\hspace{-1pt}$}
~\\[-4pt]\noindent{\textbf{\b{Defining Contours}:}}\\[-12pt]
\vspace{-05pt}\eq{\fwbox{00pt}{\hspace{-15pt}\begin{array}{@{}l@{}}\begin{array}{@{}c@{}c@{}c@{}c@{}c@{}c}~\\[-15pt]
\fwboxR{0pt}{\raisebox{2pt}{$\left\{\rule{0pt}{24pt}\right.$}\hspace{-8.5pt}}\hspace{5pt}\cfig{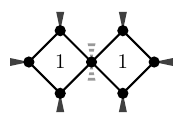}\hspace{0pt}\fwboxL{0pt}{,}&\hspace{0pt}\cfig{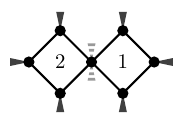}\hspace{0pt}\fwboxL{0pt}{,}&\hspace{0pt}\cfig{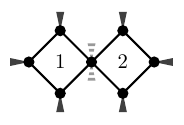}\hspace{0pt}\fwboxL{0pt}{,}&\hspace{0pt}\cfig{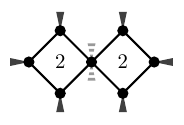}\hspace{2pt}\fwboxL{0pt}{\hspace{-4.5pt}\raisebox{2pt}{$\left.\rule{0pt}{24pt}\right\}$}}\\[-14pt]
\phantom{{}_{1}}\Omega_1&\phantom{{}_{2}}\Omega_2&\phantom{{}_{3}}\Omega_3&\phantom{{}_{4}}\Omega_4\\[-7pt]
\end{array}
\end{array}}\label{omega404}\vspace{2pt}
}
\noindent\rule{430pt}{0.5pt}

\newpage
\renewcommand{\thesubsection}{\fwboxR{2pt}{\text{[}}\r{2}\g{1}\b{2}]\hspace{-8pt}}
\renewcommand{\theequation}{\thesubsection\hspace{7pt}~}
\subsection{{Double Triangles}}\vspace{-4pt}

The \emph{double-triangle} integrals are defined to be of the form
\vspace{-4pt}\eq{\fwboxL{370pt}{\hspace{-10pt}\fig{-2.5pt}{1}{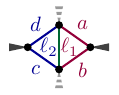}\;\bigger{\Leftrightarrow}\;\frac{\mathfrak{n}_{[\r{2}\g{1}\b{2}]}[\r{a},\!\r{b};\!\b{c},\!\b{d};\!\inf]}{\x{\r{\ell_1}}{\r{a}}\x{\r{\ell_1}}{\r{b}}\x{\r{\ell_1}}{\inf}\x{\r{\ell_1}}{\b{\ell_2}}\x{\r{\ell_1}}{\b{c}}\x{\b{\ell_2}}{\b{d}}\x{\b{\ell_2}}{\inf}}.}\label{212_masters_defined}\vspace{-10pt}}
\paragraph*{\r{Prescriptive Contours Defining Master Integrals:}}~\\[-12.5pt]
\noindent\rule{430pt}{0.5pt}\\[-14.75pt]
\noindent\rule{430pt}{0.5pt}
\dataTableQA{2}{0}{2}{0}
\noindent\rule{430pt}{0.5pt}
\dataTableQA{1}{0}{1}{2}
\noindent\rule{430pt}{0.5pt}
\dataTableQAB{1}{0}{2}{1}
\noindent\rule{430pt}{0.5pt}
\dataTableQA{1}{0}{2}{2}
\noindent\rule{430pt}{0.5pt}
\dataTableQA{1}{1}{1}{1}
\noindent\rule{430pt}{0.5pt}
\newpage
\noindent\rule{430pt}{0.5pt}
\dataTableQA{2}{0}{2}{1}
\noindent\rule{430pt}{0.5pt}
\dataTableQA{1}{1}{2}{1}
\noindent\rule{430pt}{0.5pt}
\dataTableQA{2}{1}{2}{1}
\noindent\rule{430pt}{0.5pt}

\newpage
\renewcommand{\thesubsection}{\fwboxR{2pt}{\text{[}}\r{3}\g{1}\b{2}]\hspace{-8pt}}
\renewcommand{\theequation}{\thesubsection\hspace{7pt}~}
\subsection{{Box-Triangles}}\vspace{-4pt}

The \emph{box-triangle} integrals are defined to be of the form
\vspace{-4pt}\eq{\fwboxL{370pt}{\hspace{-10pt}\fig{-2.5pt}{1}{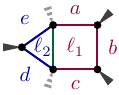}\;\bigger{\Leftrightarrow}\;\frac{\mathfrak{n}^{\hspace{3pt}i}_{[\r{3}\g{1}\b{2}]}[\r{\ell_1}\hspace{-1pt};\!\r{a},\!\r{b},\!\r{c};\!\b{d},\!\b{e};\!\inf]}{\x{\r{\ell_1}}{\r{a}}\x{\r{\ell_1}}{\r{b}}\x{\r{\ell_1}}{\r{c}}\x{\r{\ell_1}}{\inf}\x{\r{\ell_1}}{\b{\ell_2}}\x{\b{\ell_2}}{\b{d}}\x{\b{\ell_2}}{\b{e}}\x{\b{\ell_2}}{\inf}}.}\label{312_masters_defined}\vspace{-10pt}}
\paragraph*{\r{Prescriptive Contours Defining Master Integrals:}}~\\[-12.5pt]
\noindent\rule{430pt}{0.5pt}\\[-14.75pt]
\noindent\rule{430pt}{0.5pt}
\dataTableQB{11}{0}{1}{1}
\noindent\rule{430pt}{0.5pt}
\dataTableQB{11}{0}{2}{0}
\noindent\rule{430pt}{0.5pt}
\dataTableQB{11}{0}{2}{1}
\noindent\rule{430pt}{0.5pt}
\dataTableQB{11}{1}{1}{1}
\noindent\rule{430pt}{0.5pt}
\dataTableQB{12}{0}{1}{1}
\noindent\rule{430pt}{0.5pt}
\newpage
\noindent\rule{430pt}{0.5pt}
\dataTableQB{12}{0}{2}{0}
\noindent\rule{430pt}{0.5pt}
\dataTableQB{12}{1}{1}{0}
\noindent\rule{430pt}{0.5pt}
\dataTableQB{11}{1}{2}{1}
\noindent\rule{430pt}{0.5pt}
\dataTableQB{12}{0}{2}{1}
\noindent\rule{430pt}{0.5pt}
\dataTableQB{12}{1}{1}{1}
\noindent\rule{430pt}{0.5pt}
\dataTableQB{12}{1}{2}{0}
\noindent\rule{430pt}{0.5pt}
\newpage
\noindent\rule{430pt}{0.5pt}
\dataTableQB{22}{0}{1}{1}
\noindent\rule{430pt}{0.5pt}
\noindent\rule{430pt}{0.5pt}
\dataTableQB{22}{0}{2}{0}
\noindent\rule{430pt}{0.5pt}
\dataTableQB{12}{1}{2}{1}
\noindent\rule{430pt}{0.5pt}
\dataTableQB{22}{0}{2}{1}
\noindent\rule{430pt}{0.5pt}
\dataTableQB{22}{1}{1}{1}
\noindent\rule{430pt}{0.5pt}
\dataTableQB{22}{1}{2}{1}
\noindent\rule{430pt}{0.5pt}

\newpage
\renewcommand{\thesubsection}{\fwboxR{2pt}{\text{[}}\r{3}\g{1}\b{3}]\hspace{-8pt}}
\renewcommand{\theequation}{\thesubsection\hspace{7pt}~}
\subsection{{Double-Boxes}}\vspace{-4pt}

The \emph{double-box} integrals are defined to be of the form
\vspace{-4pt}\eq{\fwboxL{370pt}{\hspace{-10pt}\fig{-2.5pt}{1}{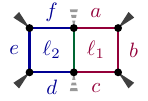}\;\bigger{\Leftrightarrow}\;\frac{\mathfrak{n}^{\hspace{6pt}i}_{[\r{3}\g{1}\b{3}]}[\r{\ell_1}\hspace{-1pt},\!\b{\ell_2}\hspace{-1pt};\!\r{a},\!\r{b},\!\r{c};\!\b{d},\!\b{e},\!\b{f};\!\inf]}{\x{\r{\ell_1}}{\r{a}}\x{\r{\ell_1}}{\r{b}}\x{\r{\ell_1}}{\r{c}}\x{\r{\ell_1}}{\inf}\x{\r{\ell_1}}{\b{\ell_2}}\x{\b{\ell_2}}{\b{d}}\x{\b{\ell_2}}{\b{e}}\x{\b{\ell_2}}{\b{f}}\x{\b{\ell_2}}{\inf}}}\label{313_masters_defined}\vspace{-10pt}}
\paragraph*{\r{Prescriptive Contours Defining Master Integrals:}}~\\[-12.5pt]
\noindent\rule{430pt}{0.5pt}\\[-14.75pt]
\noindent\rule{430pt}{0.5pt}
\dataTableQC{11}{0}{11}{0}
\noindent\rule{430pt}{0.5pt}
\dataTableQC{11}{0}{11}{1}
\noindent\rule{430pt}{0.5pt}
\newpage
\noindent\rule{430pt}{0.5pt}
\dataTableQC{11}{0}{12}{0}
\noindent\rule{430pt}{0.5pt}
\dataTableQC{11}{0}{12}{1}
\noindent\rule{430pt}{0.5pt}
\newpage
\noindent\rule{430pt}{0.5pt}
\dataTableQC{11}{0}{21}{1}
\noindent\rule{430pt}{0.5pt}
\dataTableQCB{11}{0}{22}{0}
\noindent\rule{430pt}{0.5pt}
\newpage
\noindent\rule{430pt}{0.5pt}
\dataTableQCC{11}{1}{11}{1}
\noindent\rule{430pt}{0.5pt}
\dataTableQC{12}{0}{12}{0}
\noindent\rule{430pt}{0.5pt}
\newpage
\noindent\rule{430pt}{0.5pt}
\dataTableQC{12}{0}{21}{0}
\noindent\rule{430pt}{0.5pt}
\dataTableQCB{11}{0}{22}{1}
\noindent\rule{430pt}{0.5pt}
\newpage
\noindent\rule{430pt}{0.5pt}
\dataTableQCC{11}{1}{12}{1}
\noindent\rule{430pt}{0.5pt}
\dataTableQC{12}{0}{12}{1}
\noindent\rule{430pt}{0.5pt}
\newpage
\noindent\rule{430pt}{0.5pt}
\dataTableQC{12}{0}{21}{1}
\noindent\rule{430pt}{0.5pt}
\dataTableQCC{12}{0}{22}{0}
\noindent\rule{430pt}{0.5pt}
\newpage
\noindent\rule{430pt}{0.5pt}
\dataTableQC{12}{1}{21}{0}
\noindent\rule{430pt}{0.5pt}
\dataTableQC{11}{1}{22}{1}
\noindent\rule{430pt}{0.5pt}
\newpage
\noindent\rule{430pt}{0.5pt}
\dataTableQCB{12}{0}{22}{1}
\noindent\rule{430pt}{0.5pt}
\dataTableQCC{12}{1}{12}{1}
\noindent\rule{430pt}{0.5pt}
\newpage
\noindent\rule{430pt}{0.5pt}
\dataTableQCC{12}{1}{21}{1}
\noindent\rule{430pt}{0.5pt}
\dataTableQCB{12}{1}{22}{0}
\noindent\rule{430pt}{0.5pt}
\newpage
\noindent\rule{430pt}{0.5pt}
\dataTableQCD{22}{0}{22}{0}
\noindent\rule{430pt}{0.5pt}
\dataTableQC{12}{1}{22}{1}
\noindent\rule{430pt}{0.5pt}
\newpage
\noindent\rule{430pt}{0.5pt}
\dataTableQCE{22}{0}{22}{1}
\noindent\rule{430pt}{0.5pt}
\dataTableQC{22}{1}{22}{1}
\noindent\rule{430pt}{0.5pt}

\newpage
\renewcommand{\thesubsection}{\fwboxR{2pt}{\text{[}}\r{4}\g{1}\b{2}]\hspace{-8pt}}
\renewcommand{\theequation}{\thesubsection\hspace{7pt}~}
\subsection{{Penta-Triangles}}\vspace{-4pt}

The \emph{penta-triangle} integrals are defined to be of the form
\vspace{-4pt}\eq{\fwboxL{370pt}{\hspace{-10pt}\fig{-2.5pt}{1}{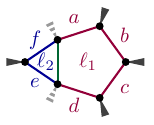}\;\bigger{\Leftrightarrow}\;\frac{\mathfrak{n}^i_{[\r{4}\g{1}\b{2}]}[\r{\ell_1}\hspace{-1pt};\!\r{a},\!\r{b},\!\r{c},\!\r{d};\!\b{e},\!\b{f};\!\inf]}{\x{\r{\ell_1}}{\inf}\x{\r{\ell_1}}{\r{a}}\x{\r{\ell_1}}{\r{b}}\x{\r{\ell_1}}{\r{c}}\x{\r{\ell_1}}{\r{d}}\x{\r{\ell_1}}{\b{\ell_2}}\x{\b{\ell_2}}{\b{e}}\x{\b{\ell_2}}{\b{f}}\x{\b{\ell_2}}{\inf}}.}\label{412_masters_defined}\vspace{-10pt}}
\paragraph*{\r{Prescriptive Contours Defining Master Integrals:}}~\\[-12.5pt]
\noindent\rule{430pt}{0.5pt}\\[-14.75pt]
\noindent\rule{430pt}{0.5pt}
\dataTableQD{222}{0}{1}{1}
\noindent\rule{430pt}{0.5pt}
\dataTableQD{222}{0}{2}{0}
\noindent\rule{430pt}{0.5pt}
\dataTableQD{222}{0}{2}{1}
\noindent\rule{430pt}{0.5pt}
\dataTableQD{222}{1}{1}{1}
\noindent\rule{430pt}{0.5pt}
\dataTableQD{222}{1}{2}{1}
\noindent\rule{430pt}{0.5pt}

\newpage
\renewcommand{\thesubsection}{\fwboxR{2pt}{\text{[}}\r{4}\g{1}\b{3}]\hspace{-8pt}}
\renewcommand{\theequation}{\thesubsection\hspace{7pt}~}
\subsection{{Penta-Boxes}}\vspace{-4pt}

The \emph{penta-triangle} integrals are defined to be of the form
\vspace{-4pt}\eq{\fwboxL{370pt}{\hspace{-10pt}\fig{-2.5pt}{1}{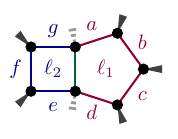}\;\bigger{\Leftrightarrow}\;\frac{\mathfrak{n}^{\hspace{4pt}i}_{[\r{4}\g{1}\b{3}]}[\r{\ell_1}\hspace{-1pt},\b{\ell_2}\hspace{-1pt};\!\r{a},\!\r{b},\!\r{c},\!\r{d};\!\b{e},\!\b{f},\!\b{g};\!\inf]}{\x{\r{\ell_1}}{\r{a}}\x{\r{\ell_1}}{\r{b}}\x{\r{\ell_1}}{\r{c}}\x{\r{\ell_1}}{\r{d}}\x{\r{\ell_1}}{\inf}\x{\r{\ell_1}}{\b{\ell_2}}\x{\b{\ell_2}}{\b{e}}\x{\b{\ell_2}}{\b{f}}\x{\b{\ell_2}}{\b{g}}\x{\b{\ell_2}}{\inf}}.}\label{413_masters_defined}\vspace{-10pt}}
\paragraph*{\r{Prescriptive Contours Defining Master Integrals:}}~\\[-12.5pt]
\noindent\rule{430pt}{0.5pt}\\[-14.75pt]
\noindent\rule{430pt}{0.5pt}
\stepcounter{topologyIndex}\setcounter{topology}{413}\renewcommand{\theequation}{$\hspace{-0pt}I^{\fwbox{18pt}{[\r{4}\g{1}\b{3}]}}_{\fwbox{0pt}{}}\hspace{0pt}$}
\vspace{-12pt}\eq{\hspace{-120pt}\hspace{-0pt}\raisebox{-0pt}{$\cfig{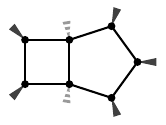}$}\hspace{5pt}\raisebox{-2pt}{$\cfig{413_table}$}\hspace{20pt}\hspace{-100pt}\vspace{-22pt}\label{int412}}
\renewcommand{\theequation}{$\Omega^{\fwbox{18pt}{[\r{4}\g{1}\b{3}]}}_{\fwbox{0pt}{}}\hspace{-1pt}$}
~\\[-8pt]\noindent{\textbf{\b{Defining Contours}:}}\\[-12pt]
\vspace{-05pt}\eq{\fwbox{00pt}{\hspace{05pt}\begin{array}{@{}l@{}}\begin{array}{@{}c@{}c@{}c@{}c@{}c@{}c}~\\[-15pt]
\fwboxR{0pt}{\raisebox{2pt}{$\left\{\rule{0pt}{24pt}\right.$}\hspace{-8.5pt}}\hspace{5pt}\cfig{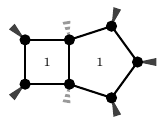}\hspace{0pt}\fwboxL{0pt}{,}&\hspace{0pt}\cfig{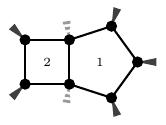}\hspace{0pt}\fwboxL{0pt}{,}&\hspace{0pt}\cfig{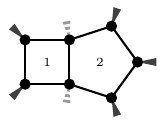}\hspace{0pt}\fwboxL{0pt}{,}&\hspace{0pt}\cfig{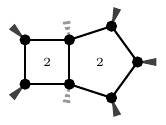}\hspace{2pt}\fwboxL{0pt}{\hspace{-4.5pt}\raisebox{2pt}{$\left.\rule{0pt}{24pt}\right\}$}}\\[-10pt]
\phantom{{}_{1}}\Omega_1&\phantom{{}_{2}}\Omega_2&\phantom{{}_{3}}\Omega_3&\phantom{{}_{4}}\Omega_4\\[-7pt]
\end{array}
\end{array}}\label{omega413}\vspace{-0pt}
}
\noindent\rule{430pt}{0.5pt}

\newpage

\begin{thebibliography}{100}

\bibitem{Mastrolia:2018uzb}
P.~Mastrolia and S.~Mizera, ``{Feynman Integrals and Intersection Theory},''
  \href{http://dx.doi.org/10.1007/JHEP02(2019)139}{{\em JHEP} {\bf 02} (2019)
  139}, \href{http://arxiv.org/abs/1810.03818}{{ arXiv:1810.03818 [hep-th]}}.

\bibitem{Frellesvig:2019kgj}
H.~Frellesvig, F.~Gasparotto, S.~Laporta, M.~K. Mandal, P.~Mastrolia,
  L.~Mattiazzi, and S.~Mizera, ``{Decomposition of Feynman Integrals on the
  Maximal Cut by Intersection Numbers},''
  \href{http://dx.doi.org/10.1007/JHEP05(2019)153}{{\em JHEP} {\bf 05} (2019)
  153}, \href{http://arxiv.org/abs/1901.11510}{{ arXiv:1901.11510 [hep-ph]}}.

\bibitem{Frellesvig:2019uqt}
H.~Frellesvig, F.~Gasparotto, M.~K. Mandal, P.~Mastrolia, L.~Mattiazzi, and
  S.~Mizera, ``{Vector Space of Feynman Integrals and Multivariate Intersection
  Numbers},'' \href{http://dx.doi.org/10.1103/PhysRevLett.123.201602}{{\em
  Phys. Rev. Lett.} {\bf 123} (2019) no. 20, 201602},
  \href{http://arxiv.org/abs/1907.02000}{{ arXiv:1907.02000 [hep-th]}}.

\bibitem{Frellesvig:2020qot}
H.~Frellesvig, F.~Gasparotto, S.~Laporta, M.~K. Mandal, P.~Mastrolia,
  L.~Mattiazzi, and S.~Mizera, ``{Decomposition of Feynman Integrals by
  Multivariate Intersection Numbers},''
  \href{http://dx.doi.org/10.1007/JHEP03(2021)027}{{\em JHEP} {\bf 03} (2021)
  027}, \href{http://arxiv.org/abs/2008.04823}{{ arXiv:2008.04823 [hep-th]}}.

\bibitem{Weinzierl:2020gda}
S.~Weinzierl, ``{Applications of Intersection Numbers in Physics},'' in {\em
  {MathemAmplitudes 2019: Intersection Theory and Feynman Integrals}}.
\newblock 11, 2020.
\newblock \href{http://arxiv.org/abs/2011.02865}{{ arXiv:2011.02865 [hep-th]}}.

\bibitem{Caron-Huot:2021iev}
S.~Caron-Huot and A.~Pokraka, ``{Duals of Feynman Integrals. Part II.
  Generalized Unitarity},''
  \href{http://dx.doi.org/10.1007/JHEP04(2022)078}{{\em JHEP} {\bf 04} (2022)
  078}, \href{http://arxiv.org/abs/2112.00055}{{ arXiv:2112.00055 [hep-th]}}.

\bibitem{Caron-Huot:2021xqj}
S.~Caron-Huot and A.~Pokraka, ``{Duals of Feynman Integrals. Part I.
  Differential Equations},''
  \href{http://dx.doi.org/10.1007/JHEP12(2021)045}{{\em JHEP} {\bf 12} (2021)
  045}, \href{http://arxiv.org/abs/2104.06898}{{ arXiv:2104.06898 [hep-th]}}.

\bibitem{lu2025intersection}
M.~Lu, Z.~Wang, and L.~L. Yang, ``{Intersection Theory, Relative Cohomology and
  the Feynman Parametrization},'' {\em Journal of High Energy Physics} {\bf
  2025} (2025) no. 5, 158.

\bibitem{Gluza:2010ws}
J.~Gluza, K.~Kajda, and D.~A. Kosower, ``{Towards a Basis for Planar Two-Loop
  Integrals},'' \href{http://dx.doi.org/10.1103/PhysRevD.83.045012}{{\em Phys.
  Rev.} {\bf D83} (2011)  045012},
\href{http://arxiv.org/abs/1009.0472}{{ arXiv:1009.0472 [hep-th]}}.

\bibitem{Caron-Huot:2012awx}
S.~Caron-Huot and K.~J. Larsen, ``{Uniqueness of Two-Loop Master Contours},''
  \href{http://dx.doi.org/10.1007/JHEP10(2012)026}{{\em JHEP} {\bf 10} (2012)
  026}, \href{http://arxiv.org/abs/1205.0801}{{ arXiv:1205.0801 [hep-ph]}}.

\bibitem{Henn:2013pwa}
J.~M. Henn, ``{Multiloop Integrals in Dimensional Regularization Made
  Simple},'' \href{http://dx.doi.org/10.1103/PhysRevLett.110.251601}{{\em Phys.
  Rev. Lett.} {\bf 110} (2013) no. 25, 251601},
\href{http://arxiv.org/abs/1304.1806}{{ arXiv:1304.1806 [hep-th]}}.

\bibitem{Henn:2014qga}
J.~M. Henn, ``{Lectures on Differential Equations for Feynman Integrals},''
  \href{http://dx.doi.org/10.1088/1751-8113/48/15/153001}{{\em J. Phys.} {\bf
  A48} (2015) no. 15, 153001},
\href{http://arxiv.org/abs/1412.2296}{{ arXiv:1412.2296 [hep-ph]}}.

\bibitem{Drummond:2010mb}
J.~M. Drummond and J.~M. Henn, ``{Simple Loop Integrals and Amplitudes in
  $\mathcal{N}\!=\!4$ SYM},''
  \href{http://dx.doi.org/10.1007/JHEP05(2011)105}{{\em JHEP} {\bf 1105} (2011)
   105},
\href{http://arxiv.org/abs/1008.2965}{{ arXiv:1008.2965 [hep-th]}}.

\bibitem{Drummond:2010cz}
J.~M. Drummond, J.~M. Henn, and J.~Trnka, ``{New Differential Equations for
  On-Shell Loop Integrals},''
  \href{http://dx.doi.org/10.1007/JHEP04(2011)083}{{\em JHEP} {\bf 1104} (2011)
   083},
\href{http://arxiv.org/abs/1010.3679}{{ arXiv:1010.3679 [hep-th]}}.

\bibitem{Henn:2013nsa}
J.~M. Henn, A.~V. Smirnov, and V.~A. Smirnov, ``{Evaluating Single-Scale and/or
  Non-Planar Diagrams by Differential Equations},''
  \href{http://dx.doi.org/10.1007/JHEP03(2014)088}{{\em JHEP} {\bf 03} (2014)
  088},
\href{http://arxiv.org/abs/1312.2588}{{ arXiv:1312.2588 [hep-th]}}.

\bibitem{Gehrmann:2018yef}
T.~Gehrmann, J.~M. Henn, and N.~A. Lo~Presti, ``{Pentagon Functions for
  Massless Planar Scattering Amplitudes},''
  \href{http://dx.doi.org/10.1007/JHEP10(2018)103}{{\em JHEP} {\bf 10} (2018)
  103}, \href{http://arxiv.org/abs/1807.09812}{{ arXiv:1807.09812 [hep-ph]}}.

\bibitem{Herrmann:2019upk}
E.~Herrmann and J.~Parra-Martinez, ``{Logarithmic Forms and Differential
  Equations for Feynman Integrals},''
  \href{http://dx.doi.org/10.1007/JHEP02(2020)099}{{\em JHEP} {\bf 02} (2020)
  099}, \href{http://arxiv.org/abs/1909.04777}{{ arXiv:1909.04777 [hep-th]}}.

\bibitem{ArkaniHamed:book}
N.~Arkani-Hamed, J.~L. Bourjaily, F.~Cachazo, A.~B. Goncharov, A.~Postnikov,
  and J.~Trnka, \href{http://dx.doi.org/10.1017/CBO9781316091548}{{\em
  {Grassmannian Geometry of Scattering Amplitudes}}}.
\newblock Cambridge University Press, 2016.

\bibitem{Elvang:2013cua}
H.~Elvang and Y.-t. Huang, ``{Scattering Amplitudes},''
\href{http://arxiv.org/abs/1308.1697}{{ arXiv:1308.1697 [hep-th]}}.

\bibitem{Henn:2014yza}
J.~M. Henn and J.~C. Plefka, ``{Scattering Amplitudes in Gauge Theories},''
\href{http://dx.doi.org/978-3-642-54021-9, 10.1007/978-3-642-54022-6}{{\em
  Lect. Notes Phys.} {\bf 883} (2014)  1--195}.

\bibitem{Arkani-Hamed:2014bca}
N.~Arkani-Hamed, J.~L. Bourjaily, F.~Cachazo, A.~Postnikov, and J.~Trnka,
  ``{On-Shell Structures of MHV Amplitudes Beyond the Planar Limit},''
  \href{http://dx.doi.org/10.1007/JHEP06(2015)179}{{\em JHEP} {\bf 06} (2015)
  179},
\href{http://arxiv.org/abs/1412.8475}{{ arXiv:1412.8475 [hep-th]}}.

\bibitem{Bourjaily:2016mnp}
J.~L. Bourjaily, S.~Franco, D.~Galloni, and C.~Wen, ``{Stratifying On-Shell
  Cluster Varieties: the Geometry of Non-Planar On-Shell Diagrams},''
  \href{http://dx.doi.org/10.1007/JHEP10(2016)003}{{\em JHEP} {\bf 10} (2016)
  003},
\href{http://arxiv.org/abs/1607.01781}{{ arXiv:1607.01781 [hep-th]}}.

\bibitem{Brink:1976bc}
L.~Brink, J.~H. Schwarz, and J.~Scherk, ``{Supersymmetric Yang-Mills
  Theories},'' \href{http://dx.doi.org/10.1016/0550-3213(77)90328-5}{{\em Nucl.
  Phys. B} {\bf 121} (1977)  77--92}.

\bibitem{Gliozzi:1976qd}
F.~Gliozzi, J.~Scherk, and D.~I. Olive, ``{Supersymmetry, Supergravity Theories
  and the Dual Spinor Model},''
  \href{http://dx.doi.org/10.1016/0550-3213(77)90206-1}{{\em Nucl. Phys. B}
  {\bf 122} (1977)  253--290}.

\bibitem{Arkani-Hamed:2008owk}
N.~Arkani-Hamed, F.~Cachazo, and J.~Kaplan, ``{What is the Simplest Quantum
  Field Theory?},'' \href{http://dx.doi.org/10.1007/JHEP09(2010)016}{{\em JHEP}
  {\bf 09} (2010)  016}, \href{http://arxiv.org/abs/0808.1446}{{
  arXiv:0808.1446 [hep-th]}}.

\bibitem{Bern:2006kd}
Z.~Bern, L.~J. Dixon, and R.~Roiban, ``{Is $\mathcal{N}\!=\!8$ Supergravity
  Ultraviolet Finite?},''
  \href{http://dx.doi.org/10.1016/j.physletb.2006.11.030}{{\em Phys. Lett.}
  {\bf B644} (2007)  265--271},
\href{http://arxiv.org/abs/hep-th/0611086}{{ arXiv:hep-th/0611086 [hep-th]}}.

\bibitem{BjerrumBohr:2006yw}
N.~E.~J. Bjerrum-Bohr, D.~C. Dunbar, H.~Ita, W.~B. Perkins, and K.~Risager,
  ``{The No-Triangle Hypothesis for $\mathcal{N}\!=\!8$ Supergravity},''
  \href{http://dx.doi.org/10.1088/1126-6708/2006/12/072}{{\em JHEP} {\bf 12}
  (2006)  072},
\href{http://arxiv.org/abs/hep-th/0610043}{{ arXiv:hep-th/0610043 [hep-th]}}.

\bibitem{BjerrumBohr:2008ji}
N.~Bjerrum-Bohr and P.~Vanhove, ``{Absence of Triangles in Maximal Supergravity
  Amplitudes},'' \href{http://dx.doi.org/10.1088/1126-6708/2008/10/006}{{\em
  JHEP} {\bf 0810} (2008)  006},
\href{http://arxiv.org/abs/0805.3682}{{ arXiv:0805.3682 [hep-th]}}.

\bibitem{Bourjaily:2021ewt}
J.~L. Bourjaily, C.~Langer, and K.~Patatoukos, ``{Locally-Finite Quantities in
  sYM},'' \href{http://dx.doi.org/10.1007/JHEP04(2021)298}{{\em JHEP} {\bf 04}
  (2021)  298}, \href{http://arxiv.org/abs/2102.02821}{{ arXiv:2102.02821
  [hep-th]}}.

\bibitem{Hodges:2009hk}
A.~Hodges, ``{Eliminating Spurious Poles from Gauge-Theoretic Amplitudes},''
  \href{http://dx.doi.org/10.1007/JHEP05(2013)135}{{\em JHEP} {\bf 1305} (2013)
   135},
\href{http://arxiv.org/abs/0905.1473}{{ arXiv:0905.1473 [hep-th]}}.

\bibitem{Bourjaily:2020qca}
J.~L. Bourjaily, E.~Herrmann, C.~Langer, and J.~Trnka, ``{Building Bases of
  Loop Integrands},'' \href{http://dx.doi.org/10.1007/JHEP11(2020)116}{{\em
  JHEP} {\bf 11} (2020)  116}, \href{http://arxiv.org/abs/2007.13905}{{
  arXiv:2007.13905 [hep-th]}}.

\bibitem{Melrose:1965kb}
D.~B. Melrose, ``{Reduction of Feynman Diagrams},''
  \href{http://dx.doi.org/10.1007/BF02832919}{{\em Nuovo Cim.} {\bf 40} (1965)
  181--213}.

\bibitem{Passarino:1978jh}
G.~Passarino and M.~Veltman, ``{One Loop Corrections for $e^+\,e^-$
  Annihilation Into $\mu^+\,\mu^-$ in the Weinberg Model},''
\href{http://dx.doi.org/10.1016/0550-3213(79)90234-7}{{\em Nucl. Phys.} {\bf
  B160} (1979)  151}.

\bibitem{vanNeerven:1983vr}
W.~L. van Neerven and J.~A.~M. Vermaseren, ``{Large Loop Integrals},''
\href{http://dx.doi.org/10.1016/0370-2693(84)90237-5}{{\em Phys. Lett.} {\bf
  B137} (1984)  241}.

\bibitem{Denner:2002ii}
A.~Denner and S.~Dittmaier, ``{Reduction of One Loop Tensor Five Point
  Integrals},'' \href{http://dx.doi.org/10.1016/S0550-3213(03)00184-6}{{\em
  Nucl. Phys. B} {\bf 658} (2003)  175--202},
  \href{http://arxiv.org/abs/hep-ph/0212259}{{ arXiv:hep-ph/0212259}}.

\bibitem{Denner:2005nn}
A.~Denner and S.~Dittmaier, ``{Reduction Schemes for One-Loop Tensor
  Integrals},'' \href{http://dx.doi.org/10.1016/j.nuclphysb.2005.11.007}{{\em
  Nucl. Phys. B} {\bf 734} (2006)  62--115},
  \href{http://arxiv.org/abs/hep-ph/0509141}{{ arXiv:hep-ph/0509141}}.

\bibitem{Cachazo:2008vp}
F.~Cachazo, ``{Sharpening The Leading Singularity},''
\href{http://arxiv.org/abs/0803.1988}{{ arXiv:0803.1988 [hep-th]}}.

\bibitem{Mastrolia:2010nb}
P.~Mastrolia, G.~Ossola, T.~Reiter, and F.~Tramontano, ``{Scattering Amplitudes
  from Unitarity-Based Reduction Algorithm at the Integrand-Level},''
  \href{http://dx.doi.org/10.1007/JHEP08(2010)080}{{\em JHEP} {\bf 08} (2010)
  080},
\href{http://arxiv.org/abs/1006.0710}{{ arXiv:1006.0710 [hep-ph]}}.

\bibitem{Ellis:2011cr}
R.~K. Ellis, Z.~Kunszt, K.~Melnikov, and G.~Zanderighi, ``{One-Loop
  Calculations in Quantum Field Theory: from Feynman Diagrams to Unitarity
  Cuts},'' \href{http://dx.doi.org/10.1016/j.physrep.2012.01.008}{{\em Phys.
  Rept.} {\bf 518} (2012)  141--250},
\href{http://arxiv.org/abs/1105.4319}{{ arXiv:1105.4319 [hep-ph]}}.

\bibitem{Mastrolia:2011pr}
P.~Mastrolia and G.~Ossola, ``{On the Integrand-Reduction Method for Two-Loop
  Scattering Amplitudes},''
  \href{http://dx.doi.org/10.1007/JHEP11(2011)014}{{\em JHEP} {\bf 1111} (2011)
   014},
\href{http://arxiv.org/abs/1107.6041}{{ arXiv:1107.6041 [hep-ph]}}.

\bibitem{Badger:2012dv}
S.~Badger, H.~Frellesvig, and Y.~Zhang, ``{An Integrand Reconstruction Method
  for Three-Loop Amplitudes},''
  \href{http://dx.doi.org/10.1007/JHEP08(2012)065}{{\em JHEP} {\bf 1208} (2012)
   065},
\href{http://arxiv.org/abs/1207.2976}{{ arXiv:1207.2976 [hep-ph]}}.

\bibitem{Mastrolia:2012an}
P.~Mastrolia, E.~Mirabella, G.~Ossola, and T.~Peraro, ``{Scattering Amplitudes
  from Multivariate Polynomial Division},''
  \href{http://dx.doi.org/10.1016/j.physletb.2012.09.053}{{\em Phys. Lett.}
  {\bf B718} (2012)  173--177},
\href{http://arxiv.org/abs/1205.7087}{{ arXiv:1205.7087 [hep-ph]}}.

\bibitem{Mastrolia:2013kca}
P.~Mastrolia, E.~Mirabella, G.~Ossola, and T.~Peraro, ``{Multiloop Integrand
  Reduction for Dimensionally Regulated Amplitudes},''
  \href{http://dx.doi.org/10.1016/j.physletb.2013.10.066}{{\em Phys. Lett.}
  {\bf B727} (2013)  532--535},
\href{http://arxiv.org/abs/1307.5832}{{ arXiv:1307.5832 [hep-ph]}}.

\bibitem{Badger:2013sta}
S.~Badger, H.~Frellesvig, and Y.~Zhang, ``{Multi-Loop Integrand Reduction with
  Computational Algebraic Geometry},''
  \href{http://dx.doi.org/10.1088/1742-6596/523/1/012061}{{\em J. Phys. Conf.
  Ser.} {\bf 523} (2014)  012061},
\href{http://arxiv.org/abs/1310.4445}{{ arXiv:1310.4445 [hep-ph]}}.

\bibitem{Mastrolia:2016dhn}
P.~Mastrolia, T.~Peraro, and A.~Primo, ``{Adaptive Integrand Decomposition in
  parallel and orthogonal space},''
  \href{http://dx.doi.org/10.1007/JHEP08(2016)164}{{\em JHEP} {\bf 08} (2016)
  164},
\href{http://arxiv.org/abs/1605.03157}{{ arXiv:1605.03157 [hep-ph]}}.

\bibitem{Ita:2015tya}
H.~Ita, ``{Two-Loop Integrand Decomposition into Master Integrals and Surface
  Terms},'' \href{http://dx.doi.org/10.1103/PhysRevD.94.116015}{{\em Phys.
  Rev.} {\bf D94} (2016) no. 11, 116015},
\href{http://arxiv.org/abs/1510.05626}{{ arXiv:1510.05626 [hep-th]}}.

\bibitem{Zhang:2016kfo}
Y.~Zhang, ``{Lecture Notes on Multi-loop Integral Reduction and Applied
  Algebraic Geometry},'' \href{http://arxiv.org/abs/1612.02249}{{
  arXiv:1612.02249 [hep-th]}}.
\url{http://inspirehep.net/record/1502112/files/arXiv:1612.02249.pdf}.

\bibitem{Ossola:2006us}
G.~Ossola, C.~G. Papadopoulos, and R.~Pittau, ``{Reducing Full One-Loop
  Amplitudes to Scalar Integrals at the Integrand Level},''
  \href{http://dx.doi.org/10.1016/j.nuclphysb.2006.11.012}{{\em Nucl. Phys.}
  {\bf B763} (2007)  147--169},
\href{http://arxiv.org/abs/hep-ph/0609007}{{ arXiv:hep-ph/0609007 [hep-ph]}}.

\bibitem{Mastrolia:2012wf}
P.~Mastrolia, E.~Mirabella, and T.~Peraro, ``{Integrand Reduction of One-Loop
  Scattering Amplitudes Through Laurent Series Expansion},''
  \href{http://dx.doi.org/10.1007/JHEP06(2012)095}{{\em JHEP} {\bf 1206} (2012)
   095},
\href{http://arxiv.org/abs/1203.0291}{{ arXiv:1203.0291 [hep-ph]}}.

\bibitem{Kleiss:2012yv}
R.~H. Kleiss, I.~Malamos, C.~G. Papadopoulos, and R.~Verheyen, ``{Counting to
  One: Reducibility of One- and Two-Loop Amplitudes at the Integrand Level},''
  \href{http://dx.doi.org/10.1007/JHEP12(2012)038}{{\em JHEP} {\bf 1212} (2012)
   038},
\href{http://arxiv.org/abs/1206.4180}{{ arXiv:1206.4180 [hep-ph]}}.

\bibitem{Feng:2012bm}
B.~Feng and R.~Huang, ``{The Classification of Two-Loop Integrand Basis in Pure
  Four-Dimension},'' \href{http://dx.doi.org/10.1007/JHEP02(2013)117}{{\em
  JHEP} {\bf 1302} (2013)  117},
\href{http://arxiv.org/abs/1209.3747}{{ arXiv:1209.3747 [hep-ph]}}.

\bibitem{Feynman:1949zx}
R.~P. Feynman, ``{Space-Time Approach to Quantum Electrodynamics},''
  \href{http://dx.doi.org/10.1103/PhysRev.76.769}{{\em Phys. Rev.} {\bf 76}
  (1949)  769--789}.

\bibitem{Schwinger:1949ra}
J.~S. Schwinger, ``{Quantum Electrodynamics. III: The Electromagnetic
  Properties of the Electron: Radiative Corrections to Scattering},''
  \href{http://dx.doi.org/10.1103/PhysRev.76.790}{{\em Phys. Rev.} {\bf 76}
  (1949)  790--817}.

\bibitem{barnes1908new}
E.~W. Barnes, ``{A New Development of the Theory of the Hypergeometric
  Functions},'' {\em Proceedings of the London Mathematical Society} {\bf 2}
  (1908) no. 1, 141--177.

\bibitem{mellin1910abriss}
H.~Mellin, ``{Abri\ss Einer Einheitlichen Theorie der Gamma-und der
  Hypergeometrischen Funktionen},'' {\em Mathematische Annalen} {\bf 68} (1910)
  no. 3, 305--337.

\bibitem{Smirnov:2004ym}
V.~A. Smirnov, ``{Evaluating Feynman Integrals},''
{\em Springer Tracts Mod. Phys.} {\bf 211} (2004)  1--244.

\bibitem{Smirnov:2006ry}
V.~A. Smirnov, {\em {Feynman Integral Calculus}}.
\newblock {Springer-Verlag},
2006.
\newblock

\bibitem{Smirnov:2012gma}
V.~A. Smirnov, ``{Analytic Tools for Feynman integrals},''
\href{http://dx.doi.org/10.1007/978-3-642-34886-0}{{\em Springer Tracts Mod.
  Phys.} {\bf 250} (2012)  1--296}.

\bibitem{Kosower:2011ty}
D.~A. Kosower and K.~J. Larsen, ``{Maximal Unitarity at Two Loops},''
  \href{http://dx.doi.org/10.1103/PhysRevD.85.045017}{{\em Phys. Rev.} {\bf
  D85} (2012)  045017},
\href{http://arxiv.org/abs/1108.1180}{{ arXiv:1108.1180 [hep-th]}}.

\bibitem{Badger:2013gxa}
S.~Badger, H.~Frellesvig, and Y.~Zhang, ``{A Two-Loop Five-Gluon Helicity
  Amplitude in QCD},'' \href{http://dx.doi.org/10.1007/JHEP12(2013)045}{{\em
  JHEP} {\bf 12} (2013)  045},
\href{http://arxiv.org/abs/1310.1051}{{ arXiv:1310.1051 [hep-ph]}}.

\bibitem{Johansson:2013sda}
H.~Johansson, D.~A. Kosower, and K.~J. Larsen, ``{Maximal Unitarity for the
  Four-Mass Double Box},''
  \href{http://dx.doi.org/10.1103/PhysRevD.89.125010}{{\em Phys. Rev.} {\bf
  D89} (2014)  125010},
\href{http://arxiv.org/abs/1308.4632}{{ arXiv:1308.4632 [hep-th]}}.

\bibitem{Nogueira:1991ex}
P.~Nogueira, ``{Automatic Feynman Graph Generation},''
\href{http://dx.doi.org/10.1006/jcph.1993.1074}{{\em J. Comput. Phys.} {\bf
  105} (1993)  279--289}.

\bibitem{Laporta:2000dsw}
S.~Laporta, ``{High-Precision Calculation of Multiloop Feynman Integrals by
  Difference Equations},''
  \href{http://dx.doi.org/10.1142/S0217751X00002159}{{\em Int. J. Mod. Phys. A}
  {\bf 15} (2000)  5087--5159}, \href{http://arxiv.org/abs/hep-ph/0102033}{{
  arXiv:hep-ph/0102033}}.

\bibitem{Binoth:2000ps}
T.~Binoth and G.~Heinrich, ``{An Automatized Algorithm to Compute Infrared
  Divergent Multiloop Integrals},''
  \href{http://dx.doi.org/10.1016/S0550-3213(00)00429-6}{{\em Nucl. Phys.} {\bf
  B585} (2000)  741--759},
\href{http://arxiv.org/abs/hep-ph/0004013}{{ arXiv:hep-ph/0004013 [hep-ph]}}.

\bibitem{Gluza:2007rt}
J.~Gluza, K.~Kajda, and T.~Riemann, ``{{\tt AMBRE}: A {\sc Mathematica} Package
  for the Construction of Mellin-Barnes Representations for Feynman
  Integrals},'' \href{http://dx.doi.org/10.1016/j.cpc.2007.07.001}{{\em Comput.
  Phys. Commun.} {\bf 177} (2007)  879--893},
\href{http://arxiv.org/abs/0704.2423}{{ arXiv:0704.2423 [hep-ph]}}.

\bibitem{Smirnov:2008iw}
A.~V. Smirnov, ``{Algorithm \texttt{FIRE} -- Feynman Integral REduction},''
  \href{http://dx.doi.org/10.1088/1126-6708/2008/10/107}{{\em JHEP} {\bf 10}
  (2008)  107}, \href{http://arxiv.org/abs/0807.3243}{{ arXiv:0807.3243
  [hep-ph]}}.

\bibitem{Smirnov:2008py}
A.~V. Smirnov and M.~N. Tentyukov, ``{Feynman Integral Evaluation by a Sector
  decomposiTion Approach (FIESTA)},''
  \href{http://dx.doi.org/10.1016/j.cpc.2008.11.006}{{\em Comput. Phys.
  Commun.} {\bf 180} (2009)  735--746},
\href{http://arxiv.org/abs/0807.4129}{{ arXiv:0807.4129 [hep-ph]}}.

\bibitem{Peraro:2014cba}
T.~Peraro, ``{Ninja: Automated Integrand Reduction via Laurent Expansion for
  One-Loop Amplitudes},''
  \href{http://dx.doi.org/10.1016/j.cpc.2014.06.017}{{\em Comput. Phys.
  Commun.} {\bf 185} (2014)  2771--2797},
\href{http://arxiv.org/abs/1403.1229}{{ arXiv:1403.1229 [hep-ph]}}.

\bibitem{Ochman:2015fho}
M.~Ochman and T.~Riemann, ``{{\tt MBsums} - A {\sc Mathematica} Package for the
  Representation of Mellin-Barnes Integrals by Multiple Sums},''
  \href{http://dx.doi.org/10.5506/APhysPolB.46.2117}{{\em Acta Phys. Polon.}
  {\bf B46} (2015) no. 11, 2117},
\href{http://arxiv.org/abs/1511.01323}{{ arXiv:1511.01323 [hep-ph]}}.

\bibitem{vonManteuffel:2014ixa}
A.~von Manteuffel and R.~M. Schabinger, ``{A Novel Approach to Integration by
  Parts Reduction},''
  \href{http://dx.doi.org/10.1016/j.physletb.2015.03.029}{{\em Phys. Lett.}
  {\bf B744} (2015)  101--104},
\href{http://arxiv.org/abs/1406.4513}{{ arXiv:1406.4513 [hep-ph]}}.

\bibitem{Peraro:2016wsq}
T.~Peraro, ``{Scattering Amplitudes over Finite Fields and Multivariate
  Functional Reconstruction},''
  \href{http://dx.doi.org/10.1007/JHEP12(2016)030}{{\em JHEP} {\bf 12} (2016)
  030}, \href{http://arxiv.org/abs/1608.01902}{{ arXiv:1608.01902 [hep-ph]}}.

\bibitem{Abreu:2019odu}
S.~Abreu, J.~Dormans, F.~Febres~Cordero, H.~Ita, B.~Page, and V.~Sotnikov,
  ``{Analytic Form of the Planar Two-Loop Five-Parton Scattering Amplitudes in
  QCD},'' \href{http://dx.doi.org/10.1007/JHEP05(2019)084}{{\em JHEP} {\bf 05}
  (2019)  084}, \href{http://arxiv.org/abs/1904.00945}{{ arXiv:1904.00945
  [hep-ph]}}.

\bibitem{Peraro:2019okx}
T.~Peraro, \href{http://dx.doi.org/10.22323/1.375.0077}{``{Analytic Multi-Loop
  Results using Finite Fields and Dataflow Graphs with FiniteFlow},''} in {\em
  {14th International Symposium on Radiative Corrections}: {Application of
  Quantum Field Theory to Phenomenology}}.
\newblock 12, 2019.
\newblock \href{http://arxiv.org/abs/1912.03142}{{ arXiv:1912.03142 [hep-ph]}}.

\bibitem{vonHippel:2025okr}
M.~von Hippel and M.~Wilhelm, ``{Refining Integration-by-Parts Reduction of
  Feynman Integrals with Machine Learning},''
  \href{http://dx.doi.org/10.1007/JHEP05(2025)185}{{\em JHEP} {\bf 05} (2025)
  185}, \href{http://arxiv.org/abs/2502.05121}{{ arXiv:2502.05121 [hep-th]}}.

\bibitem{Peraro:2019svx}
T.~Peraro, ``{$\text{FiniteFlow}$: Multivariate Functional Reconstruction using
  Finite Fields and Dataflow Graphs},''
  \href{http://dx.doi.org/10.1007/JHEP07(2019)031}{{\em JHEP} {\bf 07} (2019)
  031}, \href{http://arxiv.org/abs/1905.08019}{{ arXiv:1905.08019 [hep-ph]}}.

\bibitem{Peraro:2019cjj}
T.~Peraro and L.~Tancredi, ``{Physical Projectors for Multi-Leg Helicity
  Amplitudes},'' \href{http://dx.doi.org/10.1007/JHEP07(2019)114}{{\em JHEP}
  {\bf 07} (2019)  114}, \href{http://arxiv.org/abs/1906.03298}{{
  arXiv:1906.03298 [hep-ph]}}.

\bibitem{kontsevich2001periods}
M.~Kontsevich and D.~Zagier, ``Periods,'' in {\em Mathematics unlimited---2001
  and beyond}, pp.~771--808.
\newblock Springer, 2001.

\bibitem{Bern:1994zx}
Z.~Bern, L.~J. Dixon, D.~C. Dunbar, and D.~A. Kosower, ``{One-Loop $n$-Point
  Gauge Theory Amplitudes, Unitarity and Collinear Limits},''
  \href{http://dx.doi.org/10.1016/0550-3213(94)90179-1}{{\em Nucl. Phys.} {\bf
  B425} (1994)  217--260},
\href{http://arxiv.org/abs/hep-ph/9403226}{{ arXiv:hep-ph/9403226}}.

\bibitem{Bern:1994cg}
Z.~Bern, L.~J. Dixon, D.~C. Dunbar, and D.~A. Kosower, ``{Fusing Gauge Theory
  Tree Amplitudes into Loop Amplitudes},''
  \href{http://dx.doi.org/10.1016/0550-3213(94)00488-Z}{{\em Nucl. Phys.} {\bf
  B435} (1995)  59--101},
\href{http://arxiv.org/abs/hep-ph/9409265}{{ arXiv:hep-ph/9409265}}.

\bibitem{Britto:2004nc}
R.~Britto, F.~Cachazo, and B.~Feng, ``{Generalized Unitarity and One-Loop
  Amplitudes in $\mathcal{N}\!=\!4$ Super-Yang-Mills},''
  \href{http://dx.doi.org/10.1016/j.nuclphysb.2005.07.014}{{\em Nucl. Phys.}
  {\bf B725} (2005)  275--305},
\href{http://arxiv.org/abs/hep-th/0412103}{{ arXiv:hep-th/0412103}}.

\bibitem{Drummond:2008bq}
J.~Drummond, J.~Henn, G.~Korchemsky, and E.~Sokatchev, ``{Generalized Unitarity
  for $\mathcal{N}\!=\!4$ Super-Amplitudes},''
  \href{http://dx.doi.org/10.1016/j.nuclphysb.2012.12.009}{{\em Nucl. Phys.}
  {\bf B869} (2013)  452--492},
\href{http://arxiv.org/abs/0808.0491}{{ arXiv:0808.0491 [hep-th]}}.

\bibitem{Anastasiou:2006jv}
C.~Anastasiou, R.~Britto, B.~Feng, Z.~Kunszt, and P.~Mastrolia,
  ``{$D$-Dimensional Unitarity Cut Method},''
  \href{http://dx.doi.org/10.1016/j.physletb.2006.12.022}{{\em Phys. Lett.}
  {\bf B645} (2007)  213--216},
\href{http://arxiv.org/abs/hep-ph/0609191}{{ arXiv:hep-ph/0609191 [hep-ph]}}.

\bibitem{ArkaniHamed:2009dn}
N.~Arkani-Hamed, F.~Cachazo, C.~Cheung, and J.~Kaplan, ``{A Duality For The
  $S$-Matrix},'' \href{http://dx.doi.org/10.1007/JHEP03(2010)020}{{\em JHEP}
  {\bf 1003} (2010)  020},
\href{http://arxiv.org/abs/0907.5418}{{ arXiv:0907.5418 [hep-th]}}.

\bibitem{ArkaniHamed:2009dg}
N.~Arkani-Hamed, J.~Bourjaily, F.~Cachazo, and J.~Trnka, ``{Unification of
  Residues and Grassmannian Dualities},''
  \href{http://dx.doi.org/10.1007/JHEP01(2011)049}{{\em JHEP} {\bf 1101} (2011)
   049},
\href{http://arxiv.org/abs/0912.4912}{{ arXiv:0912.4912 [hep-th]}}.

\bibitem{ArkaniHamed:2010gg}
N.~Arkani-Hamed, J.~L. Bourjaily, F.~Cachazo, A.~Hodges, and J.~Trnka, ``{A
  {N}ote on {P}olytopes for {S}cattering {A}mplitudes},''
  \href{http://dx.doi.org/10.1007/JHEP04(2012)081}{{\em JHEP} {\bf 1204} (2012)
   081},
\href{http://arxiv.org/abs/1012.6030}{{ arXiv:1012.6030 [hep-th]}}.

\bibitem{Bourjaily:2012gy}
J.~L. Bourjaily, ``{Positroids, Plabic Graphs, and Scattering Amplitudes in
  {\sc Mathematica}},''
\href{http://arxiv.org/abs/1212.6974}{{ arXiv:1212.6974 [hep-th]}}.

\bibitem{Arkani-Hamed:2014via}
N.~Arkani-Hamed, J.~L. Bourjaily, F.~Cachazo, and J.~Trnka, ``{Singularity
  Structure of Maximally Supersymmetric Scattering Amplitudes},''
  \href{http://dx.doi.org/10.1103/PhysRevLett.113.261603}{{\em Phys. Rev.
  Lett.} {\bf 113} (2014) no. 26, 261603},
\href{http://arxiv.org/abs/1410.0354}{{ arXiv:1410.0354 [hep-th]}}.

\bibitem{Chetyrkin:1981qh}
K.~G. Chetyrkin and F.~V. Tkachov, ``{Integration by Parts: The Algorithm to
  Calculate beta Functions in 4 Loops},''
\href{http://dx.doi.org/10.1016/0550-3213(81)90199-1}{{\em Nucl. Phys.} {\bf
  B192} (1981)  159--204}.

\bibitem{Tkachov:1981wb}
F.~V. Tkachov, ``{A Theorem on Analytical Calculability of Four Loop
  Renormalization Group Functions},''
\href{http://dx.doi.org/10.1016/0370-2693(81)90288-4}{{\em Phys. Lett.} {\bf
  100B} (1981)  65--68}.

\bibitem{Bourjaily:2017bsb}
J.~L. Bourjaily, A.~J. McLeod, M.~Spradlin, M.~von Hippel, and M.~Wilhelm,
  ``{Elliptic Double-Box Integrals: Massless Scattering Amplitudes beyond
  Polylogarithms},''
  \href{http://dx.doi.org/10.1103/PhysRevLett.120.121603}{{\em Phys. Rev.
  Lett.} {\bf 120} (2018) no. 12, 121603},
\href{http://arxiv.org/abs/1712.02785}{{ arXiv:1712.02785 [hep-th]}}.

\bibitem{Bourjaily:2018ycu}
J.~L. Bourjaily, Y.-H. He, A.~J. McLeod, M.~von Hippel, and M.~Wilhelm,
  ``{Traintracks Through Calabi-Yaus: Amplitudes Beyond Elliptic
  Polylogarithms},''
\href{http://arxiv.org/abs/1805.09326}{{ arXiv:1805.09326 [hep-th]}}.

\bibitem{Bourjaily:2018yfy}
J.~L. Bourjaily, A.~J. McLeod, M.~von Hippel, and M.~Wilhelm, ``{A (Bounded)
  Bestiary of Feynman Integral Calabi-Yau Geometries},''
  \href{http://dx.doi.org/10.1103/PhysRevLett.122.031601}{{\em Phys. Rev.
  Lett.} {\bf 122} (2019) no. 3, 031601},
  \href{http://arxiv.org/abs/1810.07689}{{ arXiv:1810.07689 [hep-th]}}.

\bibitem{Bourjaily:2019hmc}
J.~L. Bourjaily, A.~J. McLeod, C.~Vergu, M.~Volk, M.~Von~Hippel, and
  M.~Wilhelm, ``{Embedding Feynman Integral (Calabi-Yau) Geometries in Weighted
  Projective Space},'' \href{http://dx.doi.org/10.1007/JHEP01(2020)078}{{\em
  JHEP} {\bf 01} (2020)  078}, \href{http://arxiv.org/abs/1910.01534}{{
  arXiv:1910.01534 [hep-th]}}.

\bibitem{Walden:2020odh}
M.~Walden and S.~Weinzierl, ``{Numerical Evaluation of Iterated Integrals
  Related to Elliptic Feynman Integrals},''
  \href{http://arxiv.org/abs/2010.05271}{{ arXiv:2010.05271 [hep-ph]}}.

\bibitem{Kristensson:2021ani}
A.~Kristensson, M.~Wilhelm, and C.~Zhang, ``{Elliptic Double Box and Symbology
  Beyond Polylogarithms},''
  \href{http://dx.doi.org/10.1103/PhysRevLett.127.251603}{{\em Phys. Rev.
  Lett.} {\bf 127} (2021) no. 25, 251603},
  \href{http://arxiv.org/abs/2106.14902}{{ arXiv:2106.14902 [hep-th]}}.

\bibitem{Bourjaily:2022bwx}
J.~L. Bourjaily {\em et al.}, ``{Functions Beyond Multiple Polylogarithms for
  Precision Collider Physics},'' in {\em {Snowmass 2021}}.
\newblock 3, 2022.
\newblock \href{http://arxiv.org/abs/2203.07088}{{ arXiv:2203.07088 [hep-ph]}}.

\bibitem{Wilhelm:2022wow}
M.~Wilhelm and C.~Zhang, ``{Symbology for Elliptic Multiple Polylogarithms and
  the Symbol Prime},'' \href{http://arxiv.org/abs/2206.08378}{{
  arXiv:2206.08378 [hep-th]}}.

\bibitem{Morales:2022csr}
R.~Morales, A.~Spiering, M.~Wilhelm, Q.~Yang, and C.~Zhang, ``{Bootstrapping
  Elliptic Feynman Integrals Using Schubert Analysis},''
  \href{http://dx.doi.org/10.1103/PhysRevLett.131.041601}{{\em Phys. Rev.
  Lett.} {\bf 131} (2023) no. 4, 041601},
  \href{http://arxiv.org/abs/2212.09762}{{ arXiv:2212.09762 [hep-th]}}.

\bibitem{Frellesvig:2023bbf}
H.~Frellesvig, R.~Morales, and M.~Wilhelm, ``{Calabi-Yau Meets Gravity: A
  Calabi-Yau Threefold at Fifth Post-Minkowskian Order},''
  \href{http://dx.doi.org/10.1103/PhysRevLett.132.201602}{{\em Phys. Rev.
  Lett.} {\bf 132} (2024) no. 20, 201602},
  \href{http://arxiv.org/abs/2312.11371}{{ arXiv:2312.11371 [hep-th]}}.

\bibitem{Goncharov:2010jf}
A.~B. Goncharov, M.~Spradlin, C.~Vergu, and A.~Volovich, ``{Classical
  Polylogarithms for Amplitudes and Wilson Loops},''
  \href{http://dx.doi.org/10.1103/PhysRevLett.105.151605}{{\em Phys. Rev.
  Lett.} {\bf 105} (2010)  151605},
\href{http://arxiv.org/abs/1006.5703}{{ arXiv:1006.5703 [hep-th]}}.

\bibitem{Duhr:2011zq}
C.~Duhr, H.~Gangl, and J.~R. Rhodes, ``{From Polygons and Symbols to
  Polylogarithmic Functions},''
  \href{http://dx.doi.org/10.1007/JHEP10(2012)075}{{\em JHEP} {\bf 10} (2012)
  075},
\href{http://arxiv.org/abs/1110.0458}{{ arXiv:1110.0458 [math-ph]}}.

\bibitem{Duhr:2012fh}
C.~Duhr, ``{Hopf Algebras, Coproducts and Symbols: an Application to Higgs
  Boson Amplitudes},'' \href{http://dx.doi.org/10.1007/JHEP08(2012)043}{{\em
  JHEP} {\bf 1208} (2012)  043},
\href{http://arxiv.org/abs/1203.0454}{{ arXiv:1203.0454 [hep-ph]}}.

\bibitem{Golden:2013lha}
J.~Golden and M.~Spradlin, ``{The Differential of All Two-Loop MHV Amplitudes
  in $\mathcal{N}\!=\!4$ Yang-Mills Theory},''
  \href{http://dx.doi.org/10.1007/JHEP09(2013)111}{{\em JHEP} {\bf 1309} (2013)
   111},
\href{http://arxiv.org/abs/1306.1833}{{ arXiv:1306.1833 [hep-th]}}.

\bibitem{Golden:2013xva}
J.~Golden, A.~B. Goncharov, M.~Spradlin, C.~Vergu, and A.~Volovich, ``{Motivic
  Amplitudes and Cluster Coordinates},''
  \href{http://dx.doi.org/10.1007/JHEP01(2014)091}{{\em JHEP} {\bf 1401} (2014)
   091},
\href{http://arxiv.org/abs/1305.1617}{{ arXiv:1305.1617 [hep-th]}}.

\bibitem{Golden:2014pua}
J.~Golden and M.~Spradlin, ``{A Cluster Bootstrap for Two-Loop MHV
  Amplitudes},'' \href{http://dx.doi.org/10.1007/JHEP02(2015)002}{{\em JHEP}
  {\bf 1502} (2015)  002},
\href{http://arxiv.org/abs/1411.3289}{{ arXiv:1411.3289 [hep-th]}}.

\bibitem{Dixon:2020cnr}
L.~J. Dixon and Y.-T. Liu, ``{Lifting Heptagon Symbols to Functions},''
  \href{http://dx.doi.org/10.1007/JHEP10(2020)031}{{\em JHEP} {\bf 10} (2020)
  031}, \href{http://arxiv.org/abs/2007.12966}{{ arXiv:2007.12966 [hep-th]}}.

\bibitem{goncharov:hyperlogs}
A.~B. Goncharov, ``{Multiple Polylogarithms, Cyclotomy and Modular
  Complexes},''
  \href{http://dx.doi.org/http://dx.doi.org/10.4310/MRL.1998.v5.n4.a7}{{\em
  Mathematical Research Letters} {\bf 5} (1998) no. 4, 497--516}.

\bibitem{Remiddi:1999ew}
E.~Remiddi and J.~A.~M. Vermaseren, ``{Harmonic Polylogarithms},''
  \href{http://dx.doi.org/10.1142/S0217751X00000367}{{\em Int. J. Mod. Phys. A}
  {\bf 15} (2000)  725--754}, \href{http://arxiv.org/abs/hep-ph/9905237}{{
  arXiv:hep-ph/9905237}}.

\bibitem{Maitre:2005uu}
D.~Maitre, ``{\texttt{HPL}, a \textsc{Mathematica} Implementation of the
  Harmonic Polylogarithms},''
  \href{http://dx.doi.org/10.1016/j.cpc.2005.10.008}{{\em Comput. Phys.
  Commun.} {\bf 174} (2006)  222--240},
  \href{http://arxiv.org/abs/hep-ph/0507152}{{ arXiv:hep-ph/0507152}}.

\bibitem{Chen:1977oja}
K.-T. Chen, ``{Iterated Path Integrals},''
  \href{http://dx.doi.org/10.1090/S0002-9904-1977-14320-6}{{\em Bull. Am. Math.
  Soc.} {\bf 83} (1977)  831--879}.

\bibitem{Brown:2004ugm}
F.~C.~S. Brown, ``{Polylogarithmes Multiples Uniformes en Une Variable},''
  \href{http://dx.doi.org/10.1016/j.crma.2004.02.001}{{\em Compt. Rend. Math.}
  {\bf 338} (2004) no. 7, 527--532}.

\bibitem{Goncharov:2009kx}
A.~B. {Goncharov}, ``{A Simple Construction of Grassmannian Polylogarithms},''
  \href{http://dx.doi.org/10.1016/j.aim.2013.03.018}{{\em {Adv. Math.}} {\bf
  241} (2013)  79--102}, \href{http://arxiv.org/abs/0908.2238}{{
  arXiv:0908.2238 [math.AG]}}.

\bibitem{Brown:2013gia}
F.~Brown, ``{Single-Valued Motivic Periods and Multiple Zeta Values},''
  \href{http://dx.doi.org/10.1017/fms.2014.18}{{\em SIGMA} {\bf 2} (2014)
  e25}, \href{http://arxiv.org/abs/1309.5309}{{ arXiv:1309.5309 [math.NT]}}.

\bibitem{Brown:2015ztw}
F.~Brown and O.~Schnetz, ``{Single-Valued Multiple Polylogarithms and a Proof
  of the Zig{\textendash}Zag Conjecture},''
  \href{http://dx.doi.org/10.1016/j.jnt.2014.09.007}{{\em J. Number Theor.}
  {\bf 148} (2015)  478--506}.

\bibitem{James:Hegelisms}
W.~James, ``{On Some Hegelisms},'' {\em Mind} {\bf 7} (1882) no. 26, 186--208.
  \url{http://www.jstor.org/stable/2246448}.

\bibitem{Arkani-Hamed:2017ahv}
N.~Arkani-Hamed and E.~Y. Yuan, ``{One-Loop Integrals from Spherical
  Projections of Planes and Quadrics},''
  \href{http://arxiv.org/abs/1712.09991}{{ arXiv:1712.09991 [hep-th]}}.

\bibitem{Bourjaily:2019exo}
J.~L. Bourjaily, E.~Gardi, A.~J. McLeod, and C.~Vergu, ``{All-Mass $n$-gon
  Integrals in $n$ Dimensions},''
  \href{http://dx.doi.org/10.1007/JHEP08(2020)029}{{\em JHEP} {\bf 08} (2020)
  no. 08, 029}, \href{http://arxiv.org/abs/1912.11067}{{ arXiv:1912.11067
  [hep-th]}}.

\bibitem{Bourjaily:2021vyj}
J.~L. Bourjaily, N.~Kalyanapuram, C.~Langer, and K.~Patatoukos, ``{Prescriptive
  Unitarity with Elliptic Leading Singularities},''
  \href{http://arxiv.org/abs/2102.02210}{{ arXiv:2102.02210 [hep-th]}}.

\bibitem{Bourjaily:2022tep}
J.~L. Bourjaily and N.~Kalyanapuram, ``{The Stratification of Rigidity},''
  \href{http://dx.doi.org/10.1007/JHEP11(2022)084}{{\em JHEP} {\bf 11} (2022)
  084}, \href{http://arxiv.org/abs/2207.00596}{{ arXiv:2207.00596 [hep-th]}}.

\bibitem{Bourjaily:2020hjv}
J.~L. Bourjaily, N.~Kalyanapuram, C.~Langer, K.~Patatoukos, and M.~Spradlin,
  ``{An Elliptic Yangian-Invariant, `Leading Singularity'},''
  \href{http://dx.doi.org/10.1103/PhysRevLett.126.201601}{{\em Phys. Rev.
  Lett.} {\bf 126} (2021) no. 20, 201601},
  \href{http://arxiv.org/abs/2012.14438}{{ arXiv:2012.14438 [hep-th]}}.

\bibitem{Weinberg:2010fx}
S.~Weinberg, ``{Six-Dimensional Methods for Four-Dimensional Conformal Field
  Theories},'' \href{http://dx.doi.org/10.1103/PhysRevD.82.045031}{{\em Phys.
  Rev.} {\bf D82} (2010)  045031},
\href{http://arxiv.org/abs/1006.3480}{{ arXiv:1006.3480 [hep-th]}}.

\bibitem{SimmonsDuffin:2012uy}
D.~Simmons-Duffin, ``{Projectors, Shadows, and Conformal Blocks},''
  \href{http://dx.doi.org/10.1007/JHEP04(2014)146}{{\em JHEP} {\bf 04} (2014)
  146},
\href{http://arxiv.org/abs/1204.3894}{{ arXiv:1204.3894 [hep-th]}}.

\bibitem{Bourjaily:2015jna}
J.~L. Bourjaily and J.~Trnka, ``{Local Integrand Representations of All
  Two-Loop Amplitudes in Planar SYM},''
  \href{http://dx.doi.org/10.1007/JHEP08(2015)119}{{\em JHEP} {\bf 08} (2015)
  119},
\href{http://arxiv.org/abs/1505.05886}{{ arXiv:1505.05886 [hep-th]}}.

\bibitem{ArkaniHamed:2010gh}
N.~Arkani-Hamed, J.~L. Bourjaily, F.~Cachazo, and J.~Trnka, ``{Local Integrals
  for Planar Scattering Amplitudes},''
  \href{http://dx.doi.org/10.1007/JHEP06(2012)125}{{\em JHEP} {\bf 1206} (2012)
   125},
\href{http://arxiv.org/abs/1012.6032}{{ arXiv:1012.6032 [hep-th]}}.

\bibitem{Bourjaily:2013mma}
J.~L. Bourjaily, S.~Caron-Huot, and J.~Trnka, ``{Dual-Conformal Regularization
  of Infrared Loop Divergences and the {\it Chiral} Box Expansion},''
  \href{http://dx.doi.org/10.1007/JHEP01(2015)001}{{\em JHEP} {\bf 1501} (2015)
   001},
\href{http://arxiv.org/abs/1303.4734}{{ arXiv:1303.4734 [hep-th]}}.

\bibitem{Bourjaily:2017wjl}
J.~L. Bourjaily, E.~Herrmann, and J.~Trnka, ``{Prescriptive Unitarity},''
  \href{http://dx.doi.org/10.1007/JHEP06(2017)059}{{\em JHEP} {\bf 06} (2017)
  059},
\href{http://arxiv.org/abs/1704.05460}{{ arXiv:1704.05460 [hep-th]}}.

\bibitem{Bourjaily:2021hcp}
J.~L. Bourjaily, C.~Langer, and Y.~Zhang, ``{Illustrations of Integrand-Basis
  Building at Two Loops},'' \href{http://arxiv.org/abs/2112.05157}{{
  arXiv:2112.05157 [hep-th]}}.

\bibitem{Bourjaily:2019iqr}
J.~L. Bourjaily, E.~Herrmann, C.~Langer, A.~J. McLeod, and J.~Trnka,
  ``{Prescriptive Unitarity for Non-Planar Six-Particle Amplitudes at Two
  Loops},'' \href{http://dx.doi.org/10.1007/JHEP12(2019)073}{{\em JHEP} {\bf
  12} (2019)  073}, \href{http://arxiv.org/abs/1909.09131}{{ arXiv:1909.09131
  [hep-th]}}.

\bibitem{Bourjaily:2019gqu}
J.~L. Bourjaily, E.~Herrmann, C.~Langer, A.~J. McLeod, and J.~Trnka,
  ``{All-Multiplicity Nonplanar Amplitude Integrands in Maximally
  Supersymmetric Yang-Mills Theory at Two Loops},''
  \href{http://dx.doi.org/10.1103/PhysRevLett.124.111603}{{\em Phys. Rev.
  Lett.} {\bf 124} (2020) no. 11, 111603},
  \href{http://arxiv.org/abs/1911.09106}{{ arXiv:1911.09106 [hep-th]}}.

\bibitem{Bourjaily:2021iyq}
J.~L. Bourjaily, C.~Langer, and Y.~Zhang, ``{All Two-Loop, Color-Dressed,
  Six-Point Amplitude Integrands in sYM},''
  \href{http://dx.doi.org/10.1103/PhysRevD.105.105015}{{\em Phys. Rev. D} {\bf
  105} (2022) no. 10, 105015}, \href{http://arxiv.org/abs/2112.06934}{{
  arXiv:2112.06934 [hep-th]}}.

\bibitem{BCF}
R.~Britto, F.~Cachazo, and B.~Feng, ``{New Recursion Relations for Tree
  Amplitudes of Gluons},''
  \href{http://dx.doi.org/10.1016/j.nuclphysb.2005.02.030}{{\em Nucl.Phys.}
  {\bf B715} (2005)  499--522},
\href{http://arxiv.org/abs/hep-th/0412308}{{ arXiv:hep-th/0412308 [hep-th]}}.

\bibitem{BCFW}
R.~Britto, F.~Cachazo, B.~Feng, and E.~Witten, ``{Direct Proof of Tree-Level
  Recursion Relation in Yang- Mills Theory},''
  \href{http://dx.doi.org/10.1103/PhysRevLett.94.181602}{{\em Phys. Rev. Lett.}
  {\bf 94} (2005)  181602},
\href{http://arxiv.org/abs/hep-th/0501052}{{ arXiv:hep-th/0501052}}.

\bibitem{CaronHuot:2010zt}
S.~Caron-Huot, ``{Loops and Trees},''
  \href{http://dx.doi.org/10.1007/JHEP05(2011)080}{{\em JHEP} {\bf 1105} (2011)
   080},
\href{http://arxiv.org/abs/1007.3224}{{ arXiv:1007.3224 [hep-ph]}}.

\bibitem{ArkaniHamed:2010kv}
N.~Arkani-Hamed, J.~L. Bourjaily, F.~Cachazo, S.~Caron-Huot, and J.~Trnka,
  ``{The All-Loop Integrand For Scattering Amplitudes in Planar
  $\mathcal{N}\!=\!4$ SYM},''
  \href{http://dx.doi.org/10.1007/JHEP01(2011)041}{{\em JHEP} {\bf 1101} (2011)
   041},
\href{http://arxiv.org/abs/1008.2958}{{ arXiv:1008.2958 [hep-th]}}.

\bibitem{Bourjaily:2023apy}
J.~L. Bourjaily and S.~Caron-Huot, ``{Loops from Cuts},''
  \href{http://dx.doi.org/10.1103/PhysRevD.108.025008}{{\em Phys. Rev. D} {\bf
  108} (2023) no. 2, 025008}, \href{http://arxiv.org/abs/2305.01673}{{
  arXiv:2305.01673 [hep-th]}}.

\bibitem{bourjaily:amplitudesInNewBasis}
J.~L. Bourjaily, ``{Manifestly Matching Every Two-Loop Cut of sYM}.'' To
  appear.

\bibitem{bourjaily:IRstructureOfamplitudes}
J.~L. Bourjaily, ``{Infrared Structure of Amplitudes at Two Loops}.'' In prep.

\bibitem{Bourjaily:2025tpr}
J.~L. Bourjaily, ``{The Supersymmetry of Cuts in Pure Gauge Theory and
  Gravity},'' \href{http://arxiv.org/abs/2512.24984}{{ arXiv:2512.24984
  [hep-th]}}.

\bibitem{Bourjaily:2021ujs}
J.~L. Bourjaily, E.~Herrmann, C.~Langer, K.~Patatoukos, J.~Trnka, and M.~Zheng,
  ``{Integrands of Less-Supersymmetric Yang-Mills at One Loop},''
  \href{http://dx.doi.org/10.1007/JHEP03(2022)126}{{\em JHEP} {\bf 03} (2022)
  126}, \href{http://arxiv.org/abs/2112.06901}{{ arXiv:2112.06901 [hep-th]}}.

\bibitem{Bourjaily:2026adf}
J.~L. Bourjaily, M.~Plesser, and P.~Velie, ``{Tree Amplitudes with Charged
  Matter in Pure Gauge Theory},'' \href{http://arxiv.org/abs/2604.18672}{{
  arXiv:2604.18672 [hep-th]}}.

\bibitem{Kotikov:1990kg}
A.~V. Kotikov, ``{Differential Equations Method: New Technique for Massive
  Feynman Diagrams Calculation},''
\href{http://dx.doi.org/10.1016/0370-2693(91)90413-K}{{\em Phys. Lett.} {\bf
  B254} (1991)  158--164}.

\bibitem{Bern:1992em}
Z.~Bern, L.~J. Dixon, and D.~A. Kosower, ``{Dimensionally Regulated One Loop
  Integrals},'' \href{http://dx.doi.org/10.1016/0370-2693(93)90400-C}{{\em
  Phys. Lett.} {\bf B302} (1993)  299--308},
\href{http://arxiv.org/abs/hep-ph/9212308}{{ arXiv:hep-ph/9212308}}.

\bibitem{Gehrmann:1999as}
T.~Gehrmann and E.~Remiddi, ``{Differential Equations for Two-Loop Four-Point
  Functions},'' \href{http://dx.doi.org/10.1016/S0550-3213(00)00223-6}{{\em
  Nucl. Phys. B} {\bf 580} (2000)  485--518},
  \href{http://arxiv.org/abs/hep-ph/9912329}{{ arXiv:hep-ph/9912329}}.

\bibitem{Bourjaily:2018aeq}
J.~L. Bourjaily, A.~J. McLeod, M.~von Hippel, and M.~Wilhelm, ``{Rationalizing
  Loop Integration},'' \href{http://dx.doi.org/10.1007/JHEP08(2018)184}{{\em
  JHEP} {\bf 08} (2018)  184}, \href{http://arxiv.org/abs/1805.10281}{{
  arXiv:1805.10281 [hep-th]}}.

\bibitem{Bourjaily:2019jrk}
J.~L. Bourjaily, F.~Dulat, and E.~Panzer, ``{Manifestly Dual-Conformal Loop
  Integration},'' \href{http://dx.doi.org/10.1016/j.nuclphysb.2019.03.022}{{\em
  Nucl. Phys.} {\bf B942} (2019)  251--302},
\href{http://arxiv.org/abs/1901.02887}{{ arXiv:1901.02887 [hep-th]}}.

\bibitem{Bourjaily:2019vby}
J.~L. Bourjaily, M.~Volk, and M.~Von~Hippel, ``{Conformally Regulated Direct
  Integration of the Two-Loop Heptagon Remainder},''
  \href{http://dx.doi.org/10.1007/JHEP02(2020)095}{{\em JHEP} {\bf 02} (2020)
  095}, \href{http://arxiv.org/abs/1912.05690}{{ arXiv:1912.05690 [hep-th]}}.

\bibitem{Bourjaily:2021lnz}
J.~L. Bourjaily, Y.-H. He, A.~J. McLeod, M.~Spradlin, C.~Vergu, M.~Volk, M.~von
  Hippel, and M.~Wilhelm,
  \href{http://dx.doi.org/10.1007/978-3-030-80219-6_5}{``{Direct Integration
  for Multi-Leg Amplitudes: Tips, Tricks, and When They Fail},''} in {\em
  {Antidifferentiation and the Calculation of Feynman Amplitudes}}.
\newblock 3, 2021.
\newblock \href{http://arxiv.org/abs/2103.15423}{{ arXiv:2103.15423 [hep-th]}}.

\bibitem{Spiering:2024sea}
A.~Spiering, M.~Wilhelm, and C.~Zhang, ``{All Planar Two-Loop Amplitudes in
  Maximally Supersymmetric Yang-Mills Theory},''
  \href{http://dx.doi.org/10.1103/PhysRevLett.134.071602}{{\em Phys. Rev.
  Lett.} {\bf 134} (2025) no. 7, 071602},
  \href{http://arxiv.org/abs/2406.15549}{{ arXiv:2406.15549 [hep-th]}}.

\bibitem{Drummond:2009fd}
J.~M. Drummond, J.~M. Henn, and J.~Plefka, ``{Yangian Symmetry of Scattering
  Amplitudes in $\mathcal{N}\!=\!4$ Super Yang-Mills Theory},''
  \href{http://dx.doi.org/10.1088/1126-6708/2009/05/046}{{\em JHEP} {\bf 05}
  (2009)  046},
\href{http://arxiv.org/abs/0902.2987}{{ arXiv:0902.2987 [hep-th]}}.

\bibitem{Bourjaily:2018omh}
J.~L. Bourjaily, E.~Herrmann, and J.~Trnka, ``{Amplitudes at Infinity},''
  \href{http://dx.doi.org/10.1103/PhysRevD.99.066006}{{\em Phys. Rev. D} {\bf
  99} (2019) no. 6, 066006}, \href{http://arxiv.org/abs/1812.11185}{{
  arXiv:1812.11185 [hep-th]}}.

\bibitem{Cramer1750}
G.~Cramer, {\em {Introduction A L'Analyse Des Lignes Courbes Alg{\'ebriques}}}.
\newblock Chez les Fr{\`eres} Cramer \& Cl. Philibert, Gen{\`e}ve, 1750.

\bibitem{Gram1883}
J.~Gram, ``{\"{U}ber die Entwickelung Reeller Functionen in Reihen Mittelst der
  Methode der Kleinsten Quadrate},'' {\em Journal f{\"u}r die reine und
  angewandte Mathematik} {\bf 94} (1883)  41--73.
  \url{http://eudml.org/doc/148523}.

\bibitem{Bourjaily:2010wh}
J.~L. Bourjaily, ``{Efficient Tree-Amplitudes in $\mathcal{N}\!=\!4$: Automatic
  BCFW Recursion in {\sc Mathematica}},''
\href{http://arxiv.org/abs/1011.2447}{{ arXiv:1011.2447 [hep-ph]}}.

\bibitem{Bourjaily:2023uln}
J.~L. Bourjaily, ``{Computational Tools for Trees in Gauge Theory and
  Gravity},'' \href{http://arxiv.org/abs/2312.17745}{{ arXiv:2312.17745
  [hep-th]}}.

\end{thebibliography}
\addtocontents{toc}{\protect~\\[-32pt]}\addcontentsline{toc}{section}{References}\addtocontents{toc}{\protect~\\[-40pt]\mbox{\vspace{0pt}}\protect\par}
\addtocontents{toc}{\protect~\\[-32pt]}
\providecommand{\href}[2]{#2}\begingroup\raggedright\endgroup

\end{document}